\documentclass[%
 reprint,
superscriptaddress,
 amsmath,amssymb,
 aps,
floatfix,
]{revtex4-1}

\usepackage[version=3]{mhchem} 
\usepackage{float}
\usepackage{graphicx}
\usepackage{dcolumn}
\usepackage{bm}

\usepackage{xcolor}

\begin{document}

\preprint{APS/123-QED}
\title{Tunable Semiconductors: Control over Carrier States and Excitations in Layered Hybrid Organic-Inorganic Perovskites}

\author{Chi Liu}
\affiliation
{Department of Chemistry, Duke University, Durham, North Carolina 27708, USA}
\author{William Huhn}
\author{Ke-Zhao Du}
\affiliation
{Department of Mechanical Engineering and Materials Science, Duke University, Durham, North Carolina 27708, USA}
\author{Alvaro Vazquez-Mayagoitia}
\affiliation
{Argonne Leadership Computing Facility, 9700 S. Cass Avenue, Lemont IL 60439, USA}
\author{David Dirkes}
\affiliation
{Department of Chemistry, University of North Carolina, Chapel Hill, North Carolina 27599, USA}
\author{Wei You}
\affiliation
{Department of Chemistry, University of North Carolina, Chapel Hill, North Carolina 27599, USA}
\author{Yosuke Kanai}
\affiliation
{Department of Chemistry, University of North Carolina, Chapel Hill, North Carolina 27599, USA}
\author{David B. Mitzi}
\affiliation
{Department of Mechanical Engineering and Materials Science, Duke University, Durham, North Carolina 27708, USA}
\affiliation
{Department of Chemistry, Duke University, Durham, North Carolina 27708, USA}
\email
{david.mitzi@duke.edu}
\author{Volker Blum}
\affiliation
{Department of Mechanical Engineering and Materials Science, Duke University, Durham, North Carolina 27708, USA}
\affiliation
{Department of Chemistry, Duke University, Durham, North Carolina 27708, USA}
\email
{volker.blum@duke.edu}




\date{\today}

\begin{abstract}
  For a class of 2D hybrid organic-inorganic perovskite semiconductors based on $\pi$-conjugated organic cations, we predict quantitatively how varying the organic and inorganic component allows control over the nature, energy and localization of carrier states in a quantum-well-like fashion. Our first-principles predictions, based on large-scale hybrid density-functional theory with spin-orbit coupling, show that the interface between the organic and inorganic parts within a single hybrid can be modulated systematically, enabling us to select between different type-I and type-II energy level alignments. Energy levels, recombination properties and transport behavior of electrons and holes thus become tunable by choosing specific organic functionalizations and juxtaposing them with suitable inorganic components. 
\end{abstract}

\maketitle

\begin{figure}
\centering
\includegraphics[scale=1]{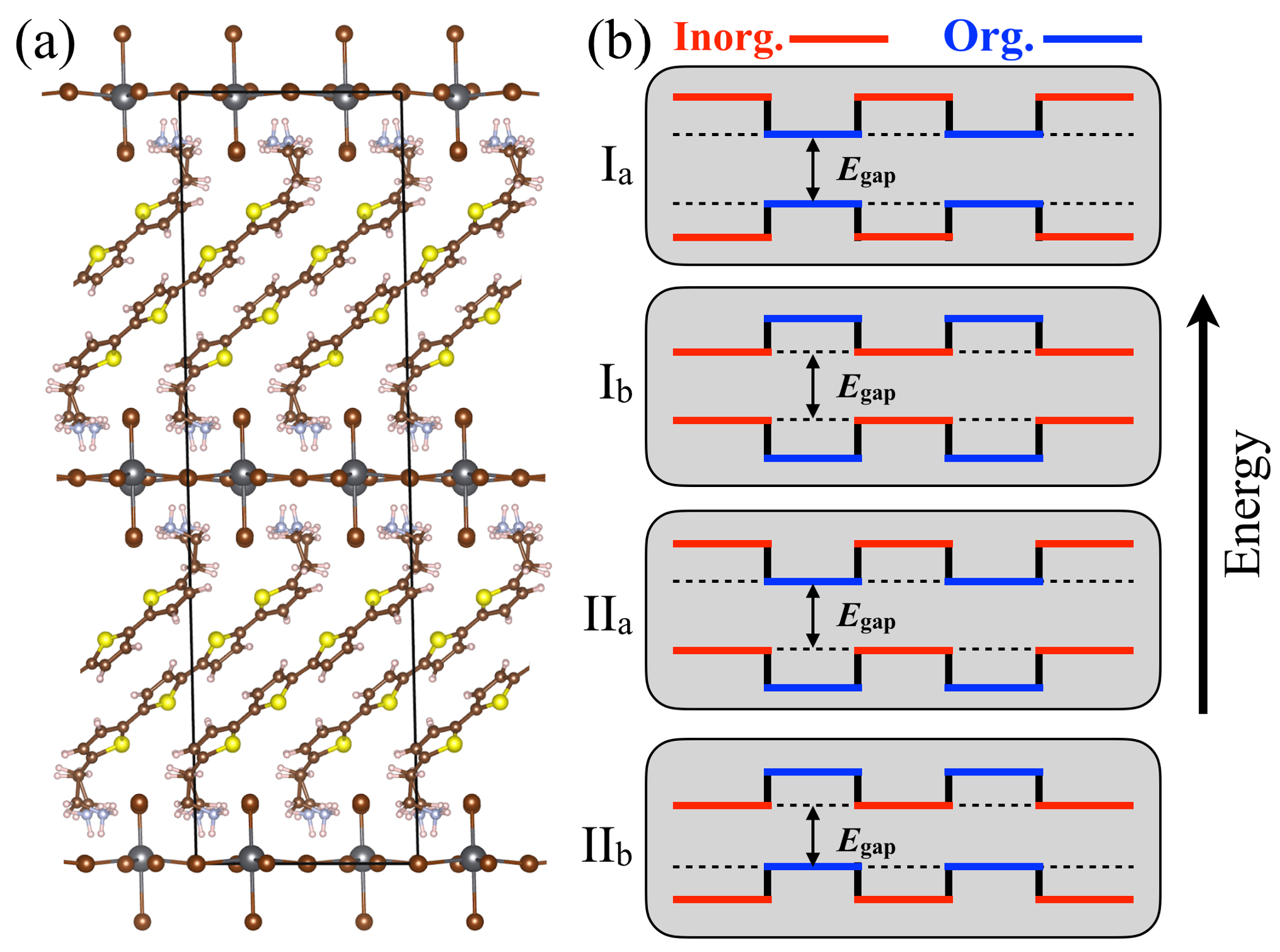}
\caption{Structure of AE4T\ce{PbBr4} fully relaxed by DFT-PBE+TS taking the experimental (X-ray diffraction) structure\cite{Mitzi1999AE4T} as the input. Possible energy level schemes (Ia, Ib, IIa, IIb) for the alternating organic-inorganic perovskite structure are shown, with the overall band gap indicated by arrows and dashed lines.}
\label{fig:AE4TPbBr4}
\end{figure}

Hybrid organic-inorganic perovskites
(HOIPs),\cite{Mitzi2016,Li2017natureFunctional} particularly  
three-dimensional (3D) HOIPs, are currently experiencing a strong
revival in interest as economically processable, optically active
semiconductor materials with excellent transport
characteristics. Their success is showcased most prominently by record
performance gains in proof-of-concept photovoltaic\cite{MAPbI3_2009,
  Heo2013, Liu2013, Jeon2014, Green2014,
  Gratzel2014,C3EE43822H,Yang1234,Yang1376,Jodlowski2017} and light-emitting
devices.\cite{Tan2014NatureNT,Wang2015ADMA,emitAdvMat15,
  Cho2015Science,emit16pnas,emit17nature,emitAdvMat17, emitTuneMaterials}  
The electronic function of 3D HOIPs can be tuned to a limited extent
by manipulating the inorganic component (from which the frontier
orbitals are derived), but the organic cations are confined by the 3D
structure and are thus necessarily small (e.g.,
methylammonium\cite{MAPbI3_2009,
  Heo2013, Liu2013, Jeon2014, Green2014,
  Gratzel2014,emit16pnas, Tan2014NatureNT,Wang2015ADMA,emitAdvMat15,Cho2015Science,emit17nature} and
formamidinium\cite{C3EE43822H,Yang1234,Yang1376, emitAdvMat17,MITZI1997376, fapbi3}), with
electronic levels that do not contribute directly to the electronic
functionality.\cite{Umebayashi2003,Chiarella2008,PRB2014Filippetti,walsh2013,Schilfgaarde2014,Motta2015}
However, the accessible chemical space of HOIPs extends well beyond
the 3D systems.\cite{Mitzi2016} In particular, the layered, so-called two-dimensional
(2D) perovskites do not place a strict length constraint on the organic
cation. In these materials, a much broader range of functional organic
molecules can be incorporated within the inorganic scaffolds, including complex functional molecules such as oligo-acene or -thiophene
derivatives.\cite{Mitzi2016, Era1998,brauna_cpl_1999_acene_PbCl,mitzi_chemmat_03_SnI_acene,EmaPrl2008,kezhao, Mitzi1999AE4T, mitzi_ic_2000_Bi_AE4T,zhu_ic_2003_2T_HOIP, era_thin_sol_film_03_poly_thiophene}
Fig. \ref{fig:AE4TPbBr4}a shows the atomic structure of a paradigmatic
example of such a 2D HOIP with active organic functionality,
bis(aminoethyl)-quaterthiophene lead bromide
AE4T\ce{PbBr4}.\cite{Mitzi1999AE4T} Similar juxtapositions of targeted
organic and inorganic components give rise to a vast, yet
systematically accessible space of possible semiconductor
materials,\cite{Mitzi2016, Li2017natureFunctional,
  Mitzi1999ProgInorgChem, Mitzi2001Dalton,
  Zhang2009} including those in which
the molecular carrier levels contribute directly to the low-lying
excitations and carrier levels.\cite{Mitzi2016, 2DhoipOrg2017chemchem,
  EmaPrl2008, Mitzi1999AE4T, brauna_cpl_1999_acene_PbCl, Mitzi1999ProgInorgChem, Mitzi2001Dalton, mitzi_chemmat_03_SnI_acene} This large space of
conceivable organic-inorganic combinations thus offers the unique
opportunity to tailor (ideally with computational guidance) materials
with particularly desirable semiconductor properties, by intentionally
controlling the spatial location and character of the electronic
carriers and optical excitations throughout the material.  

A key physical prerequisite to manipulate the
semiconductor properties of layered hybrid materials is to understand
the nature and spatial localization of their carriers and
excitations. Specifically, the question of whether 
and how exactly one can understand their properties in analogy to
quantum wells with varying confinement barriers (often assumed
\cite{Ishihara1989,Era1994APL,Zhang2009,emit16nature}) 
is subject to discussion in the
literature.\cite{Mitzi2001, prb95Exciton,even14cpc,even15jpcc} 
Fig. \ref{fig:AE4TPbBr4}b exemplifies the principle by comparing four
different, conceivable quantum-well like situations: ``Type I'', with
low-energy electrons and holes localized on the same component (either
organic, Ia or inorganic, Ib) or ``Type II'', with electrons and holes
on different components (holes on inorganic, electrons on organic, IIa,
or vice versa, IIb).
While some simple
layered HOIPs have been successfully explained in a type-Ib picture
(inorganic band edges with the organic acting as a quasi-inert
screening
medium\cite{Ishihara1989,Era1994APL,prb95Exciton,Zhang2009,emit16nature,even14cpc,even15jpcc}),
a fully quantitative 
understanding of both band gaps within and band alignment between the
materials' components is essential to recover the larger set
of possibilities in Fig.~\ref{fig:AE4TPbBr4}b. Providing this
understanding for the large, complex crystal structures at hand
constitutes a substantial challenge for current theory, both regarding
computational resources and sufficiently accurate approximations. In
this paper, we demonstrate that these challenges can now be met,
enabling us to answer questions that are central to future targeted
developments of new HOIP semiconductors: (1) Can
these structures be understood as quantum-well-like 
structures with spatially well-separated levels in the organic
vs. inorganic components, or will the electronic states be hybridized
and thus delocalized across both components? (2) What is the spatial
nature of electron/hole carriers within the structure? For instance,
do they tend to migrate to the organic or the inorganic hybrid
component in the lowest-energy configuration, drastically affecting
each carrier's transport properties (band-like inorganic
vs. hopping-like organic)? (3) To what extent can we rationally tune
the carrier and excitation properties by independently varying the
organic and inorganic components?

Whether we can uncover a new
paradigm using 2D HOIPs as ``semiconductors on demand'' with finely
tunable properties and high-precision crystalline structural control
depends on building a design principle that relates complex hybrid
atomic structures to optoelectronic properties through answering the
questions above.  
In this work, we do so through a first-principles theoretical
examination of a class of oligothiophene-based 2D HOIPs, expanding on
the AE4T\ce{PbBr4} compound shown in Fig.\ref{fig:AE4TPbBr4}a. A {\color{black} practical}
challenge for theory is the structural complexity of these 2D HOIPs
for which the unit cells are large. For instance, a (2$\times$2)
lateral supercell of the perovskite layer (Figures S1, S2, S3 in
the Supplemental Material (SM) \cite{supplemental})
in this and other structures considered in this work is needed to
cover the experimentally correct perovskite layer distortion and
molecular arrangement, leading to 424 atoms in the simulation cell for
AE4T\ce{PbBr4}. The (2$\times$2) supercell instead of the
experimentally reported (1$\times2$) structure is necessary with
regards to both accessing an energetically lower structure (Table S1
\cite{supplemental}) and also removing the disordering in the inorganic and organic
structural components in the experimental structure.
In addition, the two inorganic layers in the (1$\times$2) relaxed
structure have different Pb-Br-Pb angles, which disagrees with the
experimental structure of AE4T\ce{PbBr4} (Fig. S4 \cite{supplemental}). For
structure predictions that capture the subtle balance of different
molecular and inorganic bonding contributions, we use van der Waals
corrected semilocal density-functional theory (DFT--i.e.,
Perdew-Burke-Ernzerhof (PBE) exchange-correlation
functional,\cite{PBE} plus the Tkatchenko-Scheffler (TS) pairwise
dispersion scheme{\color{black}\cite{Tkatchenko09}}). Electronic properties require a higher
level of theory for qualitatively correct results, but the most attractive
first-principles many-body approaches such as the $GW$
approximation\cite{Ren2012NJP, reining18gw} remain out of reach for
structures of this size. For band structure predictions, we therefore
resort to the still demanding level of hybrid DFT using the
Heyd-Scuseria-Ernzerhof (HSE06) functional\cite{hse03, hse06})
including spin-orbit coupling (SOC)\cite{Huhn2017}. Importantly, and
unlike semilocal DFT,
hybrid DFT in principle contains the right
physics\cite{Perdew14032017} to capture the frontier energy levels
(valence band maximum (VBM) and conduction band minimum
(CBM)). Including SOC is critical to capture correctly the qualitative
underlying nature of carriers, changing the nature of the CBM from
``organic'' to ``inorganic'' in AE4T\ce{PbBr4} and reducing the band
gap by $\sim$ 0.3 eV (Fig. S5 \cite{supplemental}).   

All calculations are performed by the FHI-aims code,\cite{Blum2009,
  Havu2009, Ren2012NJP, LEVCHENKO201560, Igor2015NJP, Huhn2017,Ihrig2015}
using the ELSI infrastructure\cite{YU2017} and ELPA eigenvalue
solver\cite{elpa2014} for massively parallel simulations. For all
crystal geometries, we employ full relaxation of unit cell parameters
and cell-internal atomic coordinates\cite{KNUTH201533} using the
FHI-aims ``tight'' numerical defaults (Table S2 \cite{supplemental}) and k-point
grids settings of 1$\times$2$\times$2 (Table S3 \cite{supplemental}). For band
structures, FHI-aims' ``intermediate'' settings (Table S4 \cite{supplemental}) and
dense k-point grids of 3$\times$3$\times$3 are used. The exchange
mixing parameter in HSE06 was kept at 25\% and the screening parameter
at $\Omega$ = 0.11 Bohr$^{-1}$ {\color{black}\cite{Krukau06}} in order to retain a single consistent
base to compare energy band structures across different materials in
this work. We first validate this approach for the low-temperature
orthorhombic phase of \ce{MAPbI3} (Fig. S6a \cite{supplemental}). The lattice parameters predicted by
DFT-PBE+TS agree with the experimental values {\color{black}\cite{Weller2015}} to within {\color{black} 1.4\%}
(Fig. S6b \cite{supplemental}). The HSE06+SOC approach predicts a direct band gap of
1.42 eV, which underestimates the experimental value ({\color{black}1.65-1.68 eV}, Fig. S6b \cite{supplemental}){by 0.2-0.3~eV.\color{black}\cite{Kong2015,Phuong2016}}.
Details of how we constructed
the computational models for all structures considered in this work
can be found in the SM, section IX.\cite{supplemental} 
For AE4T\ce{PbBr4}
(Fig. \ref{fig:AE4TPbBr4}a), deviations of any unit cell parameters of
the resulting predicted structure compared to experiment are 1.2\% or
better, i.e., they are in excellent agreement (Table S5 \cite{supplemental}). Finally, a new crystalline sample of AE4T\ce{PbI4} was grown and
an X-ray structure refinement performed (SM section X \cite{supplemental}), also
indicating excellent agreement with the DFT-predicted structure used
in the analyses below (Table S5 \cite{supplemental}). 
\begin{figure*}
\includegraphics[width=\textwidth]{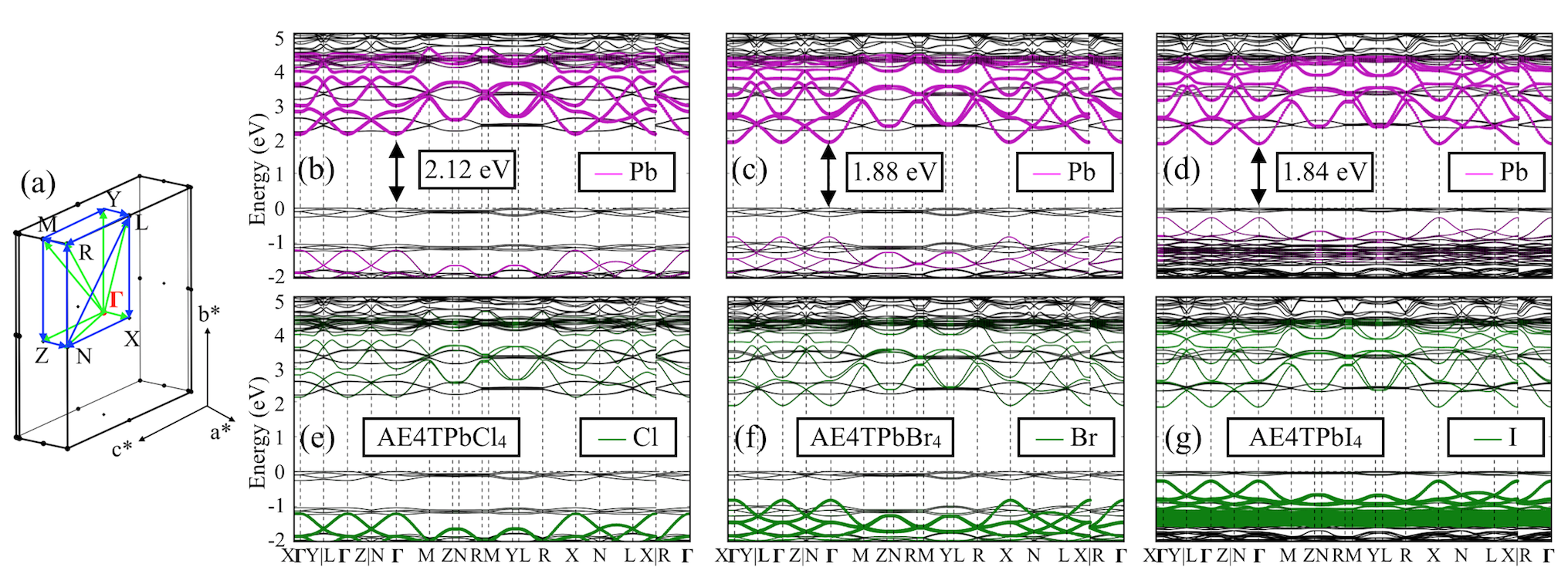}
\caption{Band structures of AE4T\ce{PbX4} (X = Cl, Br, I) calculated by DFT-HSE06+SOC with the states of Pb (b-d) and halogen (e-g) identified by projected density of states onto different species. The K path in the reciprocal space is shown in subfigure (a). {\color{black} The energy zero in (b-g) is set equal to the valence band maximum (for internal alignments relative to Pb 1$s$ levels, see Figure~\ref{fig:energy_levels_org}a below).}}
\label{fig:AE4TPbX4}
\end{figure*}

{\color{black} Turning first to energy level localization, orbital plots (see Figures
S9, S10, S11 \cite{supplemental} for example orbitals of
AE4T\ce{PbX4}, X = Cl, Br, I) show that the
states associated with inorganic and organic components are spatially
well-separated, supporting the
notion of ``quantum-well like'' states in these materials.} This answers
question (1) above and validates a discussion in terms of
separate ``inorganic'' and ``organic'' bands. 
Even {\it et al.}\cite{even14cpc, even15jpcc} have also considered 2D
HOIPs from the perspective of semiconductor quantum wells, showing
that the effective mass model may fail due to the absence of
superlattice coupling and importance of
non-parabolicity. They proposed a computational analysis in terms of
separate, neutralized organic and inorganic layers, appropriate for
type-Ib situations. In the current work, we cover the full set of materials
directly, allowing us to assess band gaps within each component as
well as the alignments of their electronic levels. Knowing the alignments enables us to
assess the full space of possible HOIP semiconductors (e.g., type-I  
and II), where both the inorganic and the organic components are
electronically active. 

The halogen atoms in the inorganic framework offer a convenient handle
for tailoring the associated electronic structure of the inorganic
component by varying it from Br to Cl and I.\cite{Mitzi1999AE4T} Full
band structures for the compound series AE4T\ce{PbX4} (X = Cl, Br, I)
are shown in Fig. \ref{fig:AE4TPbX4}. All three compounds have a
direct band gap at the $\Gamma$ point. By changing the halogen, the
dispersive bands originating from the inorganic component (Pb- and
halogen-derived states, colored lines in Figures
\ref{fig:AE4TPbX4}b-g) shift substantially with respect to the organic
bands.  In contrast, the organic-derived bands (black lines in Figures
\ref{fig:AE4TPbX4}b-g) vary only slightly among these three
compounds. {\color{black} Full and partial densities of states for these
  and other compounds in Figures S14-S15
  \cite{supplemental} corroborate the chemical makeup shown in the band
  structures.}
{\color{black} Band curvature parameters (Table S6) that correspond to
  the diagonal elements of the effective mass
  tensors\cite{tong2017,Roethlisberger2017,Feng2014} in the 
  reciprocal-space coordinate system of Fig. \ref{fig:AE4TPbX4}a
  confirm some qualitative trends emerging from the band structures:
  Uniformly flat bands (effective masses $>$20~$m_e$) perpendicular to
  the perovskite planes indicate hindered, non-bandlike
  transport. Somewhat lower effective mass tensor elements
  (2.2-11.4~$m_e$, still higher than in typical semiconductor
  materials) emerge for the holes (VBM) on the organic 
  components parallel to the planes. Low effective mass tensor
  elements, $\approx$0.2-0.5~$m_e$, for the electrons (CBM) along the
  inorganic planes, in the range typical of 3D 
  perovskites\cite{Roethlisberger2017,Feng2014} might, absent other
  detrimental factors, indicate relatively easy electron transport.}

\begin{figure}
\centering
\includegraphics[scale=1]{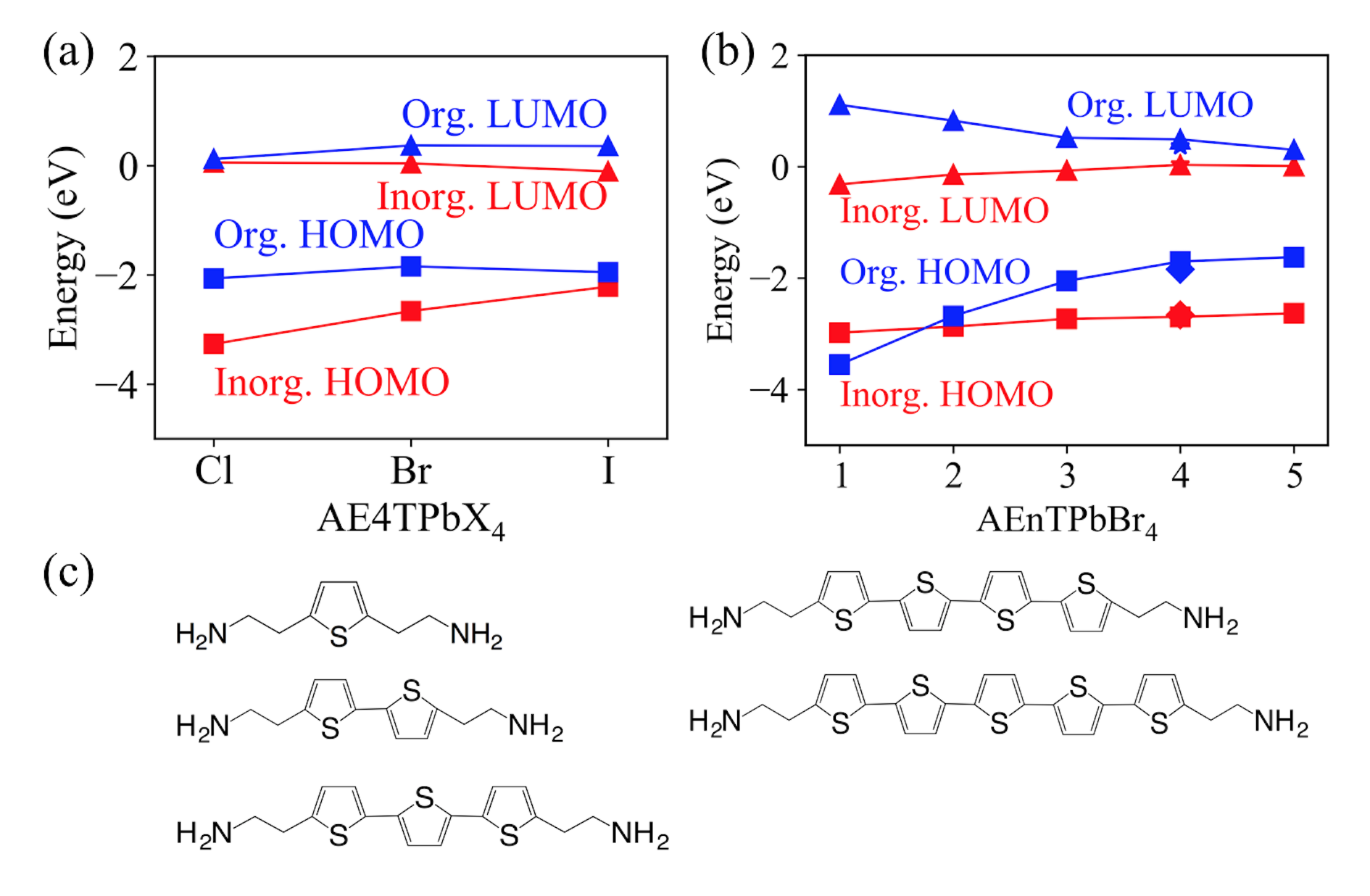}
\caption{(a) Frontier energy levels of the organic and inorganic
  components at the $\Gamma$ point among the series of AE4T\ce{PbX4}
  (X = Cl, Br, I). (b) Frontier energy levels at the $\Gamma$
  point among the series of AEnT\ce{PbBr4} (n = 1, 2, 3, 4, 5). Stars
  and diamonds indicate the energy levels of syn-anti-syn
  AE4T\ce{PbBr4} for n = 4. {\color{black} The average of Pb 1$s$ energies is chosen to align the energy levels of different compounds.} (c) Oligothiophene-based organic molecules considered in the
  all-anti configuration for varying the number n.} 
\label{fig:energy_levels_org}
\end{figure}

The trends of the ``organic'' and ``inorganic'' frontier
energy levels are shown in Figure
\ref{fig:energy_levels_org}a. {\color{black} The average of Pb 1$s$ energy levels is chosen to formally align energy levels between different HOIPs in Fig.~\ref{fig:energy_levels_org}. However, we did not study how this choice (equivalent to the absence of dipolar fields between Pb ions across an interface between two different HOIPs) pertains to real interfaces between HOIPs and the conclusions of this work do not rely on this convention.} Replacing Br by Cl increases the overall
computed band gap from 1.88 eV to 2.12 eV, whereas the substitution by
I decreases the energy gap value to 1.84 eV. While the {\it inorganic}
energy gap changes drastically from 2.70 to 3.32/2.11 eV for Cl/I
substitutions (Fig. ~\ref{fig:energy_levels_org}), the associated
change in the {\it organic} energy gap is negligibly small ($\sim$0.1
eV). However, a drastic change evident from both Figures
\ref{fig:AE4TPbX4} and \ref{fig:energy_levels_org}a is the ordering of
the levels, particularly the electron-like (CBM) states when going
from Cl to Br and I. For Br and I, the band structures indicate Type
IIb (Fig. \ref{fig:AE4TPbBr4}) like quantum well behavior, i.e.,
electrons and holes are expected to be spatially well separated on the
inorganic and organic components, respectively. In contrast, the
organic and inorganic CBM levels are predicted to lie within a few
tens of meV for the Cl-substituted compounds, i.e., they are
essentially degenerate within the uncertainties of the HSE06+SOC
treatment employed here. AE4T\ce{PbCl4} is thus between types Ia and IIb in
Fig. \ref{fig:AE4TPbBr4}b and would allow electrons to travel to
either component with reasonable ease at finite temperature. This
difference would have profound implications for the expected carrier
recombination properties of all three compounds, as evidenced, e.g.,
in photoluminescence (PL). In fact, strong quaterthiophene PL emission
at $\sim$540 nm (2.30 eV) was experimentally observed for X =
Cl,\cite{Mitzi1999AE4T} whereas the analogous PL features are
substantially quenched for X = Br, I. Our present computational result
agrees with and explains this experimental observation. While the X =
Cl compound displays a near type-Ia band alignment, the X = Br, I
compounds are clearly type-IIb. In the latter two compounds, the
energy level alignment therefore
effectively impedes the electron-hole recombination since the
electrons and the holes are preferentially located across the
interface in the inorganic and organic hybrid components,
respectively, i.e., addressing question (2) from the introduction.

{\color{black} The importance of a fully predictive, quantitative theoretical treatment is further underscored by the fact that a discussion based on qualitative factors in 1999 led to the different conclusion of type-IIa, not type-IIb alignment for this compound.\cite{Mitzi1999AE4T}}{\color{black} We note that optical excitations in absorption or emission cannot be expected to be captured based on the band structures derived in this work alone, since the typically strong excitonic effects are not included. For instance, exciton binding energies up to 540~meV have been reported on the inorganic compound in 2D perovskites\cite{Mitzi2016,Kagan2018} and an exciton binding energy of 0.4~eV has been reported in organic (not hybrid) sexithiophene thin films.\cite{Hill2000} However, the qualitative localization of carriers prior to recombination (discussed above) still provides valuable insights into their expected recombination properties.}
{\color{black} We also note that the potential implications of being able to tune levels on the organic and inorganic components independently reach beyond optical properties alone, affecting (for example) transport properties, dopability, or band offsets (and thus potential energy losses in devices) between the components.}

We finally consider the ability to tune the band gap and quantum well
nature of the structure by varying the organic component, changing the
conjugation length of the oligothiophene molecules. As shown in
Fig. \ref{fig:energy_levels_org}b, we substitute the ``all-anti''
configuration (successive S atoms on alternating sides) of
bis-ethylamine terminated oligothiophene AEnT (n = 1, 2, 3, 4, 5,
Fig. \ref{fig:energy_levels_org}c) into the scaffold of
AE4T\ce{PbBr4}. While the quaterthiophene molecule in experimental
AE4T\ce{PbBr4} (Fig.~\ref{fig:AE4TPbBr4}) adopts a syn-anti-syn
configuration\cite{Mitzi1999AE4T}, this configuration cannot be
adopted by all AEnT (n = 1, 2, 3, 4, 5) oligothiophenes. For the
purpose of having a systematic assessment, we thus restrict this part
of our study to the all-anti configuration. Note, however, that the
predicted electronic properties are only expected to be
insignificantly affected by this conformational change as shown by the
additional symbols corresponding to the syn-anti-syn conformation for
n=4 (Fig. \ref{fig:energy_levels_org}b).  
The electronic band structures reveal direct gaps for all considered
conjugation lengths of \ce{AEnTPbBr4} (Fig. S12 \cite{supplemental})
{\color{black} and band curvature trends (Table S7) are broadly consistent with those
discussed for AE4T\ce{PbX4} above}. The overall
band gap decreases as n increases, i.e., 2.66, 2.54, 1.98, 1.73 and
1.63 for n = 1, 2, 3, 4, 5, respectively. The predicted AEnT\ce{PbBr4}
compounds for n = 2-5 yield type-IIb level alignments. However, the n
= 1 compound reveals a type-Ib alignment (both CBM and VBM derived
from the inorganic component). This behavior (type Ib) is consistent
with other 2D perovskites with smaller
organic functionality, in which carriers/excitons are mainly funneled
onto the 
inorganic subcomponent.\cite{even14cpc, even15jpcc, even2013, prb95Exciton}
We thus affirmatively answer question (3) above -- i.e., the carrier
nature and neutral excitation properties and overall gap can be varied
rationally by changing the organic component or the inorganic
component in 2D HOIPs independently. 

In summary, our results show that the quantum-well model can be used
for conceptual understanding and as a useful starting point as a design
principle for the layered HOIP family of hybrid materials. The
tunability of electronic properties, exemplified by the materials
studied in this work, opens up the possibility to computationally
predict and tailor nanoscale charge separation or recombination, as
well as spatially separated charge transport within the much larger
overall class of hybrid crystalline
materials. Clearly, significant challenges would remain if theory were
applied in isolation. For example, capturing all structural 
subtleties of complex 2D HOIP arrangements is nontrivial, as is
predicting fundamental gaps with an accuracy of better than a few
tenths of an eV (the accuracy expected from the unmodified HSE06+SOC
functional{\color{black} \cite{Krukau06}, as used in this work,} for typical semiconductors\cite{hseCZTSprb2009,  
  tong2017, rangeSeparatedHybrid}) for structures of this
size. Excitingly, the combination of such predictions with subsequent
targeted experimental syntheses overcomes
these challenges, creating enormous possibilities to identify and
fine-tune entirely new layered crystalline organic-inorganic semiconductors
with deliberately selected optoelectronic/electronic properties.

\begin{acknowledgments}
This work was financially supported by the NSF under {\color{black} awards number DMR-1729297 and} DMR-1728921, as well as through the Research Triangle MRSEC (DMR-11-21107). An award of computer time was provided by the Innovative and Novel Computational Impact on Theory and Experiment (INCITE) program, and the Theta Early Science program (ESP). This research used resources of the Argonne Leadership Computing Facility (ALCF), which is a DOE Office of Science User Facility supported under Contract DE-AC02-06CH11357.
\end{acknowledgments}

%


\begin{thebibliography}{80}%
\makeatletter
\providecommand \@ifxundefined [1]{%
 \@ifx{#1\undefined}
}%
\providecommand \@ifnum [1]{%
 \ifnum #1\expandafter \@firstoftwo
 \else \expandafter \@secondoftwo
 \fi
}%
\providecommand \@ifx [1]{%
 \ifx #1\expandafter \@firstoftwo
 \else \expandafter \@secondoftwo
 \fi
}%
\providecommand \natexlab [1]{#1}%
\providecommand \enquote  [1]{``#1''}%
\providecommand \bibnamefont  [1]{#1}%
\providecommand \bibfnamefont [1]{#1}%
\providecommand \citenamefont [1]{#1}%
\providecommand \href@noop [0]{\@secondoftwo}%
\providecommand \href [0]{\begingroup \@sanitize@url \@href}%
\providecommand \@href[1]{\@@startlink{#1}\@@href}%
\providecommand \@@href[1]{\endgroup#1\@@endlink}%
\providecommand \@sanitize@url [0]{\catcode `\\12\catcode `\$12\catcode
  `\&12\catcode `\#12\catcode `\^12\catcode `\_12\catcode `\%12\relax}%
\providecommand \@@startlink[1]{}%
\providecommand \@@endlink[0]{}%
\providecommand \url  [0]{\begingroup\@sanitize@url \@url }%
\providecommand \@url [1]{\endgroup\@href {#1}{\urlprefix }}%
\providecommand \urlprefix  [0]{URL }%
\providecommand \Eprint [0]{\href }%
\providecommand \doibase [0]{http://dx.doi.org/}%
\providecommand \selectlanguage [0]{\@gobble}%
\providecommand \bibinfo  [0]{\@secondoftwo}%
\providecommand \bibfield  [0]{\@secondoftwo}%
\providecommand \translation [1]{[#1]}%
\providecommand \BibitemOpen [0]{}%
\providecommand \bibitemStop [0]{}%
\providecommand \bibitemNoStop [0]{.\EOS\space}%
\providecommand \EOS [0]{\spacefactor3000\relax}%
\providecommand \BibitemShut  [1]{\csname bibitem#1\endcsname}%
\let\auto@bib@innerbib\@empty
\bibitem [{\citenamefont {Saparov}\ and\ \citenamefont
  {Mitzi}(2016)}]{Mitzi2016}%
  \BibitemOpen
  \bibfield  {author} {\bibinfo {author} {\bibfnamefont {B.}~\bibnamefont
  {Saparov}}\ and\ \bibinfo {author} {\bibfnamefont {D.~B.}\ \bibnamefont
  {Mitzi}},\ }\href {\doibase 10.1021/acs.chemrev.5b00715} {\bibfield
  {journal} {\bibinfo  {journal} {Chem. Rev.}\ }\textbf {\bibinfo {volume}
  {116}},\ \bibinfo {pages} {4558} (\bibinfo {year} {2016})}\BibitemShut
  {NoStop}%
\bibitem [{\citenamefont {Li}\ \emph {et~al.}(2017)\citenamefont {Li},
  \citenamefont {Wang}, \citenamefont {Deschler}, \citenamefont {Gao},
  \citenamefont {Friend},\ and\ \citenamefont
  {Cheetham}}]{Li2017natureFunctional}%
  \BibitemOpen
  \bibfield  {author} {\bibinfo {author} {\bibfnamefont {W.}~\bibnamefont
  {Li}}, \bibinfo {author} {\bibfnamefont {Z.}~\bibnamefont {Wang}}, \bibinfo
  {author} {\bibfnamefont {F.}~\bibnamefont {Deschler}}, \bibinfo {author}
  {\bibfnamefont {S.}~\bibnamefont {Gao}}, \bibinfo {author} {\bibfnamefont
  {R.~H.}\ \bibnamefont {Friend}}, \ and\ \bibinfo {author} {\bibfnamefont
  {A.~K.}\ \bibnamefont {Cheetham}},\ }\href {\doibase
  10.1038/natrevmats.2016.99} {\bibfield  {journal} {\bibinfo  {journal} {Nat.
  Rev. Mater.}\ }\textbf {\bibinfo {volume} {2}},\ \bibinfo {pages} {16099}
  (\bibinfo {year} {2017})}\BibitemShut {NoStop}%
\bibitem [{\citenamefont {Kojima}\ \emph {et~al.}(2009)\citenamefont {Kojima},
  \citenamefont {Teshima}, \citenamefont {Shirai},\ and\ \citenamefont
  {Miyasaka}}]{MAPbI3_2009}%
  \BibitemOpen
  \bibfield  {author} {\bibinfo {author} {\bibfnamefont {A.}~\bibnamefont
  {Kojima}}, \bibinfo {author} {\bibfnamefont {K.}~\bibnamefont {Teshima}},
  \bibinfo {author} {\bibfnamefont {Y.}~\bibnamefont {Shirai}}, \ and\ \bibinfo
  {author} {\bibfnamefont {T.}~\bibnamefont {Miyasaka}},\ }\href {\doibase
  10.1021/ja809598r} {\bibfield  {journal} {\bibinfo  {journal} {J. Am. Chem.
  Soc.}\ }\textbf {\bibinfo {volume} {131}},\ \bibinfo {pages} {6050} (\bibinfo
  {year} {2009})}\BibitemShut {NoStop}%
\bibitem [{\citenamefont {Heo}\ \emph {et~al.}(2013)\citenamefont {Heo},
  \citenamefont {Im}, \citenamefont {Noh}, \citenamefont {Mandal},
  \citenamefont {Lim}, \citenamefont {Chang}, \citenamefont {Lee},
  \citenamefont {Kim}, \citenamefont {Sarkar}, \citenamefont {Nazeeruddin},
  \citenamefont {Gr{\"a}tzel},\ and\ \citenamefont {Seok}}]{Heo2013}%
  \BibitemOpen
  \bibfield  {author} {\bibinfo {author} {\bibfnamefont {J.~H.}\ \bibnamefont
  {Heo}}, \bibinfo {author} {\bibfnamefont {S.~H.}\ \bibnamefont {Im}},
  \bibinfo {author} {\bibfnamefont {J.~H.}\ \bibnamefont {Noh}}, \bibinfo
  {author} {\bibfnamefont {T.~N.}\ \bibnamefont {Mandal}}, \bibinfo {author}
  {\bibfnamefont {C.-S.}\ \bibnamefont {Lim}}, \bibinfo {author} {\bibfnamefont
  {J.~A.}\ \bibnamefont {Chang}}, \bibinfo {author} {\bibfnamefont {Y.~H.}\
  \bibnamefont {Lee}}, \bibinfo {author} {\bibfnamefont {H.~J.}\ \bibnamefont
  {Kim}}, \bibinfo {author} {\bibfnamefont {A.}~\bibnamefont {Sarkar}},
  \bibinfo {author} {\bibfnamefont {M.~K.}\ \bibnamefont {Nazeeruddin}},
  \bibinfo {author} {\bibfnamefont {M.}~\bibnamefont {Gr{\"a}tzel}}, \ and\
  \bibinfo {author} {\bibfnamefont {S.~I.}\ \bibnamefont {Seok}},\ }\href
  {http://dx.doi.org/10.1038/nphoton.2013.80} {\bibfield  {journal} {\bibinfo
  {journal} {Nat. Photonics}\ }\textbf {\bibinfo {volume} {7}},\ \bibinfo
  {pages} {486} (\bibinfo {year} {2013})}\BibitemShut {NoStop}%
\bibitem [{\citenamefont {Liu}\ and\ \citenamefont {Kelly}(2013)}]{Liu2013}%
  \BibitemOpen
  \bibfield  {author} {\bibinfo {author} {\bibfnamefont {D.}~\bibnamefont
  {Liu}}\ and\ \bibinfo {author} {\bibfnamefont {T.~L.}\ \bibnamefont
  {Kelly}},\ }\href {http://dx.doi.org/10.1038/nphoton.2013.342} {\bibfield
  {journal} {\bibinfo  {journal} {Nat. Photonics}\ }\textbf {\bibinfo {volume}
  {8}},\ \bibinfo {pages} {133} (\bibinfo {year} {2013})}\BibitemShut {NoStop}%
\bibitem [{\citenamefont {Jeon}\ \emph {et~al.}(2014)\citenamefont {Jeon},
  \citenamefont {Noh}, \citenamefont {Kim}, \citenamefont {Yang}, \citenamefont
  {Ryu},\ and\ \citenamefont {Seok}}]{Jeon2014}%
  \BibitemOpen
  \bibfield  {author} {\bibinfo {author} {\bibfnamefont {N.~J.}\ \bibnamefont
  {Jeon}}, \bibinfo {author} {\bibfnamefont {J.~H.}\ \bibnamefont {Noh}},
  \bibinfo {author} {\bibfnamefont {Y.~C.}\ \bibnamefont {Kim}}, \bibinfo
  {author} {\bibfnamefont {W.~S.}\ \bibnamefont {Yang}}, \bibinfo {author}
  {\bibfnamefont {S.}~\bibnamefont {Ryu}}, \ and\ \bibinfo {author}
  {\bibfnamefont {S.~I.}\ \bibnamefont {Seok}},\ }\href
  {http://dx.doi.org/10.1038/nmat4014} {\bibfield  {journal} {\bibinfo
  {journal} {Nat. Mater.}\ }\textbf {\bibinfo {volume} {13}},\ \bibinfo {pages}
  {897} (\bibinfo {year} {2014})}\BibitemShut {NoStop}%
\bibitem [{\citenamefont {Green}\ \emph {et~al.}(2014)\citenamefont {Green},
  \citenamefont {Ho-Baillie},\ and\ \citenamefont {Snaith}}]{Green2014}%
  \BibitemOpen
  \bibfield  {author} {\bibinfo {author} {\bibfnamefont {M.~A.}\ \bibnamefont
  {Green}}, \bibinfo {author} {\bibfnamefont {A.}~\bibnamefont {Ho-Baillie}}, \
  and\ \bibinfo {author} {\bibfnamefont {H.~J.}\ \bibnamefont {Snaith}},\
  }\href {\doibase 10.1038/nphoton.2014.134} {\bibfield  {journal} {\bibinfo
  {journal} {Nat. Photonics}\ }\textbf {\bibinfo {volume} {8}},\ \bibinfo
  {pages} {506} (\bibinfo {year} {2014})}\BibitemShut {NoStop}%
\bibitem [{\citenamefont {Gao}\ \emph {et~al.}(2014)\citenamefont {Gao},
  \citenamefont {Gr{\"a}tzel},\ and\ \citenamefont
  {Nazeeruddin}}]{Gratzel2014}%
  \BibitemOpen
  \bibfield  {author} {\bibinfo {author} {\bibfnamefont {P.}~\bibnamefont
  {Gao}}, \bibinfo {author} {\bibfnamefont {M.}~\bibnamefont {Gr{\"a}tzel}}, \
  and\ \bibinfo {author} {\bibfnamefont {M.~K.}\ \bibnamefont {Nazeeruddin}},\
  }\href {\doibase 10.1039/C4EE00942H} {\bibfield  {journal} {\bibinfo
  {journal} {Energy Environ. Sci.}\ }\textbf {\bibinfo {volume} {7}},\ \bibinfo
  {pages} {2448} (\bibinfo {year} {2014})}\BibitemShut {NoStop}%
\bibitem [{\citenamefont {Eperon}\ \emph {et~al.}(2014)\citenamefont {Eperon},
  \citenamefont {Stranks}, \citenamefont {Menelaou}, \citenamefont {Johnston},
  \citenamefont {Herz},\ and\ \citenamefont {Snaith}}]{C3EE43822H}%
  \BibitemOpen
  \bibfield  {author} {\bibinfo {author} {\bibfnamefont {G.~E.}\ \bibnamefont
  {Eperon}}, \bibinfo {author} {\bibfnamefont {S.~D.}\ \bibnamefont {Stranks}},
  \bibinfo {author} {\bibfnamefont {C.}~\bibnamefont {Menelaou}}, \bibinfo
  {author} {\bibfnamefont {M.~B.}\ \bibnamefont {Johnston}}, \bibinfo {author}
  {\bibfnamefont {L.~M.}\ \bibnamefont {Herz}}, \ and\ \bibinfo {author}
  {\bibfnamefont {H.~J.}\ \bibnamefont {Snaith}},\ }\href {\doibase
  10.1039/C3EE43822H} {\bibfield  {journal} {\bibinfo  {journal} {Energy
  Environ. Sci.}\ }\textbf {\bibinfo {volume} {7}},\ \bibinfo {pages} {982}
  (\bibinfo {year} {2014})}\BibitemShut {NoStop}%
\bibitem [{\citenamefont {Yang}\ \emph {et~al.}(2015)\citenamefont {Yang},
  \citenamefont {Noh}, \citenamefont {Jeon}, \citenamefont {Kim}, \citenamefont
  {Ryu}, \citenamefont {Seo},\ and\ \citenamefont {Seok}}]{Yang1234}%
  \BibitemOpen
  \bibfield  {author} {\bibinfo {author} {\bibfnamefont {W.~S.}\ \bibnamefont
  {Yang}}, \bibinfo {author} {\bibfnamefont {J.~H.}\ \bibnamefont {Noh}},
  \bibinfo {author} {\bibfnamefont {N.~J.}\ \bibnamefont {Jeon}}, \bibinfo
  {author} {\bibfnamefont {Y.~C.}\ \bibnamefont {Kim}}, \bibinfo {author}
  {\bibfnamefont {S.}~\bibnamefont {Ryu}}, \bibinfo {author} {\bibfnamefont
  {J.}~\bibnamefont {Seo}}, \ and\ \bibinfo {author} {\bibfnamefont {S.~I.}\
  \bibnamefont {Seok}},\ }\href {\doibase 10.1126/science.aaa9272} {\bibfield
  {journal} {\bibinfo  {journal} {Science}\ }\textbf {\bibinfo {volume}
  {348}},\ \bibinfo {pages} {1234} (\bibinfo {year} {2015})}\BibitemShut
  {NoStop}%
\bibitem [{\citenamefont {Yang}\ \emph {et~al.}(2017)\citenamefont {Yang},
  \citenamefont {Park}, \citenamefont {Jung}, \citenamefont {Jeon},
  \citenamefont {Kim}, \citenamefont {Lee}, \citenamefont {Shin}, \citenamefont
  {Seo}, \citenamefont {Kim}, \citenamefont {Noh},\ and\ \citenamefont
  {Seok}}]{Yang1376}%
  \BibitemOpen
  \bibfield  {author} {\bibinfo {author} {\bibfnamefont {W.~S.}\ \bibnamefont
  {Yang}}, \bibinfo {author} {\bibfnamefont {B.-W.}\ \bibnamefont {Park}},
  \bibinfo {author} {\bibfnamefont {E.~H.}\ \bibnamefont {Jung}}, \bibinfo
  {author} {\bibfnamefont {N.~J.}\ \bibnamefont {Jeon}}, \bibinfo {author}
  {\bibfnamefont {Y.~C.}\ \bibnamefont {Kim}}, \bibinfo {author} {\bibfnamefont
  {D.~U.}\ \bibnamefont {Lee}}, \bibinfo {author} {\bibfnamefont {S.~S.}\
  \bibnamefont {Shin}}, \bibinfo {author} {\bibfnamefont {J.}~\bibnamefont
  {Seo}}, \bibinfo {author} {\bibfnamefont {E.~K.}\ \bibnamefont {Kim}},
  \bibinfo {author} {\bibfnamefont {J.~H.}\ \bibnamefont {Noh}}, \ and\
  \bibinfo {author} {\bibfnamefont {S.~I.}\ \bibnamefont {Seok}},\ }\href
  {\doibase 10.1126/science.aan2301} {\bibfield  {journal} {\bibinfo  {journal}
  {Science}\ }\textbf {\bibinfo {volume} {356}},\ \bibinfo {pages} {1376}
  (\bibinfo {year} {2017})}\BibitemShut {NoStop}%
\bibitem [{\citenamefont {Jodlowski}\ \emph {et~al.}(2017)\citenamefont
  {Jodlowski}, \citenamefont {Rold{\'a}n-Carmona}, \citenamefont {Grancini},
  \citenamefont {Salado}, \citenamefont {Ralaiarisoa}, \citenamefont {Ahmad},
  \citenamefont {Koch}, \citenamefont {Camacho}, \citenamefont
  {\mbox{de~Miguel}},\ and\ \citenamefont {Nazeeruddin}}]{Jodlowski2017}%
  \BibitemOpen
  \bibfield  {author} {\bibinfo {author} {\bibfnamefont {A.~D.}\ \bibnamefont
  {Jodlowski}}, \bibinfo {author} {\bibfnamefont {C.}~\bibnamefont
  {Rold{\'a}n-Carmona}}, \bibinfo {author} {\bibfnamefont {G.}~\bibnamefont
  {Grancini}}, \bibinfo {author} {\bibfnamefont {M.}~\bibnamefont {Salado}},
  \bibinfo {author} {\bibfnamefont {M.}~\bibnamefont {Ralaiarisoa}}, \bibinfo
  {author} {\bibfnamefont {S.}~\bibnamefont {Ahmad}}, \bibinfo {author}
  {\bibfnamefont {N.}~\bibnamefont {Koch}}, \bibinfo {author} {\bibfnamefont
  {L.}~\bibnamefont {Camacho}}, \bibinfo {author} {\bibfnamefont
  {G.}~\bibnamefont {\mbox{de~Miguel}}}, \ and\ \bibinfo {author}
  {\bibfnamefont {M.~K.}\ \bibnamefont {Nazeeruddin}},\ }\href
  {https://doi.org/10.1038/s41560-017-0054-3} {\bibfield  {journal} {\bibinfo
  {journal} {Nature Energy}\ }\textbf {\bibinfo {volume} {2}},\ \bibinfo
  {pages} {972} (\bibinfo {year} {2017})}\BibitemShut {NoStop}%
\bibitem [{\citenamefont {Tan}\ \emph {et~al.}(2014)\citenamefont {Tan},
  \citenamefont {Moghaddam}, \citenamefont {Lai}, \citenamefont {Docampo},
  \citenamefont {Higler}, \citenamefont {Deschler}, \citenamefont {Price},
  \citenamefont {Sadhanala}, \citenamefont {Pazos}, \citenamefont
  {Credgington}, \citenamefont {Hanusch}, \citenamefont {Bein}, \citenamefont
  {Snaith},\ and\ \citenamefont {Friend}}]{Tan2014NatureNT}%
  \BibitemOpen
  \bibfield  {author} {\bibinfo {author} {\bibfnamefont {Z.-K.}\ \bibnamefont
  {Tan}}, \bibinfo {author} {\bibfnamefont {R.~S.}\ \bibnamefont {Moghaddam}},
  \bibinfo {author} {\bibfnamefont {M.~L.}\ \bibnamefont {Lai}}, \bibinfo
  {author} {\bibfnamefont {P.}~\bibnamefont {Docampo}}, \bibinfo {author}
  {\bibfnamefont {R.}~\bibnamefont {Higler}}, \bibinfo {author} {\bibfnamefont
  {F.}~\bibnamefont {Deschler}}, \bibinfo {author} {\bibfnamefont
  {M.}~\bibnamefont {Price}}, \bibinfo {author} {\bibfnamefont
  {A.}~\bibnamefont {Sadhanala}}, \bibinfo {author} {\bibfnamefont {L.~M.}\
  \bibnamefont {Pazos}}, \bibinfo {author} {\bibfnamefont {D.}~\bibnamefont
  {Credgington}}, \bibinfo {author} {\bibfnamefont {F.}~\bibnamefont
  {Hanusch}}, \bibinfo {author} {\bibfnamefont {T.}~\bibnamefont {Bein}},
  \bibinfo {author} {\bibfnamefont {H.~J.}\ \bibnamefont {Snaith}}, \ and\
  \bibinfo {author} {\bibfnamefont {R.~H.}\ \bibnamefont {Friend}},\ }\href
  {http://dx.doi.org/10.1038/nnano.2014.149} {\bibfield  {journal} {\bibinfo
  {journal} {Nature Nanotechnology}\ }\textbf {\bibinfo {volume} {9}},\
  \bibinfo {pages} {687} (\bibinfo {year} {2014})}\BibitemShut {NoStop}%
\bibitem [{\citenamefont {Wang}\ \emph {et~al.}(2015)\citenamefont {Wang},
  \citenamefont {Wang}, \citenamefont {Jin}, \citenamefont {Si}, \citenamefont
  {Tan}, \citenamefont {Du}, \citenamefont {Cheng}, \citenamefont {Dai},
  \citenamefont {Bai}, \citenamefont {He}, \citenamefont {Ye}, \citenamefont
  {Lai}, \citenamefont {Friend},\ and\ \citenamefont {Huang}}]{Wang2015ADMA}%
  \BibitemOpen
  \bibfield  {author} {\bibinfo {author} {\bibfnamefont {J.}~\bibnamefont
  {Wang}}, \bibinfo {author} {\bibfnamefont {N.}~\bibnamefont {Wang}}, \bibinfo
  {author} {\bibfnamefont {Y.}~\bibnamefont {Jin}}, \bibinfo {author}
  {\bibfnamefont {J.}~\bibnamefont {Si}}, \bibinfo {author} {\bibfnamefont
  {Z.-K.}\ \bibnamefont {Tan}}, \bibinfo {author} {\bibfnamefont
  {H.}~\bibnamefont {Du}}, \bibinfo {author} {\bibfnamefont {L.}~\bibnamefont
  {Cheng}}, \bibinfo {author} {\bibfnamefont {X.}~\bibnamefont {Dai}}, \bibinfo
  {author} {\bibfnamefont {S.}~\bibnamefont {Bai}}, \bibinfo {author}
  {\bibfnamefont {H.}~\bibnamefont {He}}, \bibinfo {author} {\bibfnamefont
  {Z.}~\bibnamefont {Ye}}, \bibinfo {author} {\bibfnamefont {M.~L.}\
  \bibnamefont {Lai}}, \bibinfo {author} {\bibfnamefont {R.~H.}\ \bibnamefont
  {Friend}}, \ and\ \bibinfo {author} {\bibfnamefont {W.}~\bibnamefont
  {Huang}},\ }\href {\doibase 10.1002/adma.201405217} {\bibfield  {journal}
  {\bibinfo  {journal} {Advanced Materials}\ }\textbf {\bibinfo {volume}
  {27}},\ \bibinfo {pages} {2311} (\bibinfo {year} {2015})}\BibitemShut
  {NoStop}%
\bibitem [{\citenamefont {Kim}\ \emph {et~al.}(2015)\citenamefont {Kim},
  \citenamefont {Cho}, \citenamefont {Heo}, \citenamefont {Kim}, \citenamefont
  {Myoung}, \citenamefont {Lee}, \citenamefont {Im},\ and\ \citenamefont
  {Lee}}]{emitAdvMat15}%
  \BibitemOpen
  \bibfield  {author} {\bibinfo {author} {\bibfnamefont {Y.-H.}\ \bibnamefont
  {Kim}}, \bibinfo {author} {\bibfnamefont {H.}~\bibnamefont {Cho}}, \bibinfo
  {author} {\bibfnamefont {J.~H.}\ \bibnamefont {Heo}}, \bibinfo {author}
  {\bibfnamefont {T.-S.}\ \bibnamefont {Kim}}, \bibinfo {author} {\bibfnamefont
  {N.}~\bibnamefont {Myoung}}, \bibinfo {author} {\bibfnamefont {C.-L.}\
  \bibnamefont {Lee}}, \bibinfo {author} {\bibfnamefont {S.~H.}\ \bibnamefont
  {Im}}, \ and\ \bibinfo {author} {\bibfnamefont {T.-W.}\ \bibnamefont {Lee}},\
  }\href {\doibase 10.1002/adma.201403751} {\bibfield  {journal} {\bibinfo
  {journal} {Adv. Mater.}\ }\textbf {\bibinfo {volume} {27}},\ \bibinfo {pages}
  {1248} (\bibinfo {year} {2015})}\BibitemShut {NoStop}%
\bibitem [{\citenamefont {Cho}\ \emph {et~al.}(2015)\citenamefont {Cho},
  \citenamefont {Jeong}, \citenamefont {Park}, \citenamefont {Kim},
  \citenamefont {Wolf}, \citenamefont {Lee}, \citenamefont {Heo}, \citenamefont
  {Sadhanala}, \citenamefont {Myoung}, \citenamefont {Yoo}, \citenamefont {Im},
  \citenamefont {Friend},\ and\ \citenamefont {Lee}}]{Cho2015Science}%
  \BibitemOpen
  \bibfield  {author} {\bibinfo {author} {\bibfnamefont {H.}~\bibnamefont
  {Cho}}, \bibinfo {author} {\bibfnamefont {S.-H.}\ \bibnamefont {Jeong}},
  \bibinfo {author} {\bibfnamefont {M.-H.}\ \bibnamefont {Park}}, \bibinfo
  {author} {\bibfnamefont {Y.-H.}\ \bibnamefont {Kim}}, \bibinfo {author}
  {\bibfnamefont {C.}~\bibnamefont {Wolf}}, \bibinfo {author} {\bibfnamefont
  {C.-L.}\ \bibnamefont {Lee}}, \bibinfo {author} {\bibfnamefont {J.~H.}\
  \bibnamefont {Heo}}, \bibinfo {author} {\bibfnamefont {A.}~\bibnamefont
  {Sadhanala}}, \bibinfo {author} {\bibfnamefont {N.}~\bibnamefont {Myoung}},
  \bibinfo {author} {\bibfnamefont {S.}~\bibnamefont {Yoo}}, \bibinfo {author}
  {\bibfnamefont {S.~H.}\ \bibnamefont {Im}}, \bibinfo {author} {\bibfnamefont
  {R.~H.}\ \bibnamefont {Friend}}, \ and\ \bibinfo {author} {\bibfnamefont
  {T.-W.}\ \bibnamefont {Lee}},\ }\href {\doibase 10.1126/science.aad1818}
  {\bibfield  {journal} {\bibinfo  {journal} {Science}\ }\textbf {\bibinfo
  {volume} {350}},\ \bibinfo {pages} {1222} (\bibinfo {year}
  {2015})}\BibitemShut {NoStop}%
\bibitem [{\citenamefont {Kim}\ \emph {et~al.}(2016)\citenamefont {Kim},
  \citenamefont {Cho},\ and\ \citenamefont {Lee}}]{emit16pnas}%
  \BibitemOpen
  \bibfield  {author} {\bibinfo {author} {\bibfnamefont {Y.-H.}\ \bibnamefont
  {Kim}}, \bibinfo {author} {\bibfnamefont {H.}~\bibnamefont {Cho}}, \ and\
  \bibinfo {author} {\bibfnamefont {T.-W.}\ \bibnamefont {Lee}},\ }\href
  {\doibase 10.1073/pnas.1607471113} {\bibfield  {journal} {\bibinfo  {journal}
  {Proc. Natl. Acad. Sci.}\ }\textbf {\bibinfo {volume} {113}},\ \bibinfo
  {pages} {11694} (\bibinfo {year} {2016})}\BibitemShut {NoStop}%
\bibitem [{\citenamefont {Xiao}\ \emph {et~al.}(2017)\citenamefont {Xiao},
  \citenamefont {Kerner}, \citenamefont {Zhao}, \citenamefont {Tran},
  \citenamefont {Lee}, \citenamefont {Koh}, \citenamefont {Scholes},\ and\
  \citenamefont {Rand}}]{emit17nature}%
  \BibitemOpen
  \bibfield  {author} {\bibinfo {author} {\bibfnamefont {Z.}~\bibnamefont
  {Xiao}}, \bibinfo {author} {\bibfnamefont {R.~A.}\ \bibnamefont {Kerner}},
  \bibinfo {author} {\bibfnamefont {L.}~\bibnamefont {Zhao}}, \bibinfo {author}
  {\bibfnamefont {N.~L.}\ \bibnamefont {Tran}}, \bibinfo {author}
  {\bibfnamefont {K.~M.}\ \bibnamefont {Lee}}, \bibinfo {author} {\bibfnamefont
  {T.-W.}\ \bibnamefont {Koh}}, \bibinfo {author} {\bibfnamefont {G.~D.}\
  \bibnamefont {Scholes}}, \ and\ \bibinfo {author} {\bibfnamefont {B.~P.}\
  \bibnamefont {Rand}},\ }\href {http://dx.doi.org/10.1038/nphoton.2016.269}
  {\bibfield  {journal} {\bibinfo  {journal} {Nat. Photonics}\ }\textbf
  {\bibinfo {volume} {11}},\ \bibinfo {pages} {108} (\bibinfo {year}
  {2017})}\BibitemShut {NoStop}%
\bibitem [{\citenamefont {Zhang}\ \emph {et~al.}(2017)\citenamefont {Zhang},
  \citenamefont {Liu}, \citenamefont {Wang}, \citenamefont {Zhang},
  \citenamefont {Xu}, \citenamefont {Karen}, \citenamefont {Zheng},
  \citenamefont {Liu}, \citenamefont {Chen}, \citenamefont {Wang},\ and\
  \citenamefont {Sun}}]{emitAdvMat17}%
  \BibitemOpen
  \bibfield  {author} {\bibinfo {author} {\bibfnamefont {X.}~\bibnamefont
  {Zhang}}, \bibinfo {author} {\bibfnamefont {H.}~\bibnamefont {Liu}}, \bibinfo
  {author} {\bibfnamefont {W.}~\bibnamefont {Wang}}, \bibinfo {author}
  {\bibfnamefont {J.}~\bibnamefont {Zhang}}, \bibinfo {author} {\bibfnamefont
  {B.}~\bibnamefont {Xu}}, \bibinfo {author} {\bibfnamefont {K.~L.}\
  \bibnamefont {Karen}}, \bibinfo {author} {\bibfnamefont {Y.}~\bibnamefont
  {Zheng}}, \bibinfo {author} {\bibfnamefont {S.}~\bibnamefont {Liu}}, \bibinfo
  {author} {\bibfnamefont {S.}~\bibnamefont {Chen}}, \bibinfo {author}
  {\bibfnamefont {K.}~\bibnamefont {Wang}}, \ and\ \bibinfo {author}
  {\bibfnamefont {X.~W.}\ \bibnamefont {Sun}},\ }\href {\doibase
  10.1002/adma.201606405} {\bibfield  {journal} {\bibinfo  {journal} {Adv.
  Mater.}\ }\textbf {\bibinfo {volume} {29}},\ \bibinfo {pages} {1606405}
  (\bibinfo {year} {2017})}\BibitemShut {NoStop}%
\bibitem [{\citenamefont {Adjokatse}\ \emph {et~al.}(2017)\citenamefont
  {Adjokatse}, \citenamefont {Fang},\ and\ \citenamefont
  {Loi}}]{emitTuneMaterials}%
  \BibitemOpen
  \bibfield  {author} {\bibinfo {author} {\bibfnamefont {S.}~\bibnamefont
  {Adjokatse}}, \bibinfo {author} {\bibfnamefont {H.-H.}\ \bibnamefont {Fang}},
  \ and\ \bibinfo {author} {\bibfnamefont {M.~A.}\ \bibnamefont {Loi}},\ }\href
  {\doibase https://doi.org/10.1016/j.mattod.2017.03.021} {\bibfield  {journal}
  {\bibinfo  {journal} {Mater. Today}\ }\textbf {\bibinfo {volume} {20}},\
  \bibinfo {pages} {413 } (\bibinfo {year} {2017})}\BibitemShut {NoStop}%
\bibitem [{\citenamefont {Mitzi}\ and\ \citenamefont
  {Liang}(1997)}]{MITZI1997376}%
  \BibitemOpen
  \bibfield  {author} {\bibinfo {author} {\bibfnamefont {D.~B.}\ \bibnamefont
  {Mitzi}}\ and\ \bibinfo {author} {\bibfnamefont {K.}~\bibnamefont {Liang}},\
  }\href {\doibase https://doi.org/10.1006/jssc.1997.7593} {\bibfield
  {journal} {\bibinfo  {journal} {J. Solid State Chem.}\ }\textbf {\bibinfo
  {volume} {134}},\ \bibinfo {pages} {376 } (\bibinfo {year}
  {1997})}\BibitemShut {NoStop}%
\bibitem [{\citenamefont {Han}\ \emph {et~al.}(2016)\citenamefont {Han},
  \citenamefont {Bae}, \citenamefont {Sun}, \citenamefont {Hsieh},
  \citenamefont {Yang}, \citenamefont {Rim}, \citenamefont {Zhao},
  \citenamefont {Chen}, \citenamefont {Shi}, \citenamefont {Li},\ and\
  \citenamefont {Yang}}]{fapbi3}%
  \BibitemOpen
  \bibfield  {author} {\bibinfo {author} {\bibfnamefont {Q.}~\bibnamefont
  {Han}}, \bibinfo {author} {\bibfnamefont {S.-H.}\ \bibnamefont {Bae}},
  \bibinfo {author} {\bibfnamefont {P.}~\bibnamefont {Sun}}, \bibinfo {author}
  {\bibfnamefont {Y.-T.}\ \bibnamefont {Hsieh}}, \bibinfo {author}
  {\bibfnamefont {Y.~M.}\ \bibnamefont {Yang}}, \bibinfo {author}
  {\bibfnamefont {Y.~S.}\ \bibnamefont {Rim}}, \bibinfo {author} {\bibfnamefont
  {H.}~\bibnamefont {Zhao}}, \bibinfo {author} {\bibfnamefont {Q.}~\bibnamefont
  {Chen}}, \bibinfo {author} {\bibfnamefont {W.}~\bibnamefont {Shi}}, \bibinfo
  {author} {\bibfnamefont {G.}~\bibnamefont {Li}}, \ and\ \bibinfo {author}
  {\bibfnamefont {Y.}~\bibnamefont {Yang}},\ }\href {\doibase
  10.1002/adma.201505002} {\bibfield  {journal} {\bibinfo  {journal} {Adv.
  Mater.}\ }\textbf {\bibinfo {volume} {28}},\ \bibinfo {pages} {2253}
  (\bibinfo {year} {2016})}\BibitemShut {NoStop}%
\bibitem [{\citenamefont {Umebayashi}\ \emph {et~al.}(2003)\citenamefont
  {Umebayashi}, \citenamefont {Asai}, \citenamefont {Kondo},\ and\
  \citenamefont {Nakao}}]{Umebayashi2003}%
  \BibitemOpen
  \bibfield  {author} {\bibinfo {author} {\bibfnamefont {T.}~\bibnamefont
  {Umebayashi}}, \bibinfo {author} {\bibfnamefont {K.}~\bibnamefont {Asai}},
  \bibinfo {author} {\bibfnamefont {T.}~\bibnamefont {Kondo}}, \ and\ \bibinfo
  {author} {\bibfnamefont {A.}~\bibnamefont {Nakao}},\ }\href {\doibase
  10.1103/PhysRevB.67.155405} {\bibfield  {journal} {\bibinfo  {journal} {Phys.
  Rev. B}\ }\textbf {\bibinfo {volume} {67}},\ \bibinfo {pages} {155405}
  (\bibinfo {year} {2003})}\BibitemShut {NoStop}%
\bibitem [{\citenamefont {Chiarella}\ \emph {et~al.}(2008)\citenamefont
  {Chiarella}, \citenamefont {Zappettini}, \citenamefont {Licci}, \citenamefont
  {Borriello}, \citenamefont {Cantele}, \citenamefont {Ninno}, \citenamefont
  {Cassinese},\ and\ \citenamefont {Vaglio}}]{Chiarella2008}%
  \BibitemOpen
  \bibfield  {author} {\bibinfo {author} {\bibfnamefont {F.}~\bibnamefont
  {Chiarella}}, \bibinfo {author} {\bibfnamefont {A.}~\bibnamefont
  {Zappettini}}, \bibinfo {author} {\bibfnamefont {F.}~\bibnamefont {Licci}},
  \bibinfo {author} {\bibfnamefont {I.}~\bibnamefont {Borriello}}, \bibinfo
  {author} {\bibfnamefont {G.}~\bibnamefont {Cantele}}, \bibinfo {author}
  {\bibfnamefont {D.}~\bibnamefont {Ninno}}, \bibinfo {author} {\bibfnamefont
  {A.}~\bibnamefont {Cassinese}}, \ and\ \bibinfo {author} {\bibfnamefont
  {R.}~\bibnamefont {Vaglio}},\ }\href {\doibase 10.1103/PhysRevB.77.045129}
  {\bibfield  {journal} {\bibinfo  {journal} {Phys. Rev. B}\ }\textbf {\bibinfo
  {volume} {77}},\ \bibinfo {pages} {045129} (\bibinfo {year}
  {2008})}\BibitemShut {NoStop}%
\bibitem [{\citenamefont {Filippetti}\ and\ \citenamefont
  {Mattoni}(2014)}]{PRB2014Filippetti}%
  \BibitemOpen
  \bibfield  {author} {\bibinfo {author} {\bibfnamefont {A.}~\bibnamefont
  {Filippetti}}\ and\ \bibinfo {author} {\bibfnamefont {A.}~\bibnamefont
  {Mattoni}},\ }\href {\doibase 10.1103/PhysRevB.89.125203} {\bibfield
  {journal} {\bibinfo  {journal} {Phys. Rev. B}\ }\textbf {\bibinfo {volume}
  {89}},\ \bibinfo {pages} {125203} (\bibinfo {year} {2014})}\BibitemShut
  {NoStop}%
\bibitem [{\citenamefont {Brivio}\ \emph {et~al.}(2013)\citenamefont {Brivio},
  \citenamefont {Walker},\ and\ \citenamefont {Walsh}}]{walsh2013}%
  \BibitemOpen
  \bibfield  {author} {\bibinfo {author} {\bibfnamefont {F.}~\bibnamefont
  {Brivio}}, \bibinfo {author} {\bibfnamefont {A.~B.}\ \bibnamefont {Walker}},
  \ and\ \bibinfo {author} {\bibfnamefont {A.}~\bibnamefont {Walsh}},\ }\href
  {\doibase 10.1063/1.4824147} {\bibfield  {journal} {\bibinfo  {journal} {APL
  Mater.}\ }\textbf {\bibinfo {volume} {1}},\ \bibinfo {pages} {042111}
  (\bibinfo {year} {2013})}\BibitemShut {NoStop}%
\bibitem [{\citenamefont {Brivio}\ \emph {et~al.}(2014)\citenamefont {Brivio},
  \citenamefont {Butler}, \citenamefont {Walsh},\ and\ \citenamefont {van
  Schilfgaarde}}]{Schilfgaarde2014}%
  \BibitemOpen
  \bibfield  {author} {\bibinfo {author} {\bibfnamefont {F.}~\bibnamefont
  {Brivio}}, \bibinfo {author} {\bibfnamefont {K.~T.}\ \bibnamefont {Butler}},
  \bibinfo {author} {\bibfnamefont {A.}~\bibnamefont {Walsh}}, \ and\ \bibinfo
  {author} {\bibfnamefont {M.}~\bibnamefont {van Schilfgaarde}},\ }\href
  {\doibase 10.1103/PhysRevB.89.155204} {\bibfield  {journal} {\bibinfo
  {journal} {Phys. Rev. B}\ }\textbf {\bibinfo {volume} {89}},\ \bibinfo
  {pages} {155204} (\bibinfo {year} {2014})}\BibitemShut {NoStop}%
\bibitem [{\citenamefont {Motta}\ \emph {et~al.}(2015)\citenamefont {Motta},
  \citenamefont {El-Mellouhi}, \citenamefont {Kais}, \citenamefont {Tabet},
  \citenamefont {Alharbi},\ and\ \citenamefont {Sanvito}}]{Motta2015}%
  \BibitemOpen
  \bibfield  {author} {\bibinfo {author} {\bibfnamefont {C.}~\bibnamefont
  {Motta}}, \bibinfo {author} {\bibfnamefont {F.}~\bibnamefont {El-Mellouhi}},
  \bibinfo {author} {\bibfnamefont {S.}~\bibnamefont {Kais}}, \bibinfo {author}
  {\bibfnamefont {N.}~\bibnamefont {Tabet}}, \bibinfo {author} {\bibfnamefont
  {F.}~\bibnamefont {Alharbi}}, \ and\ \bibinfo {author} {\bibfnamefont
  {S.}~\bibnamefont {Sanvito}},\ }\href {http://dx.doi.org/10.1038/ncomms8026}
  {\bibfield  {journal} {\bibinfo  {journal} {Nat. Commun.}\ }\textbf {\bibinfo
  {volume} {6}},\ \bibinfo {pages} {7026} (\bibinfo {year} {2015})}\BibitemShut
  {NoStop}%
\bibitem [{\citenamefont {Era}\ \emph {et~al.}(1998)\citenamefont {Era},
  \citenamefont {Maeda},\ and\ \citenamefont {Tsutsui}}]{Era1998}%
  \BibitemOpen
  \bibfield  {author} {\bibinfo {author} {\bibfnamefont {M.}~\bibnamefont
  {Era}}, \bibinfo {author} {\bibfnamefont {K.}~\bibnamefont {Maeda}}, \ and\
  \bibinfo {author} {\bibfnamefont {T.}~\bibnamefont {Tsutsui}},\ }\href
  {\doibase https://doi.org/10.1016/S0009-2614(98)01028-8} {\bibfield
  {journal} {\bibinfo  {journal} {Chem. Phys. Lett.}\ }\textbf {\bibinfo
  {volume} {296}},\ \bibinfo {pages} {417 } (\bibinfo {year}
  {1998})}\BibitemShut {NoStop}%
\bibitem [{\citenamefont {Braun}\ \emph {et~al.}(1999)\citenamefont {Braun},
  \citenamefont {Tuffentsammer}, \citenamefont {Wachtel},\ and\ \citenamefont
  {Wolf}}]{brauna_cpl_1999_acene_PbCl}%
  \BibitemOpen
  \bibfield  {author} {\bibinfo {author} {\bibfnamefont {M.}~\bibnamefont
  {Braun}}, \bibinfo {author} {\bibfnamefont {W.}~\bibnamefont
  {Tuffentsammer}}, \bibinfo {author} {\bibfnamefont {H.}~\bibnamefont
  {Wachtel}}, \ and\ \bibinfo {author} {\bibfnamefont {H.}~\bibnamefont
  {Wolf}},\ }\href@noop {} {\bibfield  {journal} {\bibinfo  {journal} {Chem.
  Phys. Lett.}\ }\textbf {\bibinfo {volume} {303}},\ \bibinfo {pages} {157 }
  (\bibinfo {year} {1999})}\BibitemShut {NoStop}%
\bibitem [{\citenamefont {Xu}\ and\ \citenamefont
  {Mitzi}(2003)}]{mitzi_chemmat_03_SnI_acene}%
  \BibitemOpen
  \bibfield  {author} {\bibinfo {author} {\bibfnamefont {Z.}~\bibnamefont
  {Xu}}\ and\ \bibinfo {author} {\bibfnamefont {D.~B.}\ \bibnamefont {Mitzi}},\
  }\href {\doibase 10.1021/cm034267j} {\bibfield  {journal} {\bibinfo
  {journal} {Chem. Mater.}\ }\textbf {\bibinfo {volume} {15}},\ \bibinfo
  {pages} {3632} (\bibinfo {year} {2003})}\BibitemShut {NoStop}%
\bibitem [{\citenamefont {Ema}\ \emph {et~al.}(2008)\citenamefont {Ema},
  \citenamefont {Inomata}, \citenamefont {Kato}, \citenamefont {Kunugita},\
  and\ \citenamefont {Era}}]{EmaPrl2008}%
  \BibitemOpen
  \bibfield  {author} {\bibinfo {author} {\bibfnamefont {K.}~\bibnamefont
  {Ema}}, \bibinfo {author} {\bibfnamefont {M.}~\bibnamefont {Inomata}},
  \bibinfo {author} {\bibfnamefont {Y.}~\bibnamefont {Kato}}, \bibinfo {author}
  {\bibfnamefont {H.}~\bibnamefont {Kunugita}}, \ and\ \bibinfo {author}
  {\bibfnamefont {M.}~\bibnamefont {Era}},\ }\href {\doibase
  10.1103/PhysRevLett.100.257401} {\bibfield  {journal} {\bibinfo  {journal}
  {Phys. Rev. Lett.}\ }\textbf {\bibinfo {volume} {100}},\ \bibinfo {pages}
  {257401} (\bibinfo {year} {2008})}\BibitemShut {NoStop}%
\bibitem [{\citenamefont {Du}\ \emph {et~al.}(2017)\citenamefont {Du},
  \citenamefont {Tu}, \citenamefont {Zhang}, \citenamefont {Han}, \citenamefont
  {Liu}, \citenamefont {Zauscher},\ and\ \citenamefont {Mitzi}}]{kezhao}%
  \BibitemOpen
  \bibfield  {author} {\bibinfo {author} {\bibfnamefont {K.}~\bibnamefont
  {Du}}, \bibinfo {author} {\bibfnamefont {Q.}~\bibnamefont {Tu}}, \bibinfo
  {author} {\bibfnamefont {X.}~\bibnamefont {Zhang}}, \bibinfo {author}
  {\bibfnamefont {Q.}~\bibnamefont {Han}}, \bibinfo {author} {\bibfnamefont
  {J.}~\bibnamefont {Liu}}, \bibinfo {author} {\bibfnamefont {S.}~\bibnamefont
  {Zauscher}}, \ and\ \bibinfo {author} {\bibfnamefont {D.~B.}\ \bibnamefont
  {Mitzi}},\ }\href {http://dx.doi.org/10.1021/acs.inorgchem.7b01094}
  {\bibfield  {journal} {\bibinfo  {journal} {Inorg. Chem.}\ }\textbf {\bibinfo
  {volume} {56}},\ \bibinfo {pages} {9291} (\bibinfo {year}
  {2017})}\BibitemShut {NoStop}%
\bibitem [{\citenamefont {Mitzi}\ \emph {et~al.}(1999)\citenamefont {Mitzi},
  \citenamefont {Chondroudis},\ and\ \citenamefont {Kagan}}]{Mitzi1999AE4T}%
  \BibitemOpen
  \bibfield  {author} {\bibinfo {author} {\bibfnamefont {D.~B.}\ \bibnamefont
  {Mitzi}}, \bibinfo {author} {\bibfnamefont {K.}~\bibnamefont {Chondroudis}},
  \ and\ \bibinfo {author} {\bibfnamefont {C.~R.}\ \bibnamefont {Kagan}},\
  }\href {\doibase 10.1021/ic991048k} {\bibfield  {journal} {\bibinfo
  {journal} {Inorg. Chem.}\ }\textbf {\bibinfo {volume} {38}},\ \bibinfo
  {pages} {6246} (\bibinfo {year} {1999})}\BibitemShut {NoStop}%
\bibitem [{\citenamefont {Mitzi}(2000)}]{mitzi_ic_2000_Bi_AE4T}%
  \BibitemOpen
  \bibfield  {author} {\bibinfo {author} {\bibfnamefont {D.~B.}\ \bibnamefont
  {Mitzi}},\ }\href {\doibase 10.1021/ic000794i} {\bibfield  {journal}
  {\bibinfo  {journal} {Inorg. Chem.}\ }\textbf {\bibinfo {volume} {39}},\
  \bibinfo {pages} {6107} (\bibinfo {year} {2000})}\BibitemShut {NoStop}%
\bibitem [{\citenamefont {Zhu}\ \emph {et~al.}(2003)\citenamefont {Zhu},
  \citenamefont {Mercier}, \citenamefont {Fr\`ere}, \citenamefont {Blanchard},
  \citenamefont {Roncali}, \citenamefont {Allain}, \citenamefont {Pasquier},\
  and\ \citenamefont {Riou}}]{zhu_ic_2003_2T_HOIP}%
  \BibitemOpen
  \bibfield  {author} {\bibinfo {author} {\bibfnamefont {X.-H.}\ \bibnamefont
  {Zhu}}, \bibinfo {author} {\bibfnamefont {N.}~\bibnamefont {Mercier}},
  \bibinfo {author} {\bibfnamefont {P.}~\bibnamefont {Fr\`ere}}, \bibinfo
  {author} {\bibfnamefont {P.}~\bibnamefont {Blanchard}}, \bibinfo {author}
  {\bibfnamefont {J.}~\bibnamefont {Roncali}}, \bibinfo {author} {\bibfnamefont
  {M.}~\bibnamefont {Allain}}, \bibinfo {author} {\bibfnamefont
  {C.}~\bibnamefont {Pasquier}}, \ and\ \bibinfo {author} {\bibfnamefont
  {A.}~\bibnamefont {Riou}},\ }\href {\doibase 10.1021/ic034235y} {\bibfield
  {journal} {\bibinfo  {journal} {Inorg. Chem.}\ }\textbf {\bibinfo {volume}
  {42}},\ \bibinfo {pages} {5330} (\bibinfo {year} {2003})}\BibitemShut
  {NoStop}%
\bibitem [{\citenamefont {Era}\ \emph {et~al.}(2003)\citenamefont {Era},
  \citenamefont {Yoneda}, \citenamefont {Sano},\ and\ \citenamefont
  {Noto}}]{era_thin_sol_film_03_poly_thiophene}%
  \BibitemOpen
  \bibfield  {author} {\bibinfo {author} {\bibfnamefont {M.}~\bibnamefont
  {Era}}, \bibinfo {author} {\bibfnamefont {S.}~\bibnamefont {Yoneda}},
  \bibinfo {author} {\bibfnamefont {T.}~\bibnamefont {Sano}}, \ and\ \bibinfo
  {author} {\bibfnamefont {M.}~\bibnamefont {Noto}},\ }\href@noop {} {\bibfield
   {journal} {\bibinfo  {journal} {Thin Solid Films}\ }\textbf {\bibinfo
  {volume} {438}},\ \bibinfo {pages} {322} (\bibinfo {year}
  {2003})}\BibitemShut {NoStop}%
\bibitem [{\citenamefont {Mitzi}(1999)}]{Mitzi1999ProgInorgChem}%
  \BibitemOpen
  \bibfield  {author} {\bibinfo {author} {\bibfnamefont {D.~B.}\ \bibnamefont
  {Mitzi}},\ }\href {\doibase 10.1002/9780470166499.ch1} {\bibfield  {journal}
  {\bibinfo  {journal} {Prog. Inorg. Chem.}\ }\textbf {\bibinfo {volume}
  {48}},\ \bibinfo {pages} {1} (\bibinfo {year} {1999})}\BibitemShut {NoStop}%
\bibitem [{\citenamefont {Mitzi}(2001)}]{Mitzi2001Dalton}%
  \BibitemOpen
  \bibfield  {author} {\bibinfo {author} {\bibfnamefont {D.~B.}\ \bibnamefont
  {Mitzi}},\ }\href {\doibase 10.1039/B007070J} {\bibfield  {journal} {\bibinfo
   {journal} {J. Chem. Soc. Dalton Trans.}\ }\textbf {\bibinfo {volume} {0}},\
  \bibinfo {pages} {1} (\bibinfo {year} {2001})}\BibitemShut {NoStop}%
\bibitem [{\citenamefont {Zhang}\ \emph {et~al.}(2009)\citenamefont {Zhang},
  \citenamefont {Lanty}, \citenamefont {Lauret}, \citenamefont {Deleporte},
  \citenamefont {Audebert},\ and\ \citenamefont {Galmiche}}]{Zhang2009}%
  \BibitemOpen
  \bibfield  {author} {\bibinfo {author} {\bibfnamefont {S.}~\bibnamefont
  {Zhang}}, \bibinfo {author} {\bibfnamefont {G.}~\bibnamefont {Lanty}},
  \bibinfo {author} {\bibfnamefont {J.-S.}\ \bibnamefont {Lauret}}, \bibinfo
  {author} {\bibfnamefont {E.}~\bibnamefont {Deleporte}}, \bibinfo {author}
  {\bibfnamefont {P.}~\bibnamefont {Audebert}}, \ and\ \bibinfo {author}
  {\bibfnamefont {L.}~\bibnamefont {Galmiche}},\ }\href {\doibase
  https://doi.org/10.1016/j.actamat.2009.03.037} {\bibfield  {journal}
  {\bibinfo  {journal} {Acta. Mater.}\ }\textbf {\bibinfo {volume} {57}},\
  \bibinfo {pages} {3301 } (\bibinfo {year} {2009})}\BibitemShut {NoStop}%
\bibitem [{\citenamefont {Cortecchia}\ \emph {et~al.}(2017)\citenamefont
  {Cortecchia}, \citenamefont {Soci}, \citenamefont {Cametti}, \citenamefont
  {Petrozza},\ and\ \citenamefont {Martí-Rujas}}]{2DhoipOrg2017chemchem}%
  \BibitemOpen
  \bibfield  {author} {\bibinfo {author} {\bibfnamefont {D.}~\bibnamefont
  {Cortecchia}}, \bibinfo {author} {\bibfnamefont {C.}~\bibnamefont {Soci}},
  \bibinfo {author} {\bibfnamefont {M.}~\bibnamefont {Cametti}}, \bibinfo
  {author} {\bibfnamefont {A.}~\bibnamefont {Petrozza}}, \ and\ \bibinfo
  {author} {\bibfnamefont {J.}~\bibnamefont {Martí-Rujas}},\ }\href {\doibase
  10.1002/cplu.201600477} {\bibfield  {journal} {\bibinfo  {journal}
  {ChemPlusChem}\ }\textbf {\bibinfo {volume} {82}},\ \bibinfo {pages} {681}
  (\bibinfo {year} {2017})}\BibitemShut {NoStop}%
\bibitem [{\citenamefont {Ishihara}\ \emph {et~al.}(1989)\citenamefont
  {Ishihara}, \citenamefont {Takahashi},\ and\ \citenamefont
  {Goto}}]{Ishihara1989}%
  \BibitemOpen
  \bibfield  {author} {\bibinfo {author} {\bibfnamefont {T.}~\bibnamefont
  {Ishihara}}, \bibinfo {author} {\bibfnamefont {J.}~\bibnamefont {Takahashi}},
  \ and\ \bibinfo {author} {\bibfnamefont {T.}~\bibnamefont {Goto}},\ }\href
  {\doibase https://doi.org/10.1016/0038-1098(89)90935-6} {\bibfield  {journal}
  {\bibinfo  {journal} {Solid State Communications}\ }\textbf {\bibinfo
  {volume} {69}},\ \bibinfo {pages} {933 } (\bibinfo {year}
  {1989})}\BibitemShut {NoStop}%
\bibitem [{\citenamefont {Era}\ \emph {et~al.}(1994)\citenamefont {Era},
  \citenamefont {Morimoto}, \citenamefont {Tsutsui},\ and\ \citenamefont
  {Saito}}]{Era1994APL}%
  \BibitemOpen
  \bibfield  {author} {\bibinfo {author} {\bibfnamefont {M.}~\bibnamefont
  {Era}}, \bibinfo {author} {\bibfnamefont {S.}~\bibnamefont {Morimoto}},
  \bibinfo {author} {\bibfnamefont {T.}~\bibnamefont {Tsutsui}}, \ and\
  \bibinfo {author} {\bibfnamefont {S.}~\bibnamefont {Saito}},\ }\href
  {\doibase 10.1063/1.112265} {\bibfield  {journal} {\bibinfo  {journal}
  {Applied Physics Letters}\ }\textbf {\bibinfo {volume} {65}},\ \bibinfo
  {pages} {676} (\bibinfo {year} {1994})}\BibitemShut {NoStop}%
\bibitem [{\citenamefont {Wang}\ \emph {et~al.}(2016)\citenamefont {Wang},
  \citenamefont {Cheng}, \citenamefont {Ge}, \citenamefont {Zhang},
  \citenamefont {Miao}, \citenamefont {Zou}, \citenamefont {Yi}, \citenamefont
  {Sun}, \citenamefont {Cao}, \citenamefont {Yang}, \citenamefont {Wei},
  \citenamefont {Guo}, \citenamefont {Ke}, \citenamefont {Yu}, \citenamefont
  {Jin}, \citenamefont {Liu}, \citenamefont {Ding}, \citenamefont {Di},
  \citenamefont {Yang}, \citenamefont {Xing}, \citenamefont {Tian},
  \citenamefont {Jin}, \citenamefont {Gao}, \citenamefont {Friend},
  \citenamefont {Wang},\ and\ \citenamefont {Huang}}]{emit16nature}%
  \BibitemOpen
  \bibfield  {author} {\bibinfo {author} {\bibfnamefont {N.}~\bibnamefont
  {Wang}}, \bibinfo {author} {\bibfnamefont {L.}~\bibnamefont {Cheng}},
  \bibinfo {author} {\bibfnamefont {R.}~\bibnamefont {Ge}}, \bibinfo {author}
  {\bibfnamefont {S.}~\bibnamefont {Zhang}}, \bibinfo {author} {\bibfnamefont
  {Y.}~\bibnamefont {Miao}}, \bibinfo {author} {\bibfnamefont {W.}~\bibnamefont
  {Zou}}, \bibinfo {author} {\bibfnamefont {C.}~\bibnamefont {Yi}}, \bibinfo
  {author} {\bibfnamefont {Y.}~\bibnamefont {Sun}}, \bibinfo {author}
  {\bibfnamefont {Y.}~\bibnamefont {Cao}}, \bibinfo {author} {\bibfnamefont
  {R.}~\bibnamefont {Yang}}, \bibinfo {author} {\bibfnamefont {Y.}~\bibnamefont
  {Wei}}, \bibinfo {author} {\bibfnamefont {Q.}~\bibnamefont {Guo}}, \bibinfo
  {author} {\bibfnamefont {Y.}~\bibnamefont {Ke}}, \bibinfo {author}
  {\bibfnamefont {M.}~\bibnamefont {Yu}}, \bibinfo {author} {\bibfnamefont
  {Y.}~\bibnamefont {Jin}}, \bibinfo {author} {\bibfnamefont {Y.}~\bibnamefont
  {Liu}}, \bibinfo {author} {\bibfnamefont {Q.}~\bibnamefont {Ding}}, \bibinfo
  {author} {\bibfnamefont {D.}~\bibnamefont {Di}}, \bibinfo {author}
  {\bibfnamefont {L.}~\bibnamefont {Yang}}, \bibinfo {author} {\bibfnamefont
  {G.}~\bibnamefont {Xing}}, \bibinfo {author} {\bibfnamefont {H.}~\bibnamefont
  {Tian}}, \bibinfo {author} {\bibfnamefont {C.}~\bibnamefont {Jin}}, \bibinfo
  {author} {\bibfnamefont {F.}~\bibnamefont {Gao}}, \bibinfo {author}
  {\bibfnamefont {R.~H.}\ \bibnamefont {Friend}}, \bibinfo {author}
  {\bibfnamefont {J.}~\bibnamefont {Wang}}, \ and\ \bibinfo {author}
  {\bibfnamefont {W.}~\bibnamefont {Huang}},\ }\href
  {http://dx.doi.org/10.1038/nphoton.2016.185} {\bibfield  {journal} {\bibinfo
  {journal} {Nat. Photonics}\ }\textbf {\bibinfo {volume} {10}},\ \bibinfo
  {pages} {699} (\bibinfo {year} {2016})}\BibitemShut {NoStop}%
\bibitem [{\citenamefont {Mitzi}\ \emph {et~al.}(2001)\citenamefont {Mitzi},
  \citenamefont {Chondroudis},\ and\ \citenamefont {Kagan}}]{Mitzi2001}%
  \BibitemOpen
  \bibfield  {author} {\bibinfo {author} {\bibfnamefont {D.~B.}\ \bibnamefont
  {Mitzi}}, \bibinfo {author} {\bibfnamefont {K.}~\bibnamefont {Chondroudis}},
  \ and\ \bibinfo {author} {\bibfnamefont {C.~R.}\ \bibnamefont {Kagan}},\
  }\href {\doibase 10.1147/rd.451.0029} {\bibfield  {journal} {\bibinfo
  {journal} {IBM J. Res. Dev.}\ }\textbf {\bibinfo {volume} {45}},\ \bibinfo
  {pages} {29} (\bibinfo {year} {2001})}\BibitemShut {NoStop}%
\bibitem [{\citenamefont {Muljarov}\ \emph {et~al.}(1995)\citenamefont
  {Muljarov}, \citenamefont {Tikhodeev}, \citenamefont {Gippius},\ and\
  \citenamefont {Ishihara}}]{prb95Exciton}%
  \BibitemOpen
  \bibfield  {author} {\bibinfo {author} {\bibfnamefont {E.~A.}\ \bibnamefont
  {Muljarov}}, \bibinfo {author} {\bibfnamefont {S.~G.}\ \bibnamefont
  {Tikhodeev}}, \bibinfo {author} {\bibfnamefont {N.~A.}\ \bibnamefont
  {Gippius}}, \ and\ \bibinfo {author} {\bibfnamefont {T.}~\bibnamefont
  {Ishihara}},\ }\href {\doibase 10.1103/PhysRevB.51.14370} {\bibfield
  {journal} {\bibinfo  {journal} {Phys. Rev. B}\ }\textbf {\bibinfo {volume}
  {51}},\ \bibinfo {pages} {14370} (\bibinfo {year} {1995})}\BibitemShut
  {NoStop}%
\bibitem [{\citenamefont {Even}\ \emph {et~al.}(2014)\citenamefont {Even},
  \citenamefont {Pedesseau},\ and\ \citenamefont {Katan}}]{even14cpc}%
  \BibitemOpen
  \bibfield  {author} {\bibinfo {author} {\bibfnamefont {J.}~\bibnamefont
  {Even}}, \bibinfo {author} {\bibfnamefont {L.}~\bibnamefont {Pedesseau}}, \
  and\ \bibinfo {author} {\bibfnamefont {C.}~\bibnamefont {Katan}},\ }\href
  {\doibase 10.1002/cphc.201402428} {\bibfield  {journal} {\bibinfo  {journal}
  {ChemPhysChem}\ }\textbf {\bibinfo {volume} {15}},\ \bibinfo {pages} {3733}
  (\bibinfo {year} {2014})}\BibitemShut {NoStop}%
\bibitem [{\citenamefont {Even}\ \emph {et~al.}(2015)\citenamefont {Even},
  \citenamefont {Pedesseau}, \citenamefont {Katan}, \citenamefont {Kepenekian},
  \citenamefont {Lauret}, \citenamefont {Sapori},\ and\ \citenamefont
  {Deleporte}}]{even15jpcc}%
  \BibitemOpen
  \bibfield  {author} {\bibinfo {author} {\bibfnamefont {J.}~\bibnamefont
  {Even}}, \bibinfo {author} {\bibfnamefont {L.}~\bibnamefont {Pedesseau}},
  \bibinfo {author} {\bibfnamefont {C.}~\bibnamefont {Katan}}, \bibinfo
  {author} {\bibfnamefont {M.}~\bibnamefont {Kepenekian}}, \bibinfo {author}
  {\bibfnamefont {J.-S.}\ \bibnamefont {Lauret}}, \bibinfo {author}
  {\bibfnamefont {D.}~\bibnamefont {Sapori}}, \ and\ \bibinfo {author}
  {\bibfnamefont {E.}~\bibnamefont {Deleporte}},\ }\href {\doibase
  10.1021/acs.jpcc.5b00695} {\bibfield  {journal} {\bibinfo  {journal} {J.
  Phys. Chem. C}\ }\textbf {\bibinfo {volume} {119}},\ \bibinfo {pages} {10161}
  (\bibinfo {year} {2015})}\BibitemShut {NoStop}%
\bibitem [{sup()}]{supplemental}%
  \BibitemOpen
  \href@noop {} {}\bibinfo {note} {See Supplemental Material for: structural details of the perovskites in our
  simulations; energetic impact of the supercell choice; impact of spin-orbit
  coupling in the reported energy band structures; computational details (basis
  sets and k-space grid convergence); validation of computational protocols by
  comparing to experimental results for MAPbI$_3$, for AE4TPbBr$_4$ and for
  AE4TPbI$_4$; experimental details for the synthesis and characterization of a
  new AE4TPbI$_4$ crystal; and additional computed orbitals, energy band
  structures, band curvature parameters, full and partial densities of states
  supporting the general findings demonstrated in the actual paper.
  Computational validation for MAPbI$_3$ also includes comparison to a more
  sophisticated many-body dispersion treatment.\cite{Ambrosetti2014} Synthesis
  details include Refs.~\cite{exp1,exp2,exp3}.
  All computational input
  and output files (raw data) are available at the NOMAD repository.\cite{NOMAD-URL}
  }\BibitemShut {Stop}%
\bibitem [{\citenamefont {Ambrosetti}\ \emph {et~al.}(2014)\citenamefont
  {Ambrosetti}, \citenamefont {Reilly}, \citenamefont {Jr.},\ and\
  \citenamefont {Tkatchenko}}]{Ambrosetti2014}%
  \BibitemOpen
  \bibfield  {author} {\bibinfo {author} {\bibfnamefont {A.}~\bibnamefont
  {Ambrosetti}}, \bibinfo {author} {\bibfnamefont {A.~M.}\ \bibnamefont
  {Reilly}}, \bibinfo {author} {\bibfnamefont {R.~A.~DiStasio}\ \bibnamefont {Jr.}},
  \ and\ \bibinfo {author} {\bibfnamefont {A.}~\bibnamefont {Tkatchenko}},\
  }\href {\doibase 10.1063/1.4865104} {\bibfield  {journal} {\bibinfo
  {journal} {J. Chem. Phys.}\ }\textbf {\bibinfo {volume} {140}},\ \bibinfo
  {pages} {18A508} (\bibinfo {year} {2014})}\BibitemShut {NoStop}%
\bibitem [{\citenamefont {Muguruma}\ \emph
  {et~al.}(1996{\natexlab{a}})\citenamefont {Muguruma}, \citenamefont {Saito},
  \citenamefont {Hiratsuka}, \citenamefont {Karube},\ and\ \citenamefont
  {Hotta}}]{exp1}%
  \BibitemOpen
  \bibfield  {author} {\bibinfo {author} {\bibfnamefont {H.}~\bibnamefont
  {Muguruma}}, \bibinfo {author} {\bibfnamefont {T.}~\bibnamefont {Saito}},
  \bibinfo {author} {\bibfnamefont {A.}~\bibnamefont {Hiratsuka}}, \bibinfo
  {author} {\bibfnamefont {I.}~\bibnamefont {Karube}}, \ and\ \bibinfo {author}
  {\bibfnamefont {S.}~\bibnamefont {Hotta}},\ }\href {\doibase
  10.1021/la9605706} {\bibfield  {journal} {\bibinfo  {journal} {Langmuir}\
  }\textbf {\bibinfo {volume} {12}},\ \bibinfo {pages} {5451} (\bibinfo {year}
  {1996}{\natexlab{a}})}\BibitemShut {NoStop}%
\bibitem [{\citenamefont {Muguruma}\ \emph
  {et~al.}(1996{\natexlab{b}})\citenamefont {Muguruma}, \citenamefont {Saito},
  \citenamefont {Sasaki}, \citenamefont {Hotta},\ and\ \citenamefont
  {Karube}}]{exp2}%
  \BibitemOpen
  \bibfield  {author} {\bibinfo {author} {\bibfnamefont {H.}~\bibnamefont
  {Muguruma}}, \bibinfo {author} {\bibfnamefont {T.}~\bibnamefont {Saito}},
  \bibinfo {author} {\bibfnamefont {S.}~\bibnamefont {Sasaki}}, \bibinfo
  {author} {\bibfnamefont {S.}~\bibnamefont {Hotta}}, \ and\ \bibinfo {author}
  {\bibfnamefont {I.}~\bibnamefont {Karube}},\ }\href@noop {} {\bibfield
  {journal} {\bibinfo  {journal} {J. Heterocyclic Chem.}\ }\textbf {\bibinfo
  {volume} {33}},\ \bibinfo {pages} {173} (\bibinfo {year}
  {1996}{\natexlab{b}})}\BibitemShut {NoStop}%
\bibitem [{\citenamefont {Muguruma}\ \emph {et~al.}(1998)\citenamefont
  {Muguruma}, \citenamefont {Kobiro},\ and\ \citenamefont {Hotta}}]{exp3}%
  \BibitemOpen
  \bibfield  {author} {\bibinfo {author} {\bibfnamefont {H.}~\bibnamefont
  {Muguruma}}, \bibinfo {author} {\bibfnamefont {K.}~\bibnamefont {Kobiro}}, \
  and\ \bibinfo {author} {\bibfnamefont {S.}~\bibnamefont {Hotta}},\ }\href
  {\doibase 10.1021/cm980041i} {\bibfield  {journal} {\bibinfo  {journal}
  {Chem. Mater.}\ }\textbf {\bibinfo {volume} {10}},\ \bibinfo {pages} {1459}
  (\bibinfo {year} {1998})}\BibitemShut {NoStop}%
\bibitem [{NOM()}]{NOMAD-URL}%
  \BibitemOpen
  \href@noop {} {}\bibinfo {note} {NOMAD Repository.
  http://doi.org/10.17172/ NOMAD/2018.09.21-1 (accessed Sep 22,
  2018).}\BibitemShut {Stop}%
\bibitem [{\citenamefont {Perdew}\ \emph {et~al.}(1996)\citenamefont {Perdew},
  \citenamefont {Burke},\ and\ \citenamefont {Ernzerhof}}]{PBE}%
  \BibitemOpen
  \bibfield  {author} {\bibinfo {author} {\bibfnamefont {J.~P.}\ \bibnamefont
  {Perdew}}, \bibinfo {author} {\bibfnamefont {K.}~\bibnamefont {Burke}}, \
  and\ \bibinfo {author} {\bibfnamefont {M.}~\bibnamefont {Ernzerhof}},\ }\href
  {\doibase 10.1103/PhysRevLett.77.3865} {\bibfield  {journal} {\bibinfo
  {journal} {Phys. Rev. Lett.}\ }\textbf {\bibinfo {volume} {77}},\ \bibinfo
  {pages} {3865} (\bibinfo {year} {1996})}\BibitemShut {NoStop}%
\bibitem [{\citenamefont {Tkatchenko}\ and\ \citenamefont
  {Scheffler}(2009)}]{Tkatchenko09}%
  \BibitemOpen
  \bibfield  {author} {\bibinfo {author} {\bibfnamefont {A.}~\bibnamefont
  {Tkatchenko}}\ and\ \bibinfo {author} {\bibfnamefont {M.}~\bibnamefont
  {Scheffler}},\ }\href {\doibase 10.1103/PhysRevLett.102.073005} {\bibfield
  {journal} {\bibinfo  {journal} {Phys. Rev. Lett.}\ }\textbf {\bibinfo
  {volume} {102}},\ \bibinfo {pages} {073005} (\bibinfo {year}
  {2009})}\BibitemShut {NoStop}%
\bibitem [{\citenamefont {Ren}\ \emph {et~al.}(2012)\citenamefont {Ren},
  \citenamefont {Rinke}, \citenamefont {Blum}, \citenamefont {Wieferink},
  \citenamefont {Tkatchenko}, \citenamefont {Sanfilippo}, \citenamefont
  {Reuter},\ and\ \citenamefont {Scheffler}}]{Ren2012NJP}%
  \BibitemOpen
  \bibfield  {author} {\bibinfo {author} {\bibfnamefont {X.}~\bibnamefont
  {Ren}}, \bibinfo {author} {\bibfnamefont {P.}~\bibnamefont {Rinke}}, \bibinfo
  {author} {\bibfnamefont {V.}~\bibnamefont {Blum}}, \bibinfo {author}
  {\bibfnamefont {J.}~\bibnamefont {Wieferink}}, \bibinfo {author}
  {\bibfnamefont {A.}~\bibnamefont {Tkatchenko}}, \bibinfo {author}
  {\bibfnamefont {A.}~\bibnamefont {Sanfilippo}}, \bibinfo {author}
  {\bibfnamefont {K.}~\bibnamefont {Reuter}}, \ and\ \bibinfo {author}
  {\bibfnamefont {M.}~\bibnamefont {Scheffler}},\ }\href
  {http://stacks.iop.org/1367-2630/14/i=5/a=053020} {\bibfield  {journal}
  {\bibinfo  {journal} {New J. Phys.}\ }\textbf {\bibinfo {volume} {14}},\
  \bibinfo {pages} {053020} (\bibinfo {year} {2012})}\BibitemShut {NoStop}%
\bibitem [{\citenamefont {Reining}(2017)}]{reining18gw}%
  \BibitemOpen
  \bibfield  {author} {\bibinfo {author} {\bibfnamefont {L.}~\bibnamefont
  {Reining}},\ }\href {\doibase 10.1002/wcms.1344} {\bibfield  {journal}
  {\bibinfo  {journal} {WIREs Comput. Mol. Sci.}\ }\textbf {\bibinfo {volume}
  {12}},\ \bibinfo {pages} {1344} (\bibinfo {year} {2017})}\BibitemShut
  {NoStop}%
\bibitem [{\citenamefont {Heyd}\ \emph {et~al.}(2003)\citenamefont {Heyd},
  \citenamefont {Scuseria},\ and\ \citenamefont {Ernzerhof}}]{hse03}%
  \BibitemOpen
  \bibfield  {author} {\bibinfo {author} {\bibfnamefont {J.}~\bibnamefont
  {Heyd}}, \bibinfo {author} {\bibfnamefont {G.~E.}\ \bibnamefont {Scuseria}},
  \ and\ \bibinfo {author} {\bibfnamefont {M.}~\bibnamefont {Ernzerhof}},\
  }\href {\doibase http://dx.doi.org/10.1063/1.1564060} {\bibfield  {journal}
  {\bibinfo  {journal} {J. Chem. Phys.}\ }\textbf {\bibinfo {volume} {118}},\
  \bibinfo {pages} {8207} (\bibinfo {year} {2003})}\BibitemShut {NoStop}%
\bibitem [{\citenamefont {Heyd}\ \emph {et~al.}(2006)\citenamefont {Heyd},
  \citenamefont {Scuseria},\ and\ \citenamefont {Ernzerhof}}]{hse06}%
  \BibitemOpen
  \bibfield  {author} {\bibinfo {author} {\bibfnamefont {J.}~\bibnamefont
  {Heyd}}, \bibinfo {author} {\bibfnamefont {G.~E.}\ \bibnamefont {Scuseria}},
  \ and\ \bibinfo {author} {\bibfnamefont {M.}~\bibnamefont {Ernzerhof}},\
  }\href {\doibase 10.1063/1.2204597} {\bibfield  {journal} {\bibinfo
  {journal} {J. Chem. Phys.}\ }\textbf {\bibinfo {volume} {124}},\ \bibinfo
  {pages} {219906} (\bibinfo {year} {2006})}\BibitemShut {NoStop}%
\bibitem [{\citenamefont {Huhn}\ and\ \citenamefont {Blum}(2017)}]{Huhn2017}%
  \BibitemOpen
  \bibfield  {author} {\bibinfo {author} {\bibfnamefont {W.~P.}\ \bibnamefont
  {Huhn}}\ and\ \bibinfo {author} {\bibfnamefont {V.}~\bibnamefont {Blum}},\
  }\href {\doibase 10.1103/PhysRevMaterials.1.033803} {\bibfield  {journal}
  {\bibinfo  {journal} {Phys. Rev. Materials}\ }\textbf {\bibinfo {volume}
  {1}},\ \bibinfo {pages} {033803} (\bibinfo {year} {2017})}\BibitemShut
  {NoStop}%
\bibitem [{\citenamefont {Perdew}\ \emph {et~al.}(2017)\citenamefont {Perdew},
  \citenamefont {Yang}, \citenamefont {Burke}, \citenamefont {Yang},
  \citenamefont {Gross}, \citenamefont {Scheffler}, \citenamefont {Scuseria},
  \citenamefont {Henderson}, \citenamefont {Zhang}, \citenamefont {Ruzsinszky},
  \citenamefont {Peng}, \citenamefont {Sun}, \citenamefont {Trushin},\ and\
  \citenamefont {G\"orling}}]{Perdew14032017}%
  \BibitemOpen
  \bibfield  {author} {\bibinfo {author} {\bibfnamefont {J.~P.}\ \bibnamefont
  {Perdew}}, \bibinfo {author} {\bibfnamefont {W.}~\bibnamefont {Yang}},
  \bibinfo {author} {\bibfnamefont {K.}~\bibnamefont {Burke}}, \bibinfo
  {author} {\bibfnamefont {Z.}~\bibnamefont {Yang}}, \bibinfo {author}
  {\bibfnamefont {E.~K.~U.}\ \bibnamefont {Gross}}, \bibinfo {author}
  {\bibfnamefont {M.}~\bibnamefont {Scheffler}}, \bibinfo {author}
  {\bibfnamefont {G.~E.}\ \bibnamefont {Scuseria}}, \bibinfo {author}
  {\bibfnamefont {T.~M.}\ \bibnamefont {Henderson}}, \bibinfo {author}
  {\bibfnamefont {I.~Y.}\ \bibnamefont {Zhang}}, \bibinfo {author}
  {\bibfnamefont {A.}~\bibnamefont {Ruzsinszky}}, \bibinfo {author}
  {\bibfnamefont {H.}~\bibnamefont {Peng}}, \bibinfo {author} {\bibfnamefont
  {J.}~\bibnamefont {Sun}}, \bibinfo {author} {\bibfnamefont {E.}~\bibnamefont
  {Trushin}}, \ and\ \bibinfo {author} {\bibfnamefont {A.}~\bibnamefont
  {G\"orling}},\ }\href {\doibase 10.1073/pnas.1621352114} {\bibfield
  {journal} {\bibinfo  {journal} {Proc. Natl. Acad. Sci.}\ }\textbf {\bibinfo
  {volume} {114}},\ \bibinfo {pages} {2801} (\bibinfo {year}
  {2017})}\BibitemShut {NoStop}%
\bibitem [{\citenamefont {Blum}\ \emph {et~al.}(2009)\citenamefont {Blum},
  \citenamefont {Gehrke}, \citenamefont {Hanke}, \citenamefont {Havu},
  \citenamefont {Havu}, \citenamefont {Ren}, \citenamefont {Reuter},\ and\
  \citenamefont {Scheffler}}]{Blum2009}%
  \BibitemOpen
  \bibfield  {author} {\bibinfo {author} {\bibfnamefont {V.}~\bibnamefont
  {Blum}}, \bibinfo {author} {\bibfnamefont {R.}~\bibnamefont {Gehrke}},
  \bibinfo {author} {\bibfnamefont {F.}~\bibnamefont {Hanke}}, \bibinfo
  {author} {\bibfnamefont {P.}~\bibnamefont {Havu}}, \bibinfo {author}
  {\bibfnamefont {V.}~\bibnamefont {Havu}}, \bibinfo {author} {\bibfnamefont
  {X.}~\bibnamefont {Ren}}, \bibinfo {author} {\bibfnamefont {K.}~\bibnamefont
  {Reuter}}, \ and\ \bibinfo {author} {\bibfnamefont {M.}~\bibnamefont
  {Scheffler}},\ }\href {\doibase 10.1016/j.cpc.2009.06.022} {\bibfield
  {journal} {\bibinfo  {journal} {Comput. Phys. Commun.}\ }\textbf {\bibinfo
  {volume} {180}},\ \bibinfo {pages} {2175} (\bibinfo {year}
  {2009})}\BibitemShut {NoStop}%
\bibitem [{\citenamefont {Havu}\ \emph {et~al.}(2009)\citenamefont {Havu},
  \citenamefont {Blum}, \citenamefont {Havu},\ and\ \citenamefont
  {Scheffler}}]{Havu2009}%
  \BibitemOpen
  \bibfield  {author} {\bibinfo {author} {\bibfnamefont {V.}~\bibnamefont
  {Havu}}, \bibinfo {author} {\bibfnamefont {V.}~\bibnamefont {Blum}}, \bibinfo
  {author} {\bibfnamefont {P.}~\bibnamefont {Havu}}, \ and\ \bibinfo {author}
  {\bibfnamefont {M.}~\bibnamefont {Scheffler}},\ }\href {\doibase
  http://dx.doi.org/10.1016/j.jcp.2009.08.008} {\bibfield  {journal} {\bibinfo
  {journal} {J. Comput. Phys.}\ }\textbf {\bibinfo {volume} {228}},\ \bibinfo
  {pages} {8367 } (\bibinfo {year} {2009})}\BibitemShut {NoStop}%
\bibitem [{\citenamefont {Levchenko}\ \emph {et~al.}(2015)\citenamefont
  {Levchenko}, \citenamefont {Ren}, \citenamefont {Wieferink}, \citenamefont
  {Johanni}, \citenamefont {Rinke}, \citenamefont {Blum},\ and\ \citenamefont
  {Scheffler}}]{LEVCHENKO201560}%
  \BibitemOpen
  \bibfield  {author} {\bibinfo {author} {\bibfnamefont {S.~V.}\ \bibnamefont
  {Levchenko}}, \bibinfo {author} {\bibfnamefont {X.}~\bibnamefont {Ren}},
  \bibinfo {author} {\bibfnamefont {J.}~\bibnamefont {Wieferink}}, \bibinfo
  {author} {\bibfnamefont {R.}~\bibnamefont {Johanni}}, \bibinfo {author}
  {\bibfnamefont {P.}~\bibnamefont {Rinke}}, \bibinfo {author} {\bibfnamefont
  {V.}~\bibnamefont {Blum}}, \ and\ \bibinfo {author} {\bibfnamefont
  {M.}~\bibnamefont {Scheffler}},\ }\href {\doibase
  https://doi.org/10.1016/j.cpc.2015.02.021} {\bibfield  {journal} {\bibinfo
  {journal} {Comput. Phys. Commun.}\ }\textbf {\bibinfo {volume} {192}},\
  \bibinfo {pages} {60 } (\bibinfo {year} {2015})}\BibitemShut {NoStop}%
\bibitem [{\citenamefont {Zhang}\ \emph {et~al.}(2013)\citenamefont {Zhang},
  \citenamefont {Ren}, \citenamefont {Rinke}, \citenamefont {Blum},\ and\
  \citenamefont {Scheffler}}]{Igor2015NJP}%
  \BibitemOpen
  \bibfield  {author} {\bibinfo {author} {\bibfnamefont {I.~Y.}\ \bibnamefont
  {Zhang}}, \bibinfo {author} {\bibfnamefont {X.}~\bibnamefont {Ren}}, \bibinfo
  {author} {\bibfnamefont {P.}~\bibnamefont {Rinke}}, \bibinfo {author}
  {\bibfnamefont {V.}~\bibnamefont {Blum}}, \ and\ \bibinfo {author}
  {\bibfnamefont {M.}~\bibnamefont {Scheffler}},\ }\href
  {http://stacks.iop.org/1367-2630/15/i=12/a=123033} {\bibfield  {journal}
  {\bibinfo  {journal} {New J. Phys.}\ }\textbf {\bibinfo {volume} {15}},\
  \bibinfo {pages} {123033} (\bibinfo {year} {2013})}\BibitemShut {NoStop}%
\bibitem [{\citenamefont {Ihrig}\ \emph {et~al.}(2015)\citenamefont {Ihrig},
  \citenamefont {Wieferink}, \citenamefont {Zhang}, \citenamefont {Ropo},
  \citenamefont {Ren}, \citenamefont {Rinke}, \citenamefont {Scheffler},\ and\
  \citenamefont {Blum}}]{Ihrig2015}%
  \BibitemOpen
  \bibfield  {author} {\bibinfo {author} {\bibfnamefont {A.~C.}\ \bibnamefont
  {Ihrig}}, \bibinfo {author} {\bibfnamefont {J.}~\bibnamefont {Wieferink}},
  \bibinfo {author} {\bibfnamefont {I.~Y.}\ \bibnamefont {Zhang}}, \bibinfo
  {author} {\bibfnamefont {M.}~\bibnamefont {Ropo}}, \bibinfo {author}
  {\bibfnamefont {X.}~\bibnamefont {Ren}}, \bibinfo {author} {\bibfnamefont
  {P.}~\bibnamefont {Rinke}}, \bibinfo {author} {\bibfnamefont
  {M.}~\bibnamefont {Scheffler}}, \ and\ \bibinfo {author} {\bibfnamefont
  {V.}~\bibnamefont {Blum}},\ }\href
  {http://stacks.iop.org/1367-2630/17/i=9/a=093020} {\bibfield  {journal}
  {\bibinfo  {journal} {New Journal of Physics}\ }\textbf {\bibinfo {volume}
  {17}},\ \bibinfo {pages} {093020} (\bibinfo {year} {2015})}\BibitemShut
  {NoStop}%
\bibitem [{\citenamefont {Yu}\ \emph {et~al.}(2018)\citenamefont {Yu},
  \citenamefont {Corsetti}, \citenamefont {García}, \citenamefont {Huhn},
  \citenamefont {Jacquelin}, \citenamefont {Jia}, \citenamefont {Lange},
  \citenamefont {Lin}, \citenamefont {Lu}, \citenamefont {Mi}, \citenamefont
  {Seifitokaldani}, \citenamefont {Vazquez-Mayagoitia}, \citenamefont {Yang},
  \citenamefont {Yang},\ and\ \citenamefont {Blum}}]{YU2017}%
  \BibitemOpen
  \bibfield  {author} {\bibinfo {author} {\bibfnamefont {V.~W.}\ \bibnamefont
  {Yu}}, \bibinfo {author} {\bibfnamefont {F.}~\bibnamefont {Corsetti}},
  \bibinfo {author} {\bibfnamefont {A.}~\bibnamefont {García}}, \bibinfo
  {author} {\bibfnamefont {W.~P.}\ \bibnamefont {Huhn}}, \bibinfo {author}
  {\bibfnamefont {M.}~\bibnamefont {Jacquelin}}, \bibinfo {author}
  {\bibfnamefont {W.}~\bibnamefont {Jia}}, \bibinfo {author} {\bibfnamefont
  {B.}~\bibnamefont {Lange}}, \bibinfo {author} {\bibfnamefont
  {L.}~\bibnamefont {Lin}}, \bibinfo {author} {\bibfnamefont {J.}~\bibnamefont
  {Lu}}, \bibinfo {author} {\bibfnamefont {W.}~\bibnamefont {Mi}}, \bibinfo
  {author} {\bibfnamefont {A.}~\bibnamefont {Seifitokaldani}}, \bibinfo
  {author} {\bibfnamefont {A.}~\bibnamefont {Vazquez-Mayagoitia}}, \bibinfo
  {author} {\bibfnamefont {C.}~\bibnamefont {Yang}}, \bibinfo {author}
  {\bibfnamefont {H.}~\bibnamefont {Yang}}, \ and\ \bibinfo {author}
  {\bibfnamefont {V.}~\bibnamefont {Blum}},\ }\href
  {http://www.sciencedirect.com/science/article/pii/S0010465517302941}
  {\bibfield  {journal} {\bibinfo  {journal} {Comput. Phys. Commun.}\ }\textbf
  {\bibinfo {volume} {222}},\ \bibinfo {pages} {267} (\bibinfo {year}
  {2018})}\BibitemShut {NoStop}%
\bibitem [{\citenamefont {Marek}\ \emph {et~al.}(2014)\citenamefont {Marek},
  \citenamefont {Blum}, \citenamefont {Johanni}, \citenamefont {Havu},
  \citenamefont {Lang}, \citenamefont {Auckenthaler}, \citenamefont {Heinecke},
  \citenamefont {Bungartz},\ and\ \citenamefont {Lederer}}]{elpa2014}%
  \BibitemOpen
  \bibfield  {author} {\bibinfo {author} {\bibfnamefont {A.}~\bibnamefont
  {Marek}}, \bibinfo {author} {\bibfnamefont {V.}~\bibnamefont {Blum}},
  \bibinfo {author} {\bibfnamefont {R.}~\bibnamefont {Johanni}}, \bibinfo
  {author} {\bibfnamefont {V.}~\bibnamefont {Havu}}, \bibinfo {author}
  {\bibfnamefont {B.}~\bibnamefont {Lang}}, \bibinfo {author} {\bibfnamefont
  {T.}~\bibnamefont {Auckenthaler}}, \bibinfo {author} {\bibfnamefont
  {A.}~\bibnamefont {Heinecke}}, \bibinfo {author} {\bibfnamefont {H.-J.}\
  \bibnamefont {Bungartz}}, \ and\ \bibinfo {author} {\bibfnamefont
  {H.}~\bibnamefont {Lederer}},\ }\href
  {http://stacks.iop.org/0953-8984/26/i=21/a=213201} {\bibfield  {journal}
  {\bibinfo  {journal} {J. Phys. Condens. Matter}\ }\textbf {\bibinfo {volume}
  {26}},\ \bibinfo {pages} {213201} (\bibinfo {year} {2014})}\BibitemShut
  {NoStop}%
\bibitem [{\citenamefont {Knuth}\ \emph {et~al.}(2015)\citenamefont {Knuth},
  \citenamefont {Carbogno}, \citenamefont {Atalla}, \citenamefont {Blum},\ and\
  \citenamefont {Scheffler}}]{KNUTH201533}%
  \BibitemOpen
  \bibfield  {author} {\bibinfo {author} {\bibfnamefont {F.}~\bibnamefont
  {Knuth}}, \bibinfo {author} {\bibfnamefont {C.}~\bibnamefont {Carbogno}},
  \bibinfo {author} {\bibfnamefont {V.}~\bibnamefont {Atalla}}, \bibinfo
  {author} {\bibfnamefont {V.}~\bibnamefont {Blum}}, \ and\ \bibinfo {author}
  {\bibfnamefont {M.}~\bibnamefont {Scheffler}},\ }\href {\doibase
  https://doi.org/10.1016/j.cpc.2015.01.003} {\bibfield  {journal} {\bibinfo
  {journal} {Comput. Phys. Commun.}\ }\textbf {\bibinfo {volume} {190}},\
  \bibinfo {pages} {33 } (\bibinfo {year} {2015})}\BibitemShut {NoStop}%
\bibitem [{\citenamefont {Krukau}\ \emph {et~al.}(2006)\citenamefont {Krukau},
  \citenamefont {Vydrov}, \citenamefont {Izmaylov},\ and\ \citenamefont
  {Scuseria}}]{Krukau06}%
  \BibitemOpen
  \bibfield  {author} {\bibinfo {author} {\bibfnamefont {A.~V.}\ \bibnamefont
  {Krukau}}, \bibinfo {author} {\bibfnamefont {O.~A.}\ \bibnamefont {Vydrov}},
  \bibinfo {author} {\bibfnamefont {A.~F.}\ \bibnamefont {Izmaylov}}, \ and\
  \bibinfo {author} {\bibfnamefont {G.~E.}\ \bibnamefont {Scuseria}},\ }\href
  {\doibase 10.1063/1.2404663} {\bibfield  {journal} {\bibinfo  {journal} {J.
  Chem. Phys.}\ }\textbf {\bibinfo {volume} {125}},\ \bibinfo {pages} {224106}
  (\bibinfo {year} {2006})}\BibitemShut {NoStop}%
\bibitem [{\citenamefont {Weller}\ \emph {et~al.}(2015)\citenamefont {Weller},
  \citenamefont {Weber}, \citenamefont {Henry}, \citenamefont {Di~Pumpo},\ and\
  \citenamefont {Hansen}}]{Weller2015}%
  \BibitemOpen
  \bibfield  {author} {\bibinfo {author} {\bibfnamefont {M.~T.}\ \bibnamefont
  {Weller}}, \bibinfo {author} {\bibfnamefont {O.~J.}\ \bibnamefont {Weber}},
  \bibinfo {author} {\bibfnamefont {P.~F.}\ \bibnamefont {Henry}}, \bibinfo
  {author} {\bibfnamefont {A.~M.}\ \bibnamefont {Di~Pumpo}}, \ and\ \bibinfo
  {author} {\bibfnamefont {T.~C.}\ \bibnamefont {Hansen}},\ }\href {\doibase
  10.1039/C4CC09944C} {\bibfield  {journal} {\bibinfo  {journal} {Chem.
  Commun.}\ }\textbf {\bibinfo {volume} {51}},\ \bibinfo {pages} {4180}
  (\bibinfo {year} {2015})}\BibitemShut {NoStop}%
\bibitem [{\citenamefont {Kong}\ \emph {et~al.}(2015)\citenamefont {Kong},
  \citenamefont {Ye}, \citenamefont {Qi}, \citenamefont {Zhang}, \citenamefont
  {Wang}, \citenamefont {Rahimi-Iman},\ and\ \citenamefont {Wu}}]{Kong2015}%
  \BibitemOpen
  \bibfield  {author} {\bibinfo {author} {\bibfnamefont {W.}~\bibnamefont
  {Kong}}, \bibinfo {author} {\bibfnamefont {Z.}~\bibnamefont {Ye}}, \bibinfo
  {author} {\bibfnamefont {Z.}~\bibnamefont {Qi}}, \bibinfo {author}
  {\bibfnamefont {B.}~\bibnamefont {Zhang}}, \bibinfo {author} {\bibfnamefont
  {M.}~\bibnamefont {Wang}}, \bibinfo {author} {\bibfnamefont {A.}~\bibnamefont
  {Rahimi-Iman}}, \ and\ \bibinfo {author} {\bibfnamefont {H.}~\bibnamefont
  {Wu}},\ }\href {\doibase 10.1039/C5CP02605A} {\bibfield  {journal} {\bibinfo
  {journal} {Phys. Chem. Chem. Phys.}\ }\textbf {\bibinfo {volume} {17}},\
  \bibinfo {pages} {16405} (\bibinfo {year} {2015})}\BibitemShut {NoStop}%
\bibitem [{\citenamefont {Phuong}\ \emph {et~al.}(2016)\citenamefont {Phuong},
  \citenamefont {Yamada}, \citenamefont {Nagai}, \citenamefont {Maruyama},
  \citenamefont {Wakamiya},\ and\ \citenamefont {Kanemitsu}}]{Phuong2016}%
  \BibitemOpen
  \bibfield  {author} {\bibinfo {author} {\bibfnamefont {L.~Q.}\ \bibnamefont
  {Phuong}}, \bibinfo {author} {\bibfnamefont {Y.}~\bibnamefont {Yamada}},
  \bibinfo {author} {\bibfnamefont {M.}~\bibnamefont {Nagai}}, \bibinfo
  {author} {\bibfnamefont {N.}~\bibnamefont {Maruyama}}, \bibinfo {author}
  {\bibfnamefont {A.}~\bibnamefont {Wakamiya}}, \ and\ \bibinfo {author}
  {\bibfnamefont {Y.}~\bibnamefont {Kanemitsu}},\ }\href {\doibase
  10.1021/acs.jpclett.6b00781} {\bibfield  {journal} {\bibinfo  {journal} {J.
  Phys. Chem. Lett.}\ }\textbf {\bibinfo {volume} {7}},\ \bibinfo {pages}
  {2316} (\bibinfo {year} {2016})}\BibitemShut {NoStop}%
\bibitem [{\citenamefont {Zhu}\ \emph {et~al.}(2017)\citenamefont {Zhu},
  \citenamefont {Huhn}, \citenamefont {Wessler}, \citenamefont {Shin},
  \citenamefont {Saparov}, \citenamefont {Mitzi},\ and\ \citenamefont
  {Blum}}]{tong2017}%
  \BibitemOpen
  \bibfield  {author} {\bibinfo {author} {\bibfnamefont {T.}~\bibnamefont
  {Zhu}}, \bibinfo {author} {\bibfnamefont {W.~P.}\ \bibnamefont {Huhn}},
  \bibinfo {author} {\bibfnamefont {G.~C.}\ \bibnamefont {Wessler}}, \bibinfo
  {author} {\bibfnamefont {D.}~\bibnamefont {Shin}}, \bibinfo {author}
  {\bibfnamefont {B.}~\bibnamefont {Saparov}}, \bibinfo {author} {\bibfnamefont
  {D.~B.}\ \bibnamefont {Mitzi}}, \ and\ \bibinfo {author} {\bibfnamefont
  {V.}~\bibnamefont {Blum}},\ }\href {\doibase 10.1021/acs.chemmater.7b02638}
  {\bibfield  {journal} {\bibinfo  {journal} {Chem. Mater.}\ }\textbf {\bibinfo
  {volume} {29}},\ \bibinfo {pages} {7868} (\bibinfo {year}
  {2017})}\BibitemShut {NoStop}%
\bibitem [{\citenamefont {Ashari-Astani}\ \emph {et~al.}(2017)\citenamefont
  {Ashari-Astani}, \citenamefont {Meloni}, \citenamefont {Salavati},
  \citenamefont {Palermo}, \citenamefont {Gr\"atzel},\ and\ \citenamefont
  {Roethlisberger}}]{Roethlisberger2017}%
  \BibitemOpen
  \bibfield  {author} {\bibinfo {author} {\bibfnamefont {N.}~\bibnamefont
  {Ashari-Astani}}, \bibinfo {author} {\bibfnamefont {S.}~\bibnamefont
  {Meloni}}, \bibinfo {author} {\bibfnamefont {A.~H.}\ \bibnamefont
  {Salavati}}, \bibinfo {author} {\bibfnamefont {G.}~\bibnamefont {Palermo}},
  \bibinfo {author} {\bibfnamefont {M.}~\bibnamefont {Gr\"atzel}}, \ and\
  \bibinfo {author} {\bibfnamefont {U.}~\bibnamefont {Roethlisberger}},\ }\href
  {\doibase 10.1021/acs.jpcc.7b04898} {\bibfield  {journal} {\bibinfo
  {journal} {The Journal of Physical Chemistry C}\ }\textbf {\bibinfo {volume}
  {121}},\ \bibinfo {pages} {23886} (\bibinfo {year} {2017})}\BibitemShut
  {NoStop}%
\bibitem [{\citenamefont {Feng}\ and\ \citenamefont {Xiao}(2014)}]{Feng2014}%
  \BibitemOpen
  \bibfield  {author} {\bibinfo {author} {\bibfnamefont {J.}~\bibnamefont
  {Feng}}\ and\ \bibinfo {author} {\bibfnamefont {B.}~\bibnamefont {Xiao}},\
  }\href {\doibase 10.1021/jz500480m} {\bibfield  {journal} {\bibinfo
  {journal} {The Journal of Physical Chemistry Letters}\ }\textbf {\bibinfo
  {volume} {5}},\ \bibinfo {pages} {1278} (\bibinfo {year} {2014})}\BibitemShut
  {NoStop}%
\bibitem [{\citenamefont {Straus}\ and\ \citenamefont
  {Kagan}(2018)}]{Kagan2018}%
  \BibitemOpen
  \bibfield  {author} {\bibinfo {author} {\bibfnamefont {D.~B.}\ \bibnamefont
  {Straus}}\ and\ \bibinfo {author} {\bibfnamefont {C.~R.}\ \bibnamefont
  {Kagan}},\ }\href {\doibase 10.1021/acs.jpclett.8b00201} {\bibfield
  {journal} {\bibinfo  {journal} {J. Phys. Chem. Lett.}\ }\textbf {\bibinfo
  {volume} {9}},\ \bibinfo {pages} {1434} (\bibinfo {year} {2018})}\BibitemShut
  {NoStop}%
\bibitem [{\citenamefont {Hill}\ \emph {et~al.}(2000)\citenamefont {Hill},
  \citenamefont {Kahn}, \citenamefont {Soos},\ and\ \citenamefont
  {\mbox{Pascal,~Jr}}}]{Hill2000}%
  \BibitemOpen
  \bibfield  {author} {\bibinfo {author} {\bibfnamefont {I.}~\bibnamefont
  {Hill}}, \bibinfo {author} {\bibfnamefont {A.}~\bibnamefont {Kahn}}, \bibinfo
  {author} {\bibfnamefont {Z.}~\bibnamefont {Soos}}, \ and\ \bibinfo {author}
  {\bibfnamefont {R.}~\bibnamefont {\mbox{Pascal,~Jr}}},\ }\href {\doibase
  https://doi.org/10.1016/S0009-2614(00)00882-4} {\bibfield  {journal}
  {\bibinfo  {journal} {Chem. Phys. Lett.}\ }\textbf {\bibinfo {volume}
  {327}},\ \bibinfo {pages} {181} (\bibinfo {year} {2000})}\BibitemShut
  {NoStop}%
\bibitem [{\citenamefont {Even}\ \emph {et~al.}(2013)\citenamefont {Even},
  \citenamefont {Pedesseau}, \citenamefont {Jancu},\ and\ \citenamefont
  {Katan}}]{even2013}%
  \BibitemOpen
  \bibfield  {author} {\bibinfo {author} {\bibfnamefont {J.}~\bibnamefont
  {Even}}, \bibinfo {author} {\bibfnamefont {L.}~\bibnamefont {Pedesseau}},
  \bibinfo {author} {\bibfnamefont {J.-M.}\ \bibnamefont {Jancu}}, \ and\
  \bibinfo {author} {\bibfnamefont {C.}~\bibnamefont {Katan}},\ }\href
  {\doibase 10.1021/jz401532q} {\bibfield  {journal} {\bibinfo  {journal} {J.
  Phys. Chem. Lett.}\ }\textbf {\bibinfo {volume} {4}},\ \bibinfo {pages}
  {2999} (\bibinfo {year} {2013})}\BibitemShut {NoStop}%
\bibitem [{\citenamefont {Paier}\ \emph {et~al.}(2009)\citenamefont {Paier},
  \citenamefont {Asahi}, \citenamefont {Nagoya},\ and\ \citenamefont
  {Kresse}}]{hseCZTSprb2009}%
  \BibitemOpen
  \bibfield  {author} {\bibinfo {author} {\bibfnamefont {J.}~\bibnamefont
  {Paier}}, \bibinfo {author} {\bibfnamefont {R.}~\bibnamefont {Asahi}},
  \bibinfo {author} {\bibfnamefont {A.}~\bibnamefont {Nagoya}}, \ and\ \bibinfo
  {author} {\bibfnamefont {G.}~\bibnamefont {Kresse}},\ }\href {\doibase
  10.1103/PhysRevB.79.115126} {\bibfield  {journal} {\bibinfo  {journal} {Phys.
  Rev. B}\ }\textbf {\bibinfo {volume} {79}},\ \bibinfo {pages} {115126}
  (\bibinfo {year} {2009})}\BibitemShut {NoStop}%
\bibitem [{\citenamefont {Skone}\ \emph {et~al.}(2016)\citenamefont {Skone},
  \citenamefont {Govoni},\ and\ \citenamefont {Galli}}]{rangeSeparatedHybrid}%
  \BibitemOpen
  \bibfield  {author} {\bibinfo {author} {\bibfnamefont {J.~H.}\ \bibnamefont
  {Skone}}, \bibinfo {author} {\bibfnamefont {M.}~\bibnamefont {Govoni}}, \
  and\ \bibinfo {author} {\bibfnamefont {G.}~\bibnamefont {Galli}},\ }\href
  {\doibase 10.1103/PhysRevB.93.235106} {\bibfield  {journal} {\bibinfo
  {journal} {Phys. Rev. B}\ }\textbf {\bibinfo {volume} {93}},\ \bibinfo
  {pages} {235106} (\bibinfo {year} {2016})}\BibitemShut {NoStop}%
\end{thebibliography}

\end{document}


\title{Supporting Material \\[2.0ex] Tunable Semiconductors: Control over Carrier States and Excitations in Layered Hybrid Organic-Inorganic Perovskites}

\author{Chi Liu}
\affiliation
{Department of Chemistry, Duke University, Durham, North Carolina 27708, USA}
\author{William Huhn}
\author{Ke-Zhao Du}
\affiliation
{Department of Mechanical Engineering and Materials Science, Duke University, Durham, North Carolina 27708, USA}
\author{Alvaro Vazquez-Mayagoitia}
\affiliation
{Argonne Leadership Computing Facility, 9700 S. Cass Avenue, Lemont IL 60439, USA}
\author{David Dirkes}
\affiliation
{Department of Chemistry, University of North Carolina, Chapel Hill, North Carolina 27599, USA}
\author{Wei You}
\affiliation
{Department of Chemistry, University of North Carolina, Chapel Hill, North Carolina 27599, USA}
\author{Yosuke Kanai}
\affiliation
{Department of Chemistry, University of North Carolina, Chapel Hill, North Carolina 27599, USA}
\author{David B. Mitzi}
\affiliation
{Department of Mechanical Engineering and Materials Science, Duke University, Durham, North Carolina 27708, USA}
\affiliation
{Department of Chemistry, Duke University, Durham, North Carolina 27708, USA}
\email
{david.mitzi@duke.edu}
\author{Volker Blum}
\affiliation
{Department of Mechanical Engineering and Materials Science, Duke University, Durham, North Carolina 27708, USA}
\affiliation
{Department of Chemistry, Duke University, Durham, North Carolina 27708, USA}
\email
{volker.blum@duke.edu}




\date{\today}


\maketitle

\clearpage
\section*{Short Description}
This Supplemental Material includes: structural details of the perovskites in our
  simulations; energetic impact of the supercell choice; impact of spin-orbit
  coupling in the reported energy band structures; computational details (basis
  sets and k-space grid convergence); validation of computational protocols by
  comparing to experimental results for MAPbI$_3$, for AE4TPbBr$_4$ and for
  AE4TPbI$_4$; experimental details for the synthesis and characterization of a
  new AE4TPbI$_4$ crystal; and additional computed orbitals, energy band
  structures, band curvature parameters, full and partial densities of states
  supporting the general findings demonstrated in the actual paper.
  Computational validation for MAPbI$_3$ also includes comparison to a more
  sophisticated many-body dispersion treatment.\cite{Ambrosetti2014} Synthesis
  details include Refs.~\cite{exp1,exp2,exp3}. All computational input
  and output files (raw data) are available at the NOMAD repository.\cite{NOMAD-URL}

\clearpage
  
\section{(2$\times$2) supercell of AE4TP\lowercase{b}B\lowercase{r}$\mathrm{_4}$}
\begin{figure}[H]
\centering
\includegraphics[scale=1]{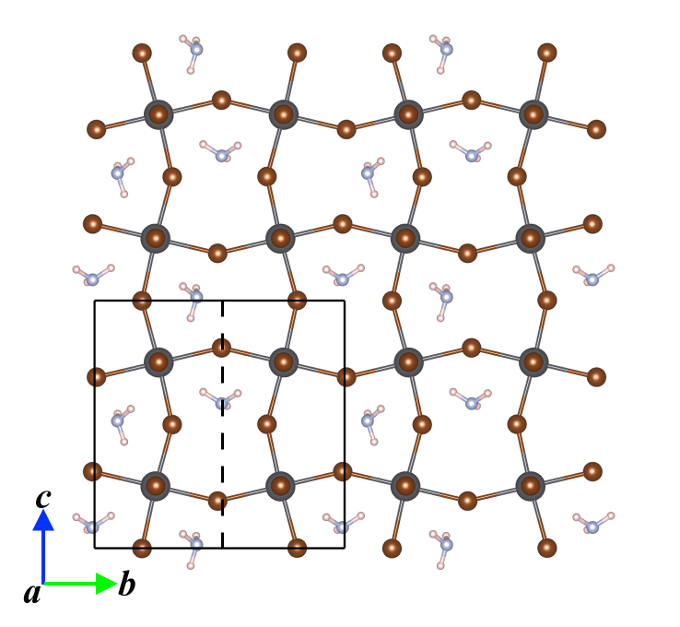}
	\caption{Experimental structure of AE4T\ce{PbBr4} viewed along the $a$-axis direction, adapted from Ref. \cite{Mitzi1999AE4T}. Only the \ce{PbBr4} framework and the \ce{NH3} are shown for clarity. Two different configurations for the lead bromide framework and the ammonium tail can be seen arranged in an alternating fashion. Solid lines indicate the (2$\times$2) cell in units of the hypothetical basic perovskite unit cell dimension, which is used as the basic structural unit in all calculations reported in the present work. The dashed line delineates the (1$\times$2) unit cell model (which contained split atomic positions) as used in the X-ray refinement of AE4T\ce{PbBr4} in Ref. \cite{Mitzi1999AE4T}.}
\label{fig:2by2}
\end{figure}

\clearpage
\section{Comparison of (2$\times$2) and (1$\times$2) structure models relaxed by DFT-PBE+TS}
\subsection{Nomenclature of Pb-X-Pb bond angles (X = Cl, Br, I)}
\begin{figure}[H]
    \centering
    \includegraphics[scale=1]{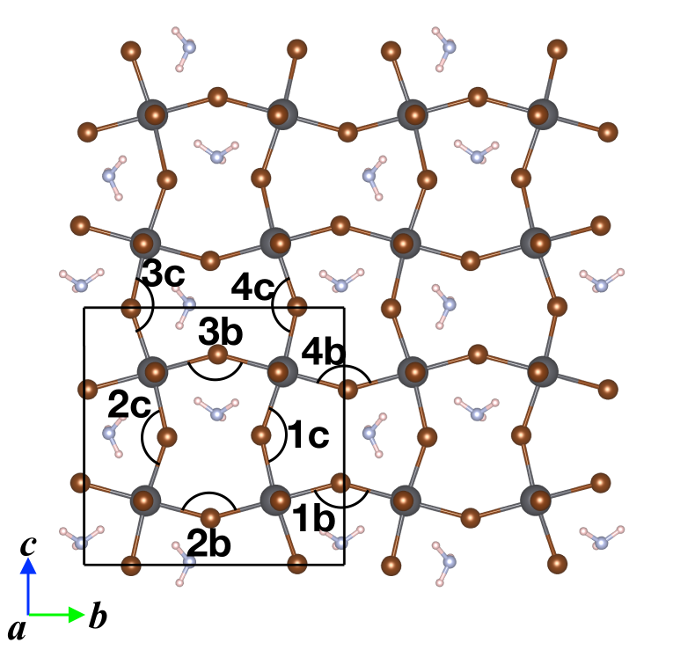}
    \caption{Illustration of the Pb-X-Pb (X = Cl, Br, I) angles
      investigated in this paper for all hybrid compounds. The PBE+TS
      relaxed structure of AE4T\ce{PbX4} is viewed from the $a$
      direction. Only the Pb-X inorganic framework and the \ce{NH3}
      tails are shown for clarity. All other hybrid compounds
      discussed in this paper can be understood similarly. Two
      distinct inorganic \ce{PbX4} layers are identified [e.g., see
        Fig. 1(a) in the main text] and labeled as layer 1 and layer 2. Each layer contains four \ce{PbX4} octahedra (those inside the black solid line). There is one Pb-X-Pb angle along the $b$ direction and one along the $c$ direction for each \ce{PbX4} octahedron. Therefore 8 distinct Pb-X-Pb angles exist for each inorganic layer, labeled as 1, 2, 3, 4 for $b$ and $c$ directions, respectively.}
    \label{fig:Pb-X-Pb_illus}
\end{figure}

\begin{figure}
    \centering
    \includegraphics[scale=1]{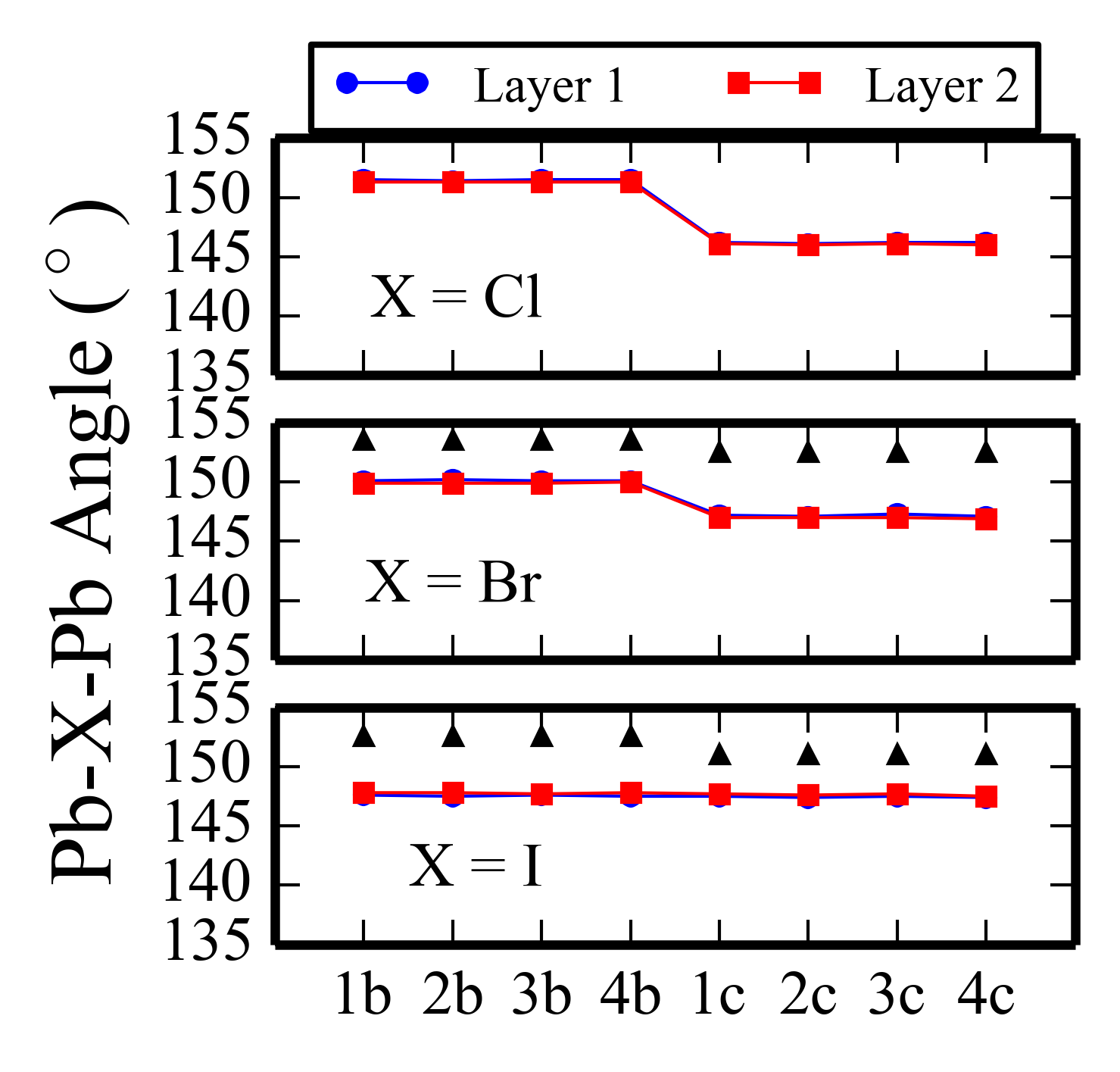}
    \caption{Pb-X-Pb (X = Cl, Br, I) angles for AE4T\ce{PbX4}(X = Cl, Br, I) HOIPs relaxed by DFT-PBE+TS using a (2$\times$2) unit cell. The experimental angle values for AE4T\ce{PbBr4} and AE4T\ce{PbI4} are indicated by black triangles. No experimental data are available for the X = Cl compound.}
    \label{fig:plot_angle_inorg}
\end{figure}

\begin{figure}
    \centering
    \includegraphics[scale=1]{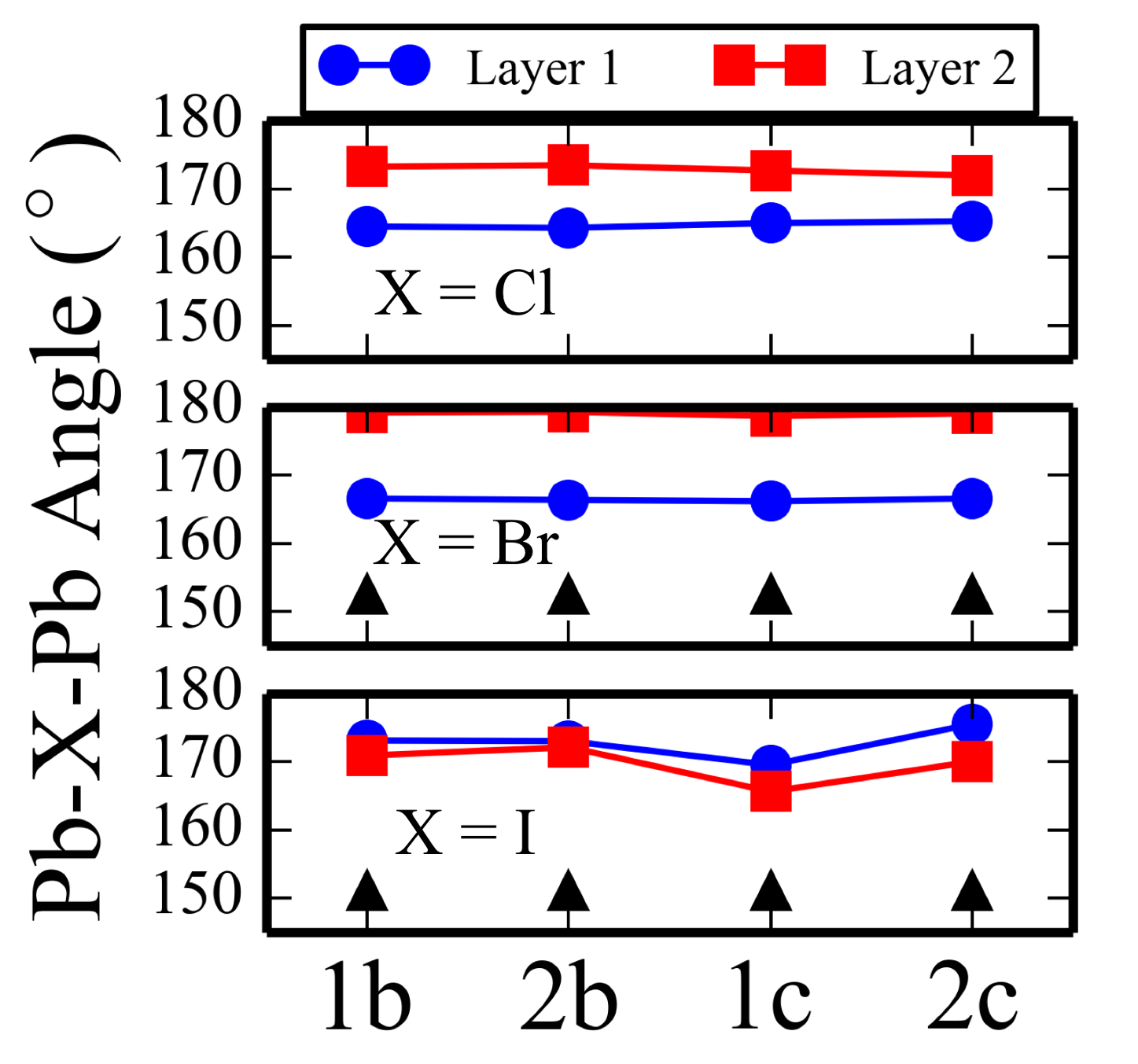}
    \caption{Pb-X-Pb (X = Cl, Br, I) angles for AE4T\ce{PbX4}(X = Cl, Br, I) HOIPs relaxed by DFT-PBE+TS using the (1$\times$2) model unit cell. The experimental angle values for AE4T\ce{PbBr4} and AE4T\ce{PbI4} are indicated by black triangles.}
    \label{fig:plot_angle_inorg1by2}
\end{figure}

\clearpage
\subsection{PBE+TS total energy}
\begin{table}[H]
    \centering
	\caption{The PBE+TS total energy difference (E(2$\times$2) - E(1$\times$2)) of the PBE+TS predicted structure using the (1$\times$2) and (2$\times$2) unit cells for the syn-anti-syn AE4T\ce{PbX4} (X = Cl, Br, I) and the all-anti AEnT\ce{PbBr4} (n = 2, 3, 4, 5). K-point grids settings of 1$\times$4$\times$2 were used for the (1$\times$2) calculation, whereas 1$\times$2$\times$2 k-point grids were used for the (2$\times$2) calculation. The FHI-aims ``tight'' numerical defaults (Table \ref{tight}) were used for both calculations.  The total energy of the syn-anti-syn AE4T\ce{PbBr4} is 2.44 eV/cell lower than that of the all-anti AE4T\ce{PbBr4} compound (both in a (2x2) configuration).}
    \begin{tabular}{ c c c c}
	\hline
	Compounds & Energy Difference (eV) \\
	\hline
	AE4T\ce{PbCl4} & -2.49\\
        AE4T\ce{PbBr4} & -2.70 \\
	AE4T\ce{PbI4} & -2.73\\
	\hline
        AE2T\ce{PbBr4} & -3.76  \\
        AE3T\ce{PbBr4} & -0.79 \\
        all-anti-AE4T\ce{PbBr4} & -1.50 \\
        AE5T\ce{PbBr4} & -0.83 \\
	\hline
    \end{tabular}\\
    \label{tab:pbe}
\end{table}



\newpage
\section{Effect of spin orbit coupling}
\begin{figure}[H]
	\centering
	\includegraphics[scale=2]{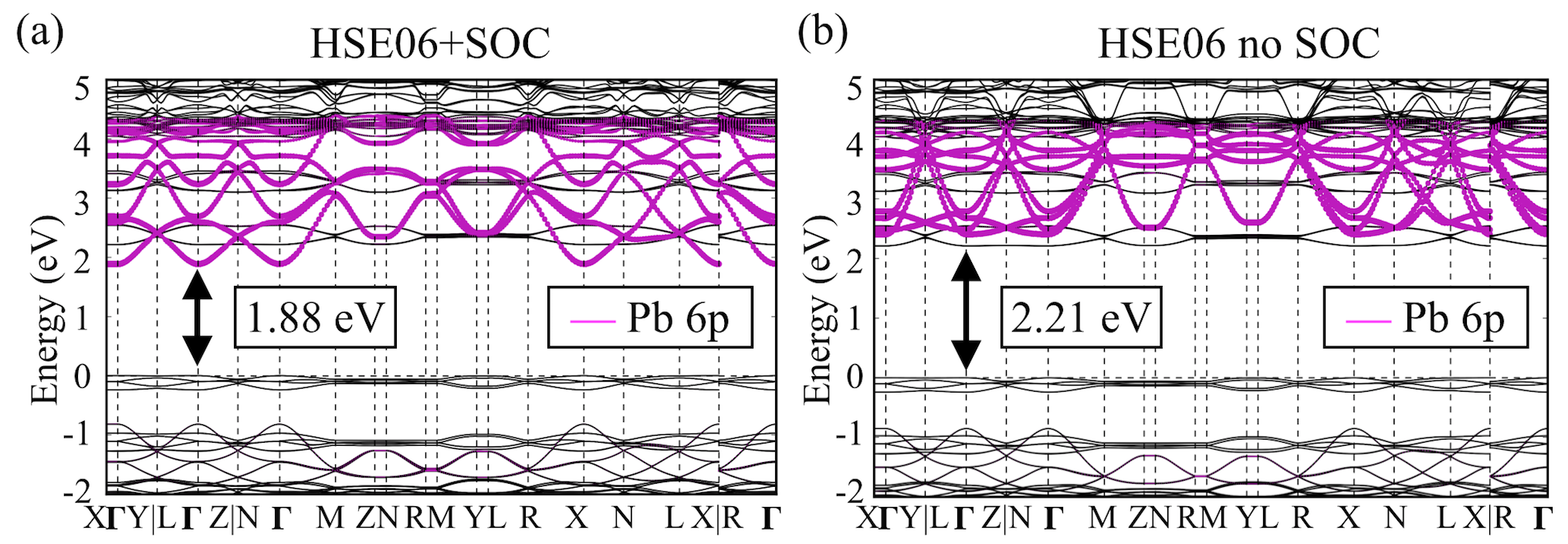}
	\caption{(a) and (b) show the electronic band structures of AE4T\ce{PbBr4} with the states of Pb-6p identified by Mulliken decompositions of the bands, calculated by DFT-HSE06+SOC and DFT-HSE06 without including SOC, respectively. The width of the colored lines is proportional to the Mulliken decomposition value of species or (in the case of Pb 6p) atomic orbitals of species they represent. Thus, wider colored lines correspond to bands having more contribution from the species or atomic orbitals of the species that the Mulliken decomposition is projected onto. The VBM and CBM are indicated by black arrows with the value of the band gap indicated.}
	\label{fig:AE4TPbBr4}
\end{figure}

\clearpage
\section{Computational basis settings: tight}
\begin{table}[H]
    \caption{Basis functions (radial functions) in the ``tight" computational settings used for the DFT-PBE+TS relaxation for H, C, N, O, S, Cl, Br, I and Pb elements.}
    {\centering
    \begin{tabular}{lllll}
        \hline
        & H & C & N & S \\
        \hline
        minimal & 1s & [He]+2s2p & [He]+2s2p & [Ne]+3s3p  \\
        \hline
        \textit{tier} 1 & H(2s,2.1) & H(2p,1.7) & H(2p,1.8) & S$\mathrm{^{2+}}$(3d) \\
        & H(2p,3.5) & H(3d,6.0) & H(3d,6.8) & H(2p,1.8) \\
        & & H(2s,4.9) & H(3s,5.8) & H(4f,7) \\
        & & & & S$\mathrm{^{2+}}$(3s) \\
        \hline
        \textit{tier} 2 & H(1s,0.85) & H(4f,9.8) & H(4f,10.8) & H(4d,6.2) \\
        & H(2p,3.7) & H(3p,5.2) & H(3p,5.8) & H(5g,10.8) \\
        & H(2s,1.2) & H(3s,4.3) & H(1s,0.8) & \\
        & H(3d,7.0) & H(5g,14.4) & H(5g,16) & \\
        & & H(3d,6.2) & H(3d,4.9) & \\
        \hline
    \end{tabular}
    
    \begin{tabular}{lllll}
        \hline
        & Cl & Br & I & Pb \\
        \hline
        minimal & [Ne]+3s3p & [Ar]+4s4p3d & [Kr]+5s5p4d & [Xe]+6s6p5d4f \\
        \hline
        \textit{tier} 1 & Cl$\mathrm{^{2+}}$(3d) & H(3d,4.6) & H(3d,4) & H(3p,2.3) \\
        & H(2p,1.9) & H(2p,1.7) & H(4f,6.4) & H(4f,7.6) \\
        & H(4f,7.4) & H(4f,7.6) & H(2p,1.6) & H(3d,3.5) \\
        & Cl$\mathrm{^{2+}}$(3s) & Br$\mathrm{^{2+}}$(4s) & I$\mathrm{^{2+}}$(5s) & H(5g,9.8)\\
        & H(5g,10.4) & & & H(3s,3.2) \\
        \hline
        \textit{tier} 2 & H(3d,3.3) & & & \\
        \hline
    \end{tabular}\\}
    The first line (``minimal") summarizes the free-atom radial functions used (noble-gas configuration of the core and quantum numbers of the additional valence radial functions). ``H(nl, z)" denotes a hydrogen-like basis function for the bare Coulomb potential z/r, including its radial and angular momentum quantum numbers, n and l. X$\mathrm{^{2+}}$(nl) denotes a n, l radial function of a doubly positive free ion of species X. This nomenclature follows the convention employed in Table 1 of Ref.~\cite{Blum2009}.
    \label{tight}
\end{table}
\clearpage
\section{K-grid convergence test}
\begin{table}[H]
    \centering
	\caption{DFT-PBE total energy convergence with different k-grid settings for reciprocal-space integrals of AE4T\ce{PbBr4}.}
    \begin{tabular}{ c c c c}
	\hline
	K-grid & $\Delta E$(eV) & $\Delta E$(meV/atom) \\
	\hline
	1$\times$2$\times$2 & 0* & 0*\\
    2$\times$4$\times$4 & -0.066 & -0.16 \\
	\hline
    \end{tabular}\\
    * reference value
    \label{tab:kgrid}
\end{table}

\clearpage
\section{Computational basis settings: intermediate}
\begin{table}[H]
    \caption{Basis functions (radial functions) in the ``intermediate" computational settings used for the DFT-HSE06+SOC calculation for H, C, N, O, S, Cl, Br, I and Pb elements.}
    {\centering
    \begin{tabular}{lllll}
        \hline
        & H & C & N & S \\
        \hline
        minimal & 1s & [He]+2s2p & [He]+2s2p & [Ne]+3s3p  \\
        \hline
        \textit{tier} 1 & H(2s,2.1) & H(2p,1.7) & H(2p,1.8) & S$\mathrm{^{2+}}$(3d) \\
        & H(2p,3.5) & H(3d,6.0) & H(3d,6.8) & H(2p,1.8) \\
        & & H(2s,4.9) & H(3s,5.8) & H(4f,7) \\
        & & & & S$\mathrm{^{2+}}$(3s) \\
        \hline
        \textit{tier} 2 & H(1s,0.85) & H(4f ,9.8) & H(4f,10.8) & H(5g,10.8)$_{aux}$ \\
        & H(3d,7.0)$_{aux}$ & H(5g,14.4)$_{aux}$ & H(5g,16)$_{aux}$ & \\
        \hline
    \end{tabular}\\
    \begin{tabular}{lllll}
	\hline
	& Cl & Br & I & Pb \\
	\hline
	Minimal & [Ne]+3s3p & [Ar]+4s4p3d & [Kr]+5s5p4d & [Xe]+6s6p5d4f \\
	\hline
	\textit{tier} 1 & Cl$\mathrm{^{2+}}$(3d) & H(3d,4.6) & H(3d,4) & H(3p,2.3) \\
	& H(2p,1.9) & H(2p,1.7) & H(4f,6.4) & H(4f,7.6) \\
	& H(4f,7.4) & H(4f,7.6) & H(2p,1.6) & H(3d,3.5) \\
	& Cl$\mathrm{^{2+}}$(3s) & Br$\mathrm{^{2+}}$(4s) & I$\mathrm{^{2+}}$(5s) & H(5g,9.8)$_{aux}$\\
	& H(5g,10.4) & & & H(3s,3.2) \\
	\hline
	\textit{tier} 2 & H(3d,3.3) & & \\
	& H(5g,10.8)$_{aux}$ & H(5g,10.4)$_{aux}$ & H(5g,9.4)$_{aux}$ & \\
	\hline
    \end{tabular}\\}
    The first line (``minimal") summarizes the free-atom radial
    functions used (noble-gas configuration of the core and quantum
    numbers of the additional valence radial functions). ``H(nl, z)"
    denotes a hydrogen-like basis function for the bare Coulomb
    potential z/r, including its radial and angular momentum quantum
    numbers, n and l. X$\mathrm{^{2+}}$(nl) denotes a n, l radial
    function of a doubly positive free ion of species X. This
    nomenclature follows the convention employed in Table 1 of
    Ref.~\cite{Blum2009}. The subscript ``\textit{aux}'' denotes a
    basis function that is \emph{not} used to expand the actual,
    generalized Kohn-Sham orbitals but that \emph{is} used during the
    creation of the auxiliary basis sets that expand the screened
    Coulomb operator of the HSE06 functional, following the method
    described in Ref.~\cite{Ihrig2015}. 
    \label{intermediate}
\end{table}

\clearpage
\section{Validation of the computational approach}
\begin{figure}[H]
    \centering
    (a) \hspace*{0.25\textwidth}
    \includegraphics[width=0.6\textwidth]{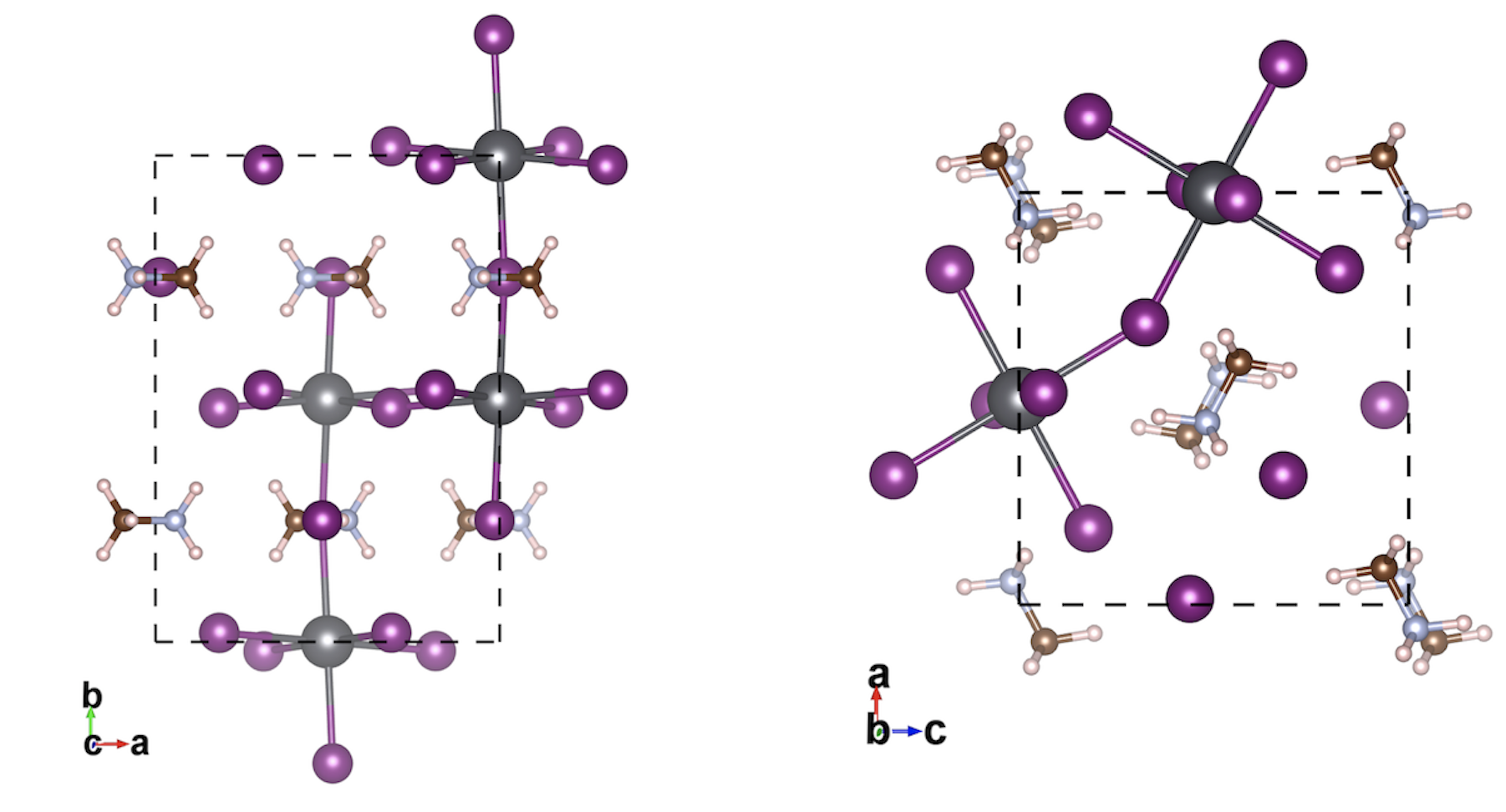} 
    \hspace*{0.03\textwidth}
    \\[1.0ex]
    (b) \hspace*{0.05\textwidth}
    {\color{black}
    \begin{tabular}{lllllll}
    \hline\hline
     & & \multicolumn{3}{c}{Lattice parameters (\AA)} & & \\[-2.0ex]
     Source & & a      & b       & c      & Volume (\AA$^3$) & E$_g$ (eV) \\ \hline
     Exp. \cite{Weller2015}  &            & 8.8657 & 12.6293 & 8.5769 & 960.3 & 1.65-1.675 \cite{Kong2015} \\
             &           & & & & & 1.68 \cite{Phuong2016} \\ \hline
     PBE     &           & 9.25   & 12.88  & 8.62   & 1027 & 1.42 (HSE06+SOC)\\
             & deviation & 4.3 \% & 2.0 \% & 0.5 \% & 7.0 \% & \\
     PBE+TS  &           & 8.99   & 12.71  & 8.48 & 968.1 & 1.42 (HSE06+SOC)\\
             & deviation & 1.4 \% & 0.6 \% & $-$1.1 \% & 0.8 \% & \\
     PBE+MBD &           & 9.00   & 12.72  & 8.48      & 971    & 1.43 (HSE06+SOC)\\
             & deviation & 1.5 \% & 0.7 \% & $-$1.2 \%    & 1.1 \% & \\
    \hline\hline
    \end{tabular}
    }
    \caption{(a) Structure of the low-temperature orthorhombic phase of \ce{MAPbI3} from two different perspectives, fully relaxed by DFT-PBE+TS. (b) Lattice parameters and unit cell volume relaxed at different levels of DFT (PBE\cite{PBE}, PBE+TS{\color{black}\cite{Tkatchenko09}} and PBE+MBD\cite{Ambrosetti2014}), their deviation from experiment {\color{black}(neutron powder diffraction at $T$=100~K in Ref. \cite{Weller2015})} and their impact on the predicted band gap by DFT-HSE06 including spin-orbit coupling (SOC) compared with the experimental value. {\color{black} The experimental band gap values shown reflect the apparent near-band-edge emission of orthorhombic \ce{MAPbI3} observed at $T$=10~K by time-resolved photoluminescence in Ref. \cite{Kong2015} and the optical transient absorption peak attributed to near-band-edge emission of orthorhombic \ce{MAPbI3} at $T$=15~K in Ref.~\cite{Phuong2016}.} PBE+MBD stands for a more recent and more sophisticated many-body dispersion (MBD) scheme compared with PBE+TS to incorporate the vdW terms.\cite{Ambrosetti2014} This comparison demonstrates that key parameters for \ce{MAPbI3} remain unaffected by switching between the two different dispersion treatments.}
    \label{fig:validation}
\end{figure}

\clearpage
\section{Optimized structures of AE4TP\lowercase{b}X$\mathrm{_4}$ (X = C\lowercase{l}, B\lowercase{r}, I)}
\begin{table}[H]
\centering
	\caption{Computationally predicted unit cell parameters of AE4T\ce{PbX4}(X = Cl, Br, I) relaxed by DFT-PBE+TS. The experimental structure refinements for AE4T\ce{PbBr4}\cite{Mitzi1999AE4T} and AE4T\ce{PbI4}(this work) are shown for comparison.}
\begin{tabular}{llllllll}
\hline
HOIPs & a({\AA}) & b({\AA}) & c({\AA}) & $\alpha$ & $\beta$ & $\gamma$ & V({\AA}$^3$) \\
\hline
		AE4T\ce{PbCl4} \\
		PBE+TS &  40.85 & 11.29 & 10.95 & 90.0$^\circ$ & 91.8$^\circ$ & 90.0$^\circ$ &  5049 \\
\hline
	AE4T\ce{PbBr4} \\
Exp.\cite{Mitzi1999AE4T} & 39.74 & 11.68 & 11.57 & 90.0$^\circ$ & 92.4$^\circ$ & 90.0$^\circ$ & 5369 \\
PBE+TS & 40.01 & 11.60 & 11.47 & 90.0$^\circ$ & 91.3$^\circ$ & 90.0$^\circ$ & 5322 \\
$\Delta$(\%) & 0.67 & -0.74 & -0.87 & 0.02 & -1.17 & 0.04 & -0.88 \\
	\hline
	AE4T\ce{PbI4} \\
	Exp.$^*$ & 38.79 & 12.18 & 12.33 & 90.0$^\circ$ & 92.3$^\circ$ & 90.0$^\circ$ & 5818 \\
	PBE+TS & 39.02 & 12.10 & 12.22 & 90.0$^\circ$ & 91.1$^\circ$ & 90.0$^\circ$ & 5771 \\
	$\Delta$(\%) & 0.59 &	-0.65&	-0.86&	0.04&	-1.31&	0.04&	-0.81 \\
	\hline
\end{tabular}\\
* see Section \ref{exp_AE4TPbI4}, ``Experimental details for AE4T\ce{PbI4}'', in this work.
\label{tab:AE4TPbBr4}
\end{table}

\clearpage
\section{Construction of the computational structure models of the HOIP compounds assessed in this work}
\begin{figure}[H]
	\centering
	\includegraphics[scale=1]{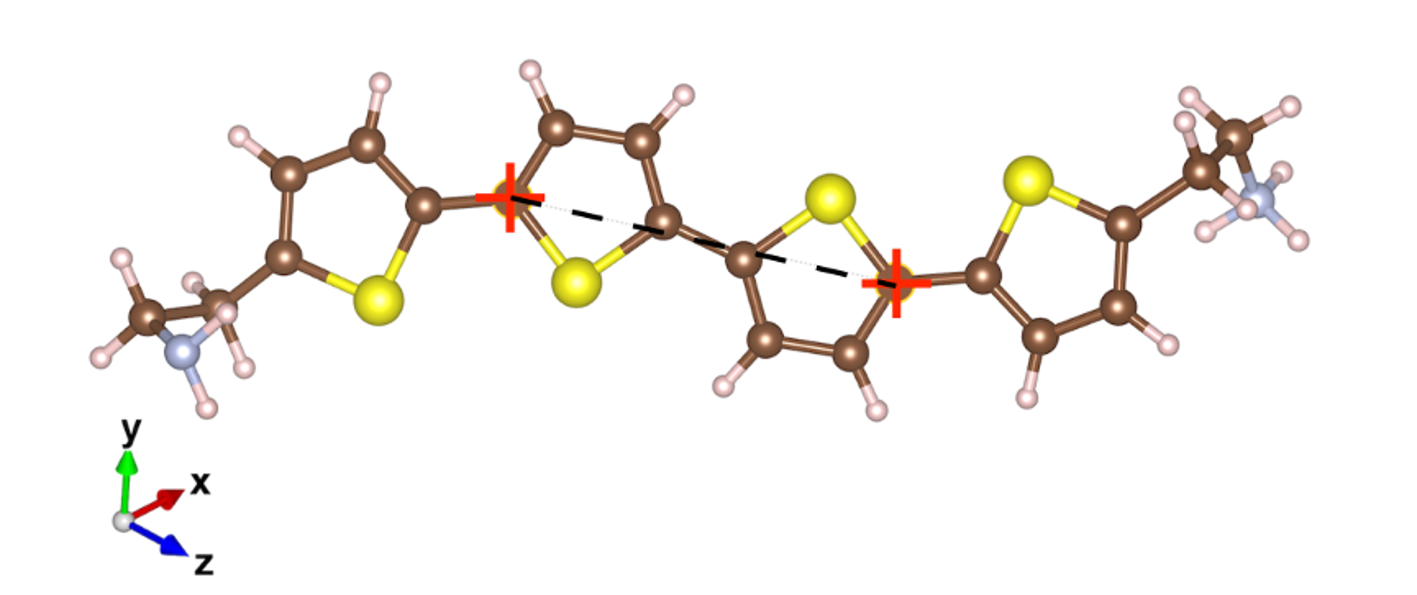}
	\caption{Structure of the AE4T molecule in the experimental structure of AE4T\ce{PbBr4} adopting the syn-anti-syn configuration.}
	\label{fig:syn-anti-syn}
\end{figure}

We perform full relaxation of both unit cell parameters and cell-internal atomic coordinates on the initial input structure for all the HOIPs investigated. Initial models for the structure optimization were constructed as follows:
\begin{itemize}
\item \textbf{AE4T\ce{PbBr4}:} Taken as the experimental structure in Ref. \cite{Mitzi1999AE4T}.
\item \textbf{AE4T\ce{PbX4} (X = Cl, I):} Obtained by taking the experimental structure of AE4T\ce{PbBr4} and replacing the Br atoms with Cl and I. The lattice parameters and the atomic coordinates of non-halogen atoms were kept the same among the three hybrid compounds for the input structure. The initial configuration of AE4T molecules was kept to be syn-anti-syn as in the experimental structure of AE4T\ce{PbBr4}.
\item \textbf{all-anti-AE4T\ce{PbBr4}:} Obtained by taking the experimental structure of AE4T\ce{PbBr4} and changing the configuration of AE4T to all-anti. The inner two thiophene rings of the syn-anti-syn AE4T were rotated by 180$^\circ$ around the two C atoms labeled by red ``+" as the rotation axis (Fig. \ref{fig:syn-anti-syn}). This rotation was performed for all the 8 AE4T molecules in the unit cell.
\end{itemize}


\clearpage
\begin{itemize} 
\item \textbf{AEnT\ce{PbBr4} (n = 1, 2, 3, 5):} Obtained by taking the experimental structure of AE4T\ce{PbBr4} and replacing the AE4T organic part by all-anti AEnT molecules. Using AE3T-\ce{PbBr4} as an example, we start from the experimental structure of AE4T\ce{PbBr4} with syn-anti-syn AE4T molecules. The oligothiophene part (4T) is removed while the inorganic Pb-Br framework together with the \ce{EtNH3} tails are kept--i.e., the configuration of the Pb-Br framework together with the \ce{EtNH3} tails is not changed during the construction. The adjacent inorganic layers are then pushed towards each other to reduce the distance between them to the extent that it is suitable for the insertion of the 3T molecules (all-anti). We then place the optimized 3T molecules into those positions where the 4T molecules originally existed (for AE4T\ce{PbBr4}). This construction process preserves the configuration between the inorganic Pb-Br framework and the \ce{EtNH3} tails (Fig. \ref{fig:2by2}) and keeps the two inorganic layers in each unit cell identical.
\end{itemize}

\clearpage
\section{Experimental details for AE4T\ce{PbI4}}
\begin{figure}[H]
    \centering
    \includegraphics[scale=1]{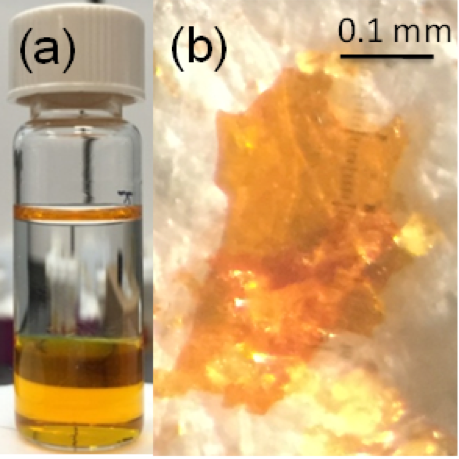}
    \caption{Crystal of AE4T\ce{PbI4}}	
    \label{fig:exp}
\end{figure}
\label{exp_AE4TPbI4}
\textbf{Chemical}. \ce{PbI2} (99.999\% trace metal basis), HI (57 wt. \% in \ce{H2O}, with hypophosphorous acid as stabilizer, assay 99.95\%) and N,N-Dimethylformamide (anhydrous, 99.8\%) were purchased from Sigma-Aldrich company. AE4T$\cdot$HI was synthesized in the lab.\cite{exp1, exp2, exp3} 2-butanol (99.5\%) was purchased from VWR International.

\textbf{Synthesis}. 2 mg \ce{PbI2} and 3 mg AE4T$\cdot$HI were dissolved in 0.7 ml DMF with a drop of HI. Then, 2 ml 2-butanol was layered on the top of the solution (Figure \ref{fig:exp}a). The target crystals came out after several days (Figure \ref{fig:exp}b).

\textbf{Characterization}. Single crystal X-ray diffraction data were collected in a Bruker D8 ADVANCE Series II at room temperature. The unit cell parameters determined from this data are a = 38.779(3) {\AA}, b = 6.0863(5) {\AA}, c = 12.3306(10) {\AA}, \( \beta \) = 92.271(4), V = 2908.0(4) {\AA}$\mathrm{^3}$, C2/c.

\clearpage
\section{\boldmath DFT-HSE06+SOC frontier orbitals for AE4TPbX$_4$ (X=Cl, Br, I)}
\begin{figure}[H]
\centering
\includegraphics[scale=2]{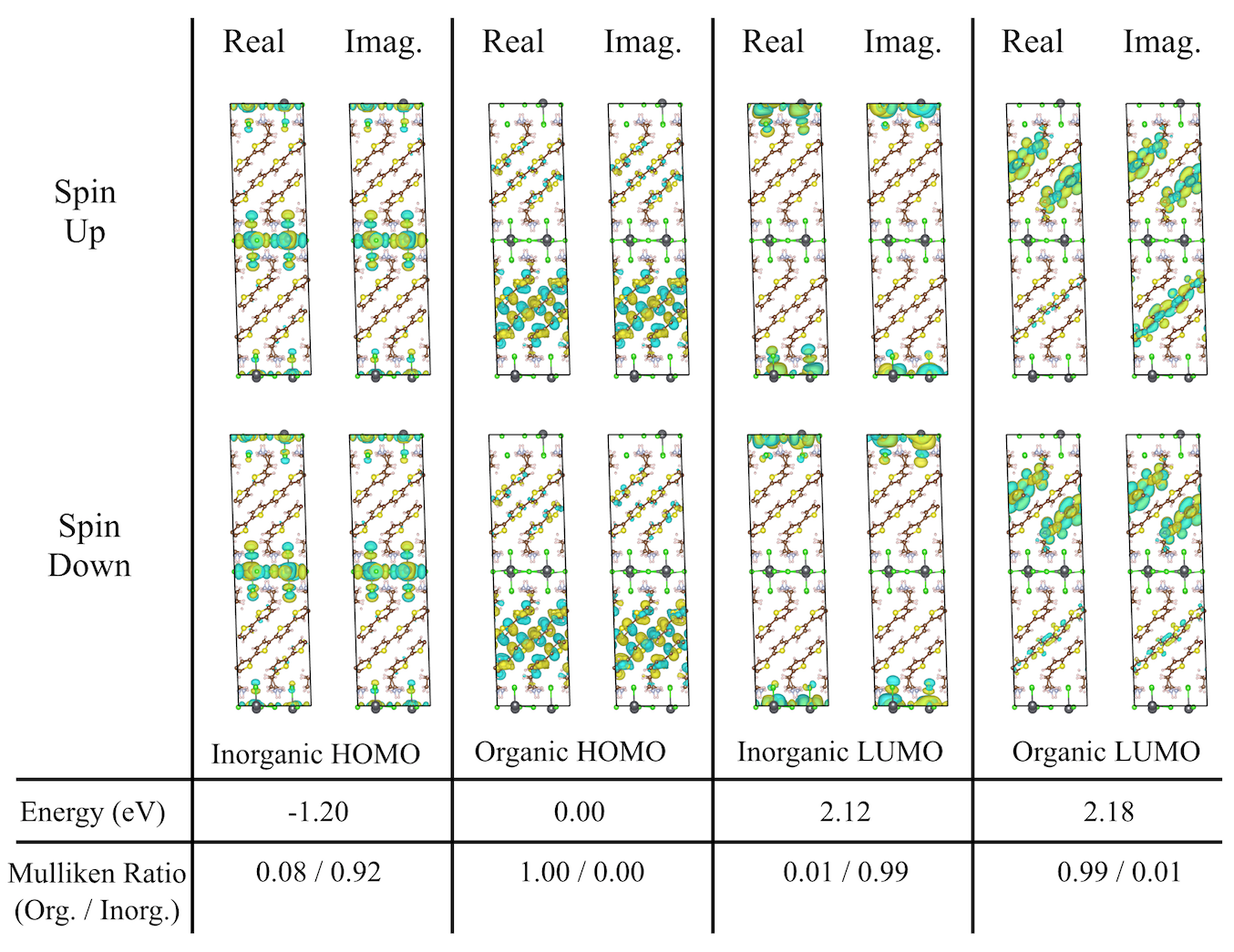}
\caption{Frontier orbitals of the organic and inorganic components of
  AE4T\ce{PbCl4} at the $\Gamma$ point, predicted by
  DFT-HSE06+SOC. The real and imaginary parts of the spin up and spin
  down channels of each spin-orbit coupled orbital are shown. Orbital
  energy values are given with respect to the overall HOMO energy as
  the energy zero. The Mulliken charge decomposition ratio of the
  organic component (C, H, N, S) over the inorganic component (Pb, Cl)
  is given. For each orbital type (organic/inorganic HOMO/LUMO), only
  one of two or more degenerate orbitals is shown. Individual orbitals
  can extend across equivalent parts of the structure in different
  ways since they can be linearly combined with their degenerate
  counterparts.} 
\end{figure}

\clearpage
\begin{figure}[H]
\centering
\includegraphics[scale=2]{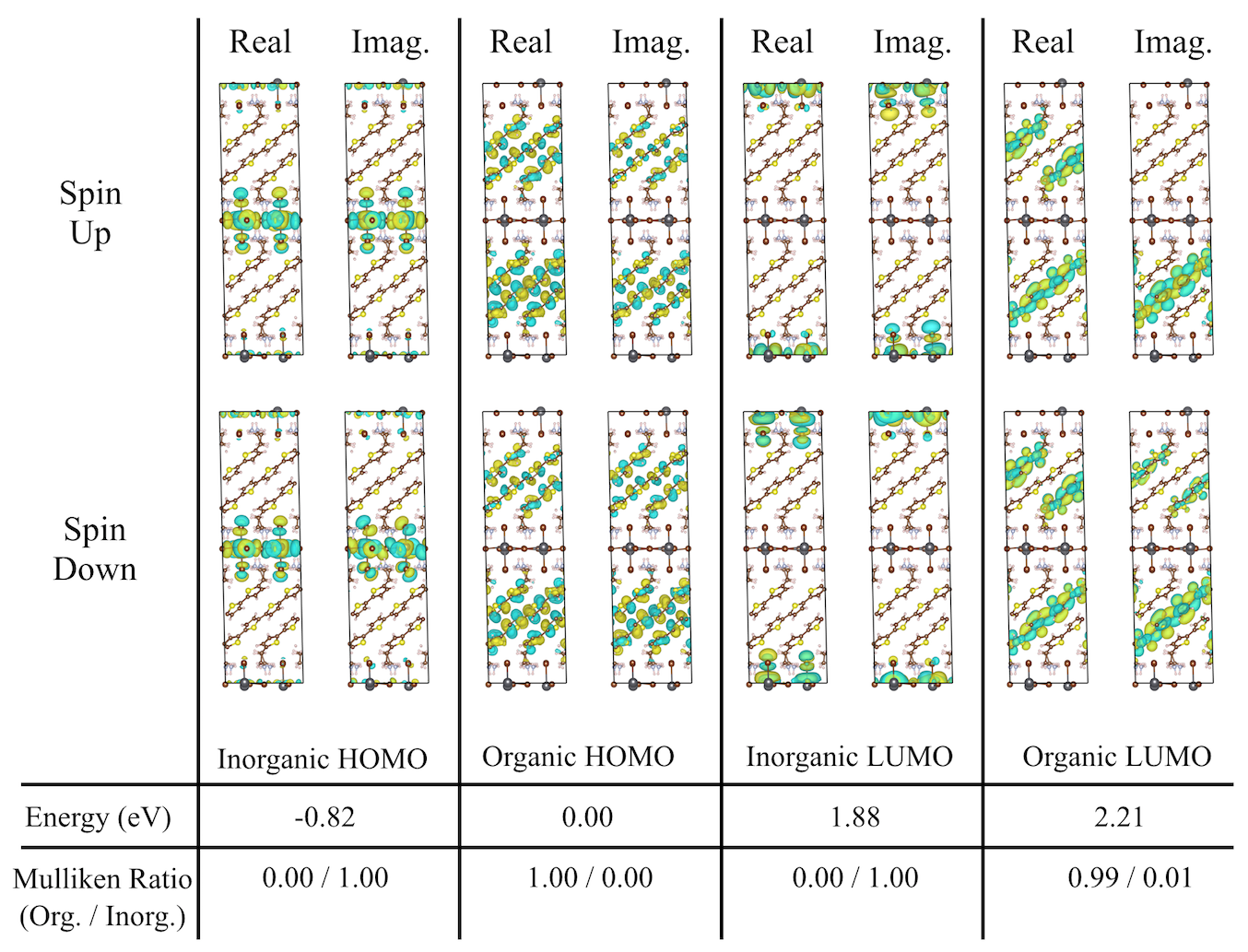}
\caption{Frontier orbitals of the organic and inorganic components of AE4T\ce{PbBr4} at the $\Gamma$ point, predicted by DFT-HSE06+SOC. The real and imaginary parts of the spin up and spin down channels of each spin-orbit coupled orbital are shown. Orbital energy values are given with respect to the overall HOMO energy as the energy zero. The Mulliken charge decomposition ratio of the organic component (C, H, N, S) over the inorganic component (Pb, Br) is given.}
\end{figure}

\clearpage
\begin{figure}[H]
\centering
\includegraphics[scale=2]{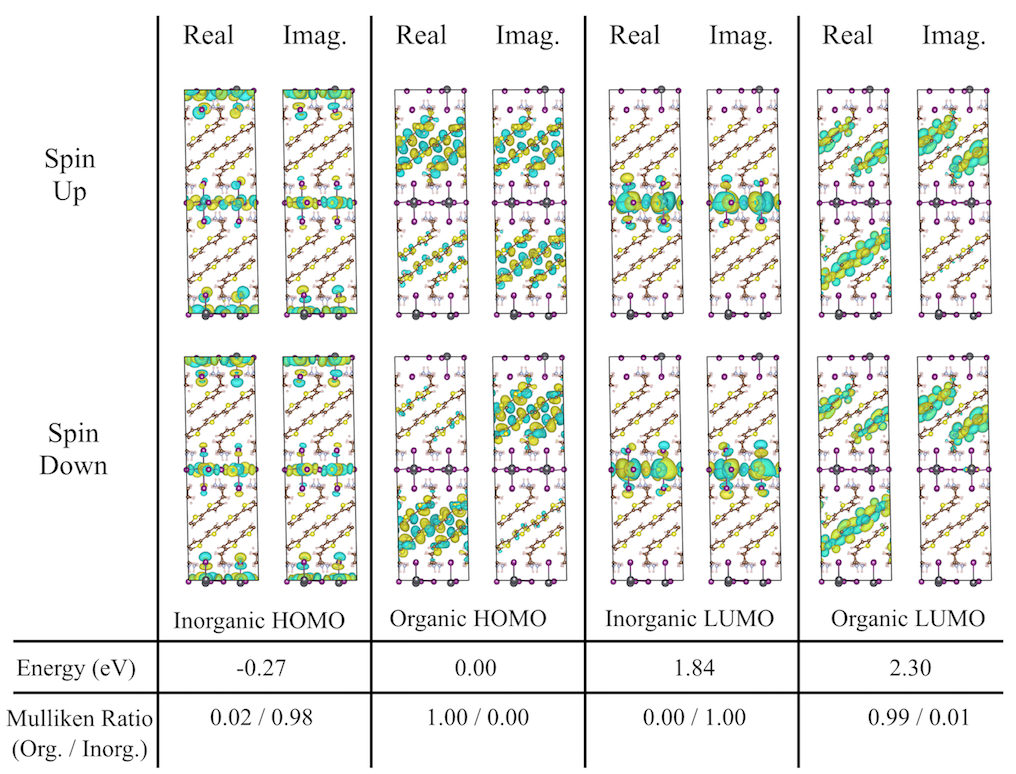}
\caption{Frontier orbitals of the organic and inorganic components of
  AE4T\ce{PbI4} at the $\Gamma$ point, predicted by DFT-HSE06+SOC. The
  real and imaginary parts of the spin up and spin down channels of
  each spin-orbit coupled orbital are shown. Orbital energy values are
  given with respect to the overall HOMO energy as the energy
  zero. The Mulliken charge decomposition ratio of the organic
  component (C, H, N, S) over the inorganic component (Pb, I) is
  given. For each orbital type (organic/inorganic HOMO/LUMO), only
  one of two or more degenerate orbitals is shown. Individual orbitals
  can extend across equivalent parts of the structure in different
  ways since they can be linearly combined with their degenerate
  counterparts.}
\end{figure}

\clearpage
\section{DFT-HSE06+SOC band structures of AE\lowercase{n}TP\lowercase{b}B\lowercase{r}$\mathrm{_4}$ (\lowercase{n} = 1, 2, 3, 4, 5) }
\begin{figure}[H]
\includegraphics[width=\textwidth]{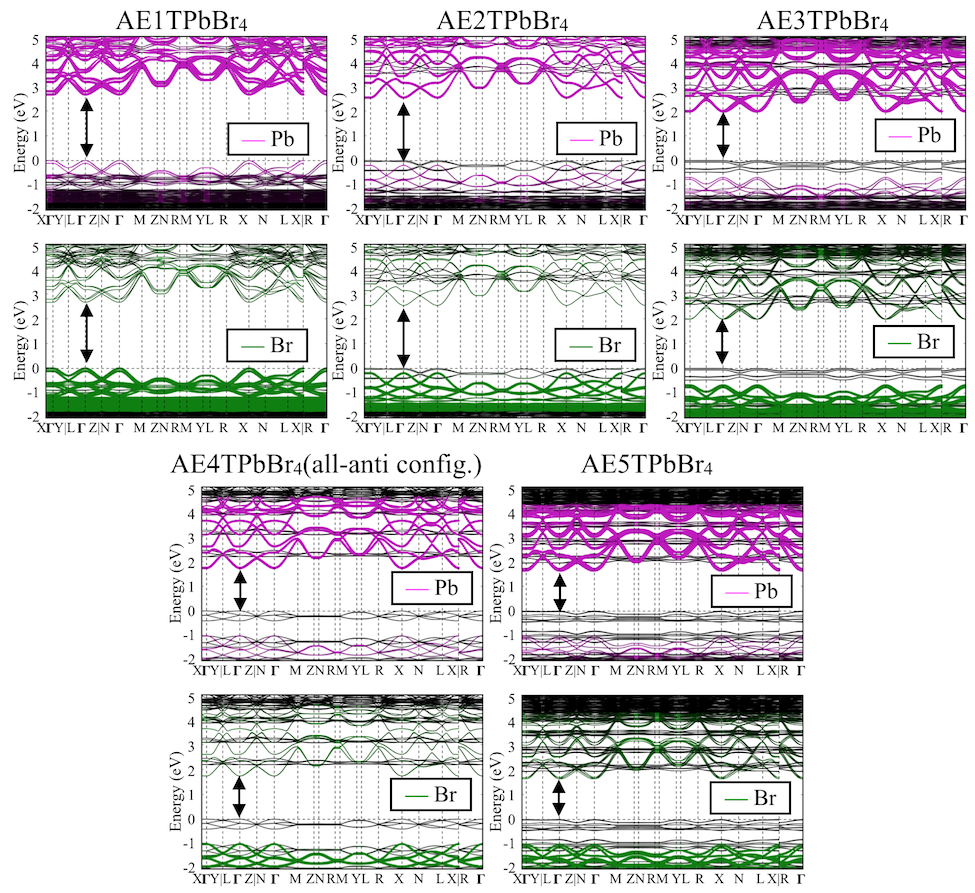}
\caption{Band structures of AEnT\ce{PbBr4} (n = 1, 2, 3, 4, 5)  with
  all-anti configuration of oligothiophene calculated by HSE06+SOC
  with the bands of Pb and Br (determined by Mulliken decompositions)
  colored on the band plot. The VBM and CBM are indicated by black
  arrows. For each orbital type (organic/inorganic HOMO/LUMO), only
  one of two or more degenerate orbitals is shown. Individual orbitals
  can extend across equivalent parts of the structure in different
  ways since they can be linearly combined with their degenerate
  counterparts.}
\label{fig:AE4TPbX4}
\end{figure}

\clearpage
\section{HOMO-LUMO Gaps of isolated oligothiophene molecules}
\begin{figure}[H]
\includegraphics[width=0.7\textwidth]{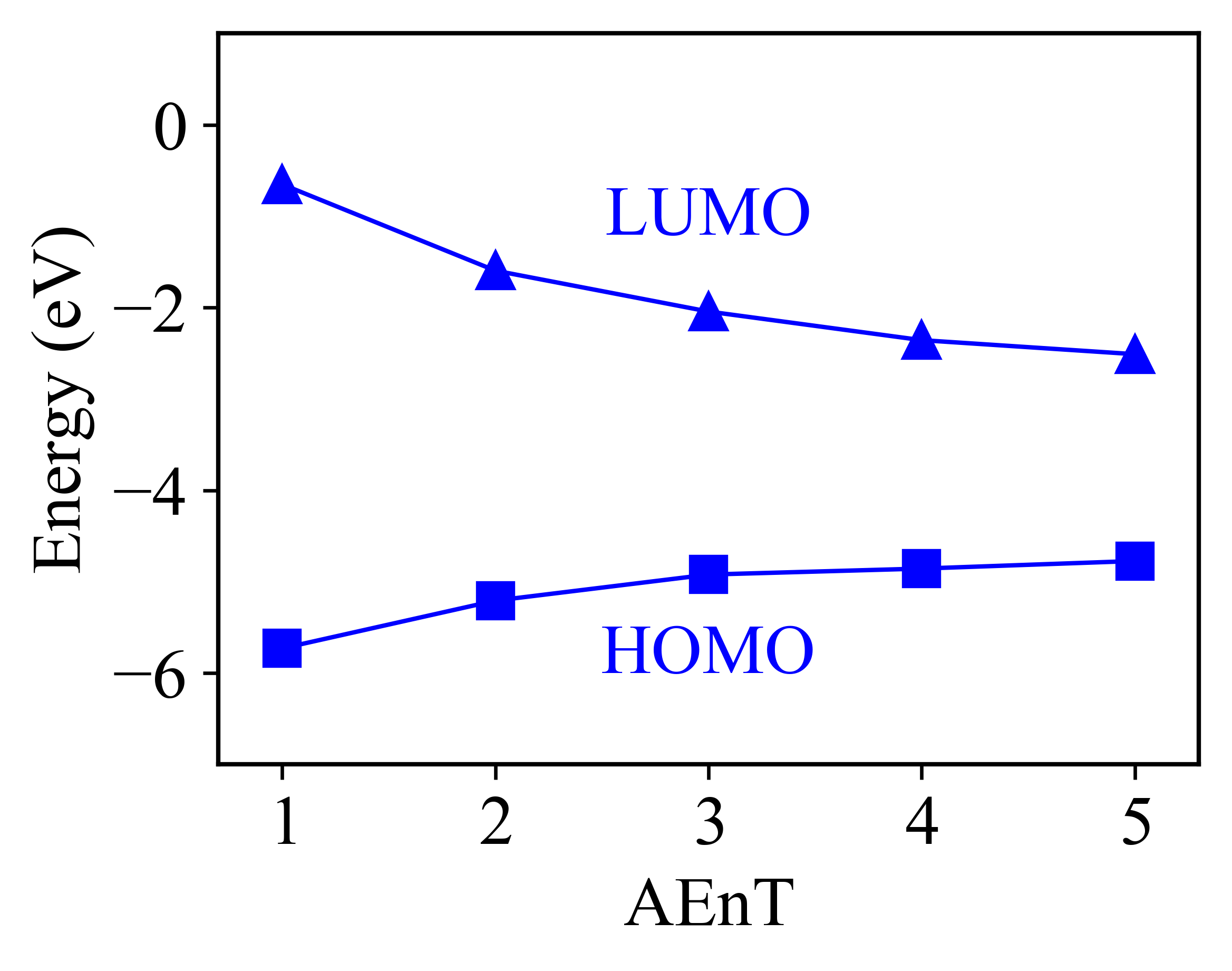}
\caption{Calculated HOMO and LUMO levels of the isolated oligothiophene molecules, $n$=1-5, with unprotonated C$_2$H$_5$NH$_2$ terminations.}
\end{figure}



\clearpage
\section{Effective Mass Parameters Derived from the Band Structures Computed in This Work}
\begin{table}[H]
\centering
\caption{Here and in Table \ref{tab:effective-masses-2}, we provide anisotropic
  band curvatures (in effective mass units, $m_e$) 
  at the $\Gamma$ point, derived from the HSE06+SOC computed
  energy band structures for all compounds addressed in this work.
  For each material, separate band curvatures are given for the HOMO
  and the LUMO of the organic and the inorganic component. The
  absolute VBM and CBM of each material are marked with $^*$ symbols
  in the table. Corresponding band structure diagrams can be found in
  Figure 2 (main paper) and in Figure S12 of this supporting
  information. The curvature parameters were determined by parabolic
  fits to individual bands at the 
  $\Gamma$ points, in the directions parallel to the coordinate axes
  [see Figure 2(a) in the main paper], with small spacings of
  the fitted points of 0.01 $a_0^{-1}$ or 0.02 $a_0^{-1}$, where $a_0$
  is the Bohr radius. The curvature parameters thus correspond to
  the diagonal elements of the effective mass tensor in this
  coordinate system, without a diagonalization of the effective mass
  tensor. Nevertheless, the diagonal elements give a clear qualitative
  indication of the band curvatures in the organic and
  inorganic components of each material considered. Large effective
  mass values correspond to essentially flat bands at the $\Gamma$
  point and are unlikely to be meaningful in the context of band-like
  transport; values above 20~$m_e$ are therefore not further detailed
  in the table.}
\begin{tabular*}{\textwidth}{llllll}
\hline\hline
Material       & Direction \hspace*{1.0em} & HOMO (inorg.) \hspace*{1.0em} & HOMO
(org.) \hspace*{1.0em} & LUMO
(inorg.) \hspace*{1.0em} & LUMO (org.) \\
\hline
  AE4TPbCl$_4$ & $\Gamma$-X &   $>$20       &    $>$20$^*$ & $>$20$^*$
         &   $>$20     \\
               & $\Gamma$-Y &   0.42        &   $>$20$^*$ & 0.32$^*$
         &   1.38      \\
               & $\Gamma$-Z &   0.46        &   2.66$^*$  & 0.43$^*$
         &   6.8           \\
\hline
  AE4TPbBr$_4$ & $\Gamma$-X &   $>$20       &    $>$20$^*$ & $>$20$^*$
         &   $>$20     \\
               & $\Gamma$-Y &   0.31        &   $>$20$^*$  & 0.26$^*$
         &   1.57      \\
               & $\Gamma$-Z &   0.33        &   2.23$^*$   & 0.31$^*$
         &   15.9      \\
\hline
  AE4TPbI$_4$ & $\Gamma$-X &   17.3         &   $>$20$^*$ & $>$20$^*$
         &   $>$20     \\
               & $\Gamma$-Y &  0.27         &   $>$20$^*$ & 0.22$^*$
         &    7.6       \\
               & $\Gamma$-Z &  0.27         &   11.4$^*$  & 0.22$^*$
         &    9.0      \\
\hline\hline
\end{tabular*}\\
\label{tab:effective-masses-1}
\end{table}

\begin{table}[H]
\centering
\centering
\caption{Here and in Table \ref{tab:effective-masses-1}, we provide anisotropic
  band curvatures (in effective mass units, $m_e$) 
  at the $\Gamma$ point, derived from the HSE06+SOC computed
  energy band structures for all compounds addressed in this work. See
  caption of Table \ref{tab:effective-masses-1} for details.} 
\begin{tabular}{llllll}
\hline\hline
Material       & Direction \hspace*{1.0em} & HOMO (inorg.) \hspace*{1.0em} & HOMO
(org.) \hspace*{1.0em} & LUMO (inorg.) \hspace*{1.0em} & LUMO (org.) \\
\hline
  AE1TPbBr$_4$ & $\Gamma$-X &  11.1$^*$     & 5.1      & $>$20$^*$
         & $>$20       \\
               & $\Gamma$-Y &  0.37$^*$     & 4.6      & 0.23$^*$
         & 0.97        \\
               & $\Gamma$-Z &  0.38$^*$     & 1.7      & 0.44$^*$
         & 1.08        \\
\hline
  AE2TPbBr$_4$ & $\Gamma$-X &  $>$20        & $>$20$^*$   & $>$20$^*$
         & 6.9         \\
               & $\Gamma$-Y &  0.33         & 7.5$^*$     & 0.23$^*$
         & 16.9        \\
               & $\Gamma$-Z &  0.37         & 1.2$^*$     & 0.36$^*$
         & 2.9         \\
\hline
  AE3TPbBr$_4$ & $\Gamma$-X &  $>$20        & $>$20$^*$   & $>$20$^*$
         & $>$20       \\
               & $\Gamma$-Y &  0.38         & $>$20$^*$   & 0.24$^*$
         & 2.8         \\
               & $\Gamma$-Z &  0.39         & 1.9$^*$     & 0.55$^*$
         & $>$20       \\
\hline
  AE4TPbBr$_4$ (all-anti) & $\Gamma$-X & 16.3 & $>$20$^*$ & $>$20$^*$
         & $>$20       \\
               & $\Gamma$-Y & 0.40          & 5.6$^*$     & 0.24$^*$
         & 2.3         \\
               & $\Gamma$-Z & 0.39          & 1.1$^*$     & 0.35$^*$
         & $>$20       \\
\hline
  AE5TPbBr$_4$ & $\Gamma$-X & $>$20         & $>$20$^*$   & $>$20$^*$
         & $>$20       \\
               & $\Gamma$-Y & 0.55          & 7.3$^*$     & 0.24$^*$
         & 5.5         \\
               & $\Gamma$-Z & 0.54          & 1.2$^*$     & 0.55$^*$
         & 2.8          \\
\hline\hline
\end{tabular}\\
\label{tab:effective-masses-2}
\end{table}

\clearpage
\section{Total and Partial Densities of States}
\begin{figure}[H]
\centering
\includegraphics[scale=1]{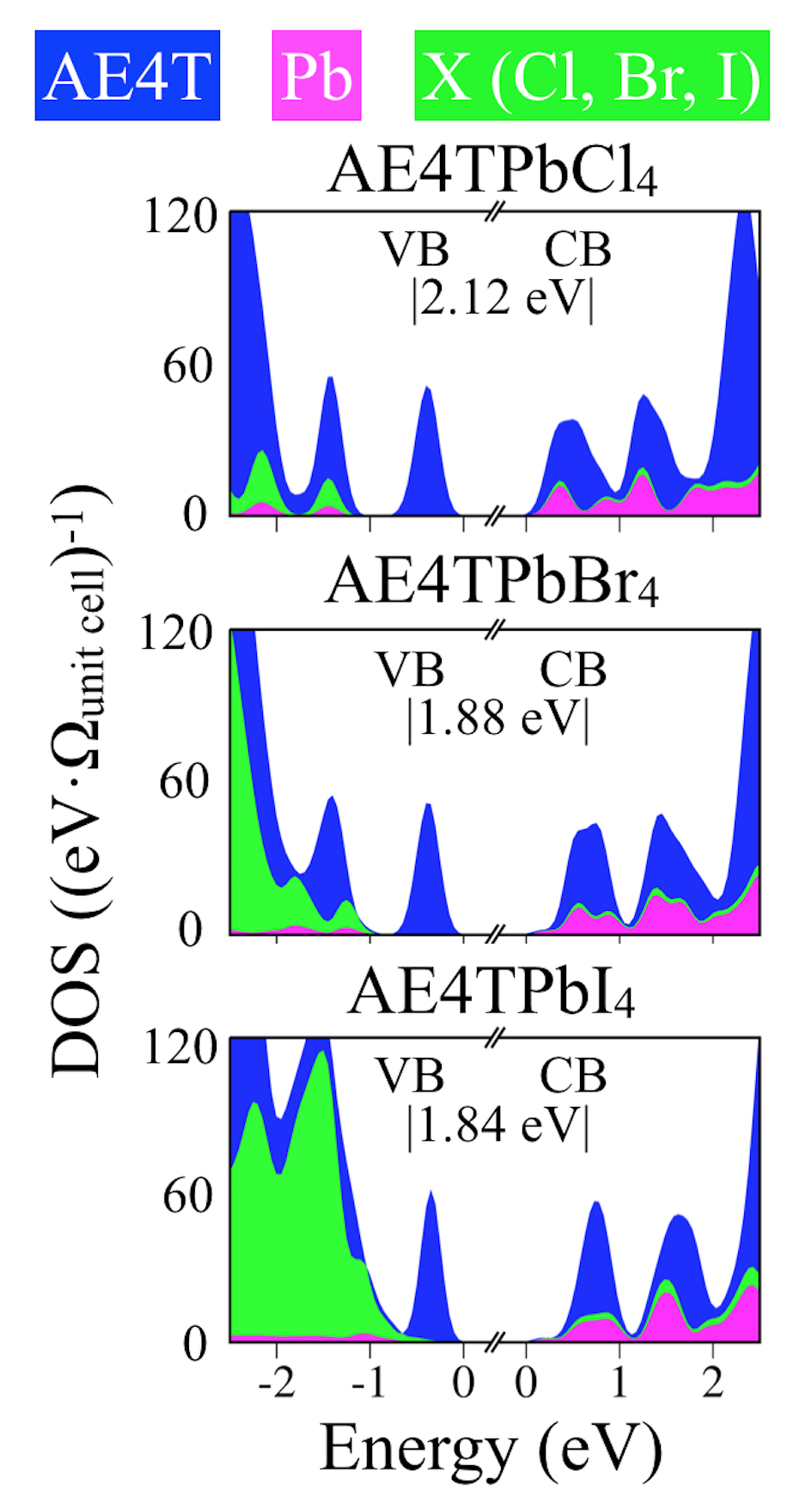}
	\caption{Total densities of states (DOS) and element-projected DOS based on the DFT-HSE06+SOC for the HOIPs AE4T\ce{PbX4}(X = Cl, Br, I). K-point grids settings of 3$\times$3$\times$3 were used in the DOS calculation. The Gaussian broadening parameter was 0.1 eV. The VBMs and CBMs are set to zero energy in each plot, with the actual band gap omitted. Band gap values are labeled in each subplot. The element-projected DOS is shown by different colors: magenta for Pb, green for halogens Cl, Br, I, and blue for organic component AE4T. The sum of all colored components is the total DOS, indicated by the outermost curve. $\mathrm{\Omega_{unit\ cell}}$ in the Y axis label represents the volume of unit cell for each compound.}
\label{fig:dosInorg}
\end{figure}

\begin{figure}[H]
\centering
\includegraphics[scale=1]{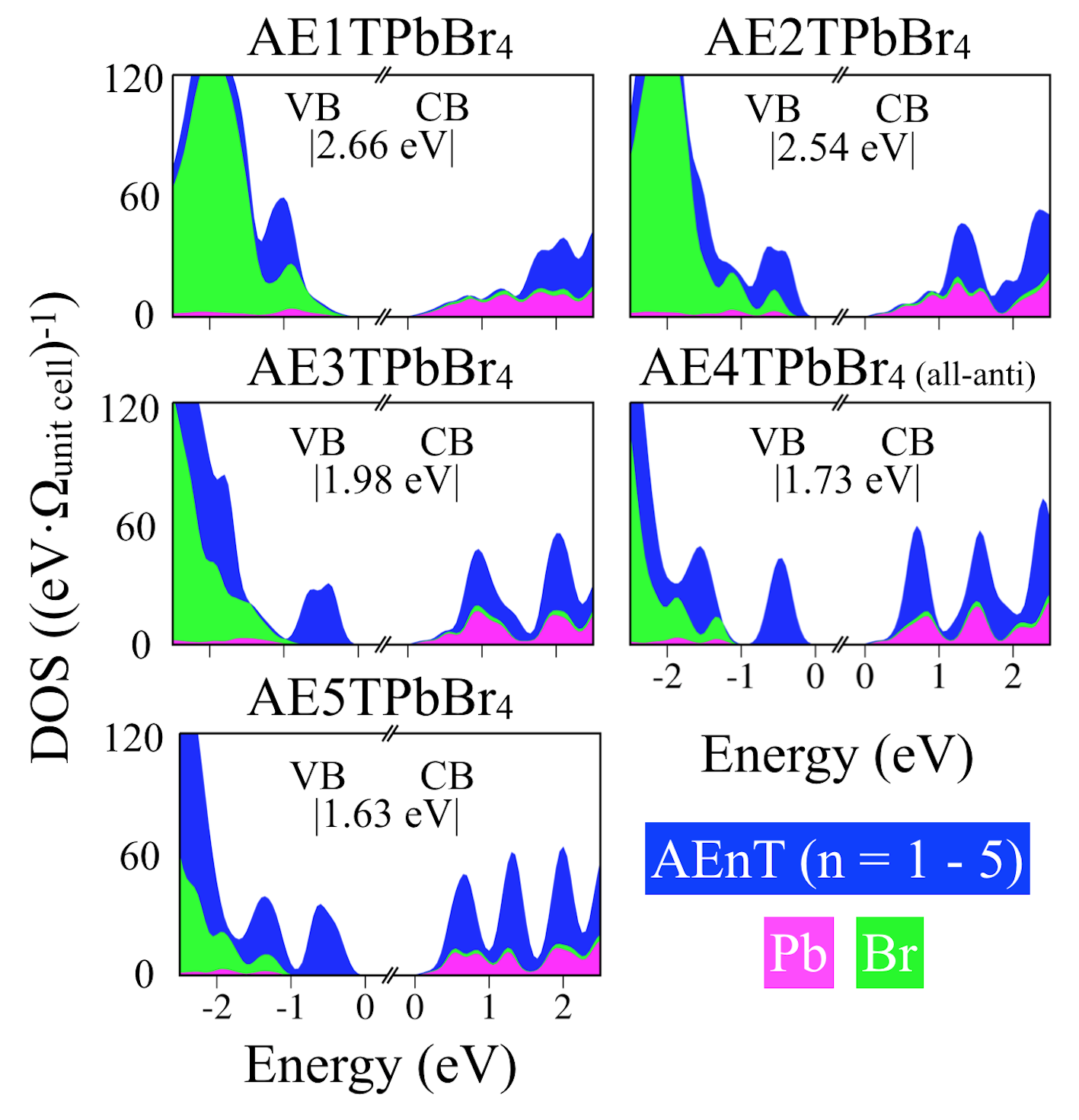}
	\caption{Total densities of states (DOS) and element-projected DOS based on the DFT-HSE06+SOC for the HOIPs AEnT\ce{PbBr4} (n = 1, 2, 3, 4, 5) with all-anti AEnT molecules. See caption of Fig. \ref{fig:dosInorg} for details.}
\label{fig:dosOrg}
\end{figure}

\clearpage

%